\documentclass[a4paper,12pt]{article}
\usepackage{latexsym}
\usepackage{amssymb,amsmath,amsthm}
\usepackage[pdftex]{graphicx,color}

\usepackage[implicit=false,
	colorlinks=true,
	urlcolor=black,
	linkcolor=green
]{hyperref}

\newcommand{\beq}{\begin{eqnarray*}}
\newcommand{\beqn}{\begin{eqnarray}}
\newcommand{\eeq}{\end{eqnarray*}}
\newcommand{\eeqn}{\end{eqnarray}}
\newcommand{\bmat}{\begin{pmatrix}}
\newcommand{\emat}{\end{pmatrix}}

\newcommand{\lbeq}[1]{\label{eq:#1}}
\newcommand{\req}[1]{\ref{eq:#1}}
\newcommand{\nn}{\nonumber}

\newcommand{\ds}{\displaystyle}
\newcommand{\ts}{\textstyle}
\newcommand{\la}{\langle}
\newcommand{\ra}{\rangle}
\newcommand{\pt}{\partial}
\newcommand{\pro}{\mathop\Pi}

\newcommand{\vep}{\varepsilon}
\newcommand{\eps}{\epsilon}
\newcommand{\vp}{\varphi}

\newcommand{\E}{{\sf E}}

\newcommand{\V}{{\sf V}}
\newcommand{\R}{{\mathbb R}}
\newcommand{\C}{{\mathbb C}}

\newcommand{\cL}{{\cal L}}

\newcommand{\cF}{{\cal F}}
\newcommand{\rint}{{\rm int}}

\newcommand{\Id}{{I\!d}}

\newcommand{\ti}{\tilde}
\newcommand{\tu}{{\tilde u}}

\newcommand{\lam}{{\lambda}}
\newcommand{\blam}{{\bar\lambda}}
\newcommand{\om}{{\omega}}

\newcommand{\bv}{{\bar v}}
\newcommand{\bw}{{\bar w}}
\newcommand{\bq}{{\bar q}}

\newcommand{\bb}{{\bar b}}

\newcommand{\bz}{{\bar z}}
\newcommand{\bU}{{\bar U}}

\newcommand{\G}{{|\Gamma|}}

\newcommand{\re}{{\rm Re}}
\newcommand{\im}{{\rm Im}}
\newcommand{\pp}{{{\phantom{+}}}}

\newcommand{\pI}{\phantom{\displaystyle\int_I^I}}

\newcommand{\pM}{\phantom{mmm}}
\newcommand{\pS}{\phantom{\displaystyle\sum}}

\begin{document}

\pagenumbering{gobble}

$$\phantom{mm}$$
\vskip 3.0cm

\centerline{\LARGE The Dynamics of the Hubbard Model }  

\bigskip
\centerline{\large  through}  

\bigskip
\centerline{\LARGE  Stochastic Calculus} 

\bigskip
\centerline{\large  and}  

\bigskip
\centerline{\LARGE   Girsanov Transformation } 

\bigskip
\bigskip
\bigskip
\bigskip
\bigskip
\centerline{Detlef Lehmann, Hochschule RheinMain, Faculty of Engineering } 
\centerline{Postfach 3251, 65022 Wiesbaden, Germany}

\medskip

\centerline{\footnotesize detlef.lehmann@hs-rm.de}
\centerline{\footnotesize \url{https://www.hs-rm.de/de/hochschule/personen/lehmann-detlef/}}

\vskip 3.0cm
\noindent{\bf Abstract:} As a typical quantum many body problem, we consider the time evolution of 
density matrix elements in the Bose-Hubbard 
model. For an arbitrary initial state, these quantities can be obtained from an SDE or stochastic differential equation system. 
To this SDE system, a Girsanov transformation can be applied. 
This has the effect that all the information from the initial state moves into the drift part, 
into the mean field part, of the transformed system. In the large $N$ limit with $g=UN$ fixed, the diffusive part of the  
trans\-formed system vanishes and as a result, the exact quantum dynamics is given by an ODE system which turns out to be 
the time dependent discrete Gross Pitaevskii equation. For the two site Bose-Hubbard model, the GP equation reduces to the 
mathematical pendulum and the difference of expected number of particles at the two lattice sites is equal to the velocity of that pendulum 
which is either oscillatory or it can have rollovers which then corresponds to the self trapping or insulating phase. As a by-product, we also 
find an equivalence of the mathematical pendulum with a quartic double well potential. Collapse and revivals 
are a more subtle phenomenom, in order to see these the diffusive part of the SDE system or quantum corrections have to be taken into account. 
This can be done with an approximation and collapse and revivals can be reproduced, numerically and also through an analytic calculation.  
Since expectation values of Fresnel or Wiener diffusion processes, we write the density matrix elements exactly in this way, can be obtained from 
parabolic second order PDEs, we also obtain various exact PDE representations. The paper has been written 
with the goal to come up with an efficient calculation scheme for quantum many body systems and as such the formalism is generic 
and applies to arbitrary dimension, arbitrary hopping matrices and, with suitable adjustments, to fermionic models.  

\newpage

$$\phantom{.}$$
\noindent{\bf Content and Outline: } 

{\footnotesize

\bigskip
\begin{itemize}
\item[{ 1.}] General Setup and SDE Representation \hfill 1
\begin{itemize}  
\item[{ 1.1}] Time Evolution of States as Fresnel Expectation Values  \hfill 2 \\
  {\tiny $\phantom{m}$ Theorem 1 \;(introduces time evolution matrix $U_t$ and its SDE) \hfill 4 }
\item[{ 1.2}] Density Matrix Elements from Stochastic Differential Equations  \hfill 7\\
  {\tiny $\phantom{m}$ Theorem 2 \;(writes density matrix elements as expectation values of diffusion processes) \hfill 10 }
\end{itemize}
\item[{ 2.}] Martingale Property and Girsanov Transformation \hfill 11 
\begin{itemize}
\item[{ 2.1}] Unitary Time Evolution as a Martingale \hfill 11 \\ 
  {\tiny $\phantom{m}$ Theorem 3 \;(shows $f(U_{x,t}U_{y,t}^+)$ is a martingale for arbitrary $f$) \hfill 13 }
\item[{ 2.2}] Girsanov Transformed SDE System for the Density Matrix Elements \hfill 14 \\ 
  {\tiny $\phantom{m}$ Theorem 4 \;(main theorem, large $N$ limit can be read off and gives GP equation) \hfill 17 }
\item[{ 2.3}] Number States \hfill 18 \\ 
  {\tiny $\phantom{m}$ Theorem 5 \;(analog of Theorem 4 for number states, Theorem 4 has coherent states) \hfill 21 }
\end{itemize}

\item[{ 3.}] PDE Representations \hfill 22
\begin{itemize}
\item[{ 3.1}] Untransformed Case, before Girsanov Transformation \hfill 22
\item[{ 3.2}] Transformed Case, after Girsanov Transformation \hfill 24
\item[{ 3.3}] The PDE Version of the Girsanov Transformation Formula \hfill 26 \\ 
  {\tiny $\phantom{m}$ Theorem 6 \;(summary PDE representations for density matrix elements) \hfill 27 }
\end{itemize}

\item[{ 4.}] An Explicit Solvable Test Case: The 0D Bose-Hubbard Model $\phantom{\ds\int}$ \hfill 28 \\ 
  {\tiny $\phantom{mmmmm}$ Checking the formalism and providing some intuition for the Fresnel measure\\
         $\phantom{mmmmm}$ and the mathematical mechanism which leads to collapse and revivals  }

\smallskip
\item[{ 5.}] The Two Site Bose-Hubbard Model  \hfill 34
\begin{itemize}
\item[{ 5.1}] Large $N$ Limit   \hfill 37 \\
  {\tiny $\phantom{m}$ Numerical Results \;(ODE versus exact diagonalization) \hfill 39,40 }\\ 
  {\tiny $\phantom{m}$ Theorem 7 \;(equivalence of mathematical pendulum and quartic double well potential) \hfill 41 }
\item[{ 5.2}] PDE Representations \hfill 42  \\
  {\tiny $\phantom{m}$ Theorem 8 \;(another PDE representation different from those of chapter 3, using the quadratic quantities as variables) \hfill 44 }
\item[{ 5.3}] Collapse and Revivals \hfill 48 \\
  {\tiny $\phantom{m}$ Numerical Results \;(analytical versus exact diagonalization) \hfill 51,52 }\\
  {\tiny $\phantom{m}$ Numerical Results \;(ODE versus exact diagonalization) \hfill 53 }
\end{itemize}

{\tiny $\phantom{}$Theorems 1 - 8 \,summarize the material of their sections }

\item[{ 6.}] Summary \hfill 54

\smallskip
\item[{ 7.}] Additional Remarks \hfill 54

\end{itemize}

\bigskip
\bigskip
\noindent{ Appendix:} Compact Summary Stochastic Calculus  \hfill 56
\begin{itemize}

\item[] { A.1}\; Standard Brownian Motion and Wiener Measure  \hfill 56
\item[] { A.2}\; Ito Formula and Stochastic Integrals  \hfill 57
\item[] { A.3}\; Kolmogorov Backward Equation and Feynman-Kac Formula \hfill 60
\item[] { A.4}\; Stochastic Calculus with Respect to Fresnel Measure \hfill 62

\end{itemize}

\bigskip
\bigskip
\bigskip
\noindent{References}  \hfill 67

}

\newpage

\pagenumbering{arabic}

\noindent{\large \bf 1. General Setup and SDE Representation } 
\numberwithin{equation}{section}
\renewcommand\thesection{1}

\bigskip
\bigskip
\bigskip
We consider the $d$-dimensional Bose-Hubbard model with Hamiltonian 
\medskip
\beqn
H&=& -J\,\sum_{\la i,j\ra} \,a_i^+a_j \;+\; {U\over 2} \,\sum_{j} a_j^+a_j^+ a_ja_j \;+\;\sum_{j} \eps_j\,\,a_j^+a_j \pI  \lbeq{1.1}
\eeqn
with bosonic annihilation and creation operators $a_j,a_j^+$ satisfying the commutation relations 
\smallskip
\beqn
[\,a_i\,,a_j^+]&=& \delta_{i,j} \pM  \lbeq{1.2}
\eeqn
As usual, $\la i,j\ra$ denotes the sum over nearest neighbors and to be specific, we choose a cubic lattice $\Gamma$ given by 
\beqn
 j\,=\,(j_1,\cdots,j_d)\;\in\;\{1,2,\cdots,L\}^d\;=:\;\Gamma \lbeq{1.3}
\eeqn
We find it convenient to work in the Bargmann-Segal representation [1] where the $a_j,a_j^+$ are realized through the operators
\smallskip
\beqn
a_j\;=\;\ts {\pt\over \pt z_j} \;,\;\;\;\; a_j^+\;=\; z_j \phantom{mm} \lbeq{1.4}
\eeqn
which act on the Hilbert space of analytic functions of $\G=L^d$ complex variables 
\beqn
\cF&:=&\Bigl\{\; f=f(\,\{z_j\}\,)\,:\,\mathbb C^\G\to \mathbb C\;\;{\rm analytic}\;\;\Bigr|\;\;\;\|f\|^2_\cF=(f,f)_{\cF}\,<\,\infty\; \Bigr\} \phantom{mm} \pI \lbeq{1.5}
\eeqn
with scalar product 
\beqn
(f,g)_\cF&:=&\ts \int_{\mathbb C^\G=\mathbb R^{2\G}}\, f(z)\,\overline{g(z)}\; d\mu(z) \pI \lbeq{1.6}\\
d\mu(z)&:=&\ts\prod_{j} \, e^{-|z_j|^2}\;\ts {d\re z_j \,d\im z_j \over \pi } \pI \lbeq{1.7}
\eeqn
In the following, sums $\,\Sigma_j\,$ or $\,\Sigma_{i,j}\,$ or products $\,\pro_j\,$ are always meant to be sums and products over all lattice sites  
if not specified otherwise. That is, we use the notation 
\beqn
\ts \sum_j\cdots &:=&\ts \sum_{j\in\Gamma} \cdots \pI  \lbeq{1.8}\\ 
\ts \prod_j\cdots &:=&\ts \prod_{j\in\Gamma} \cdots \lbeq{1.9}
\eeqn

\smallskip
and $\,\Sigma_{i,j}\,:=\,\Sigma_{i,j\in\Gamma}\,$. Actually we can allow for a general hopping matrix which should be real and symmetric,  
\beqn
\vep&:=&(\,\vep_{ij}\,)\;\;\in\;\;\R^{\G\times\G} \pM  \lbeq{1.10}
\eeqn
with $\,\vep_{ij}=\vep_{ji}\,$. With that, the final Hamiltonian, we use a small $h$ instead of a capital $H$, reads 
\beqn
h&=& h_0\,+\,h_\rint  \pI \phantom{nm}  \lbeq{1.12}
\eeqn
with a quadratic part 
\beqn
h_0&=&\ts \sum_{i,j}{\ts \,\vep_{ij}\, z_i\,{\pt\over \pt z_j} } \pI  \lbeq{1.13}
\eeqn
and a quartic part
\beqn
h_\rint&=&\ts u\,\sum_j \,\ts z_j^2\,{\pt^2\over \pt z_j^2} \pI \lbeq{1.14}
\eeqn
where we substituted the capital $U$ by a small $u$ according to (the capital $U$'s we use later for a unitary evolution matrix)
\beqn
u&:=&{U\over 2} \pM  \lbeq{1.15}
\eeqn
For a nearest neighbor hopping $J$ and trapping potentials $\eps_j$ as in (\req{1.1}) we have 
\medskip
\beqn
\vep_{ij}&=&\begin{cases}  -J &\;{\rm if}\;\;|i-j|=1 \\ 
                      +\eps_j &\;{\rm if}\;\;i=j \\ 
                       \phantom{+}0  &\;{\rm otherwise\;.}  \end{cases}   \lbeq{1.16}
\eeqn

\bigskip
\bigskip
\noindent{\bf 1.1\; Time Evolution of States as Fresnel Expectation Values}

\bigskip
\bigskip
\noindent 
The time evolution $\;e^{-ith}=e^{-it(h_0+h_\rint)}\;$ we calculate through the Trotter formula. That is, we discretize time
\beqn
t\;\;=\;\;t_k\;\;=\;\;k\, dt \lbeq{1.17}
\eeqn
and write 
\beqn
e^{-ith}\;\;=\;\;e^{-i\,kdt\,(h_0+h_\rint)}  &\approx& \bigl(\, e^{-i\,dt\,h_\rint}\; e^{-i\,dt\,h_0} \,\bigr)^k  \pI \lbeq{1.18}
\eeqn

\medskip
where the approximate equality becomes exact in the limit $dt\to 0$ which we implicitely assume from now on. 
The action of $e^{-i\,dt\,h_0}$ is given by 
\medskip
\beqn
(e^{-i\,dt\,h_0}f)(z)&=&f\bigl( e^{-i\, dt\, \vep}z\bigr) \pI \lbeq{1.19}
\eeqn

\medskip
The action of $e^{-i\,dt\,h_\rint}$ we write as follows: First, on monomials $\pro_jz_j^{n_j}$ we have 
\medskip
\beqn
e^{-i\,dt\,h_\rint}\,\pro_j z_j^{n_j}&=&e^{iudt\,\sum_jn_j}\, e^{-iudt\,\sum_jn_j^2}\; \pro_j z_j^{n_j}  \pI \lbeq{1.20}
\eeqn

\medskip
Recall the Fresnel integral 
\smallskip
\beqn
\int_{\mathbb R} e^{-i\lambda \phi} \, e^{i{\phi^2\over 2}} \,\ts{d\phi\over \sqrt{2\pi i}} &=& e^{-i{\lambda^2\over 2}}  \lbeq{1.21}
\eeqn

\smallskip
with $\,\sqrt{i}=e^{i{\pi\over 4}}\,$. We can write
\medskip 
\beqn
e^{-i\,dt\,h_\rint}\,\pro_j z_j^{n_j}&=&
 \int_{\mathbb R^\G} \,\pro_j\, \bigl(\, e^{iudt\,-\,i\sqrt{2udt}\, \phi_j} z_j \,\bigr)^{n_j} \,
   \pro_j \,  e^{i{\phi_j^2\over 2}}  \,\ts {d\phi_j \over \sqrt{2\pi i}}   \phantom{mm} \pI \lbeq{1.22}  
\eeqn

\goodbreak
Thus, for any analytic $f$ (by slight abuse of notation, we temporarily label the lattice sites with natural numbers 
from 1 to $\G$ in the middle of the next line)
\beqn
f(z)\;=\;f(\,\{z_j\}\,)&=&\sum_{n_1,...,n_\G=0}^\infty \,c_{n_1\cdots n_\G}\,z_1^{n_1}\cdots z_\G^{n_\G} 
\;\; \equiv \;\; \sum_{\{n_j\}} \, c_{\{n_j\}}\; \pro_j z_j^{n_j} \lbeq{1.23}   \pI
\eeqn
we have
\beqn
(e^{-i\,dt\,h_\rint}f)(z)&=&\sum_{\{n_j\}} \, c_{\{n_j\}}\; e^{-i\,dt\,h_\rint}\pro_j z_j^{n_j} \nn   \pI\\
&&\nn \\
&=&\sum_{\{n_j\}} \, c_{\{n_j\}}\;  \int_{\mathbb R^\G} \,\pro_j\, \bigl(\, e^{iudt\,-\,i\sqrt{2udt}\, \phi_j} z_j \,\bigr)^{n_j} \,
   \pro_j \,  e^{i{\phi_j^2\over 2}}  \,\ts {d\phi_j \over \sqrt{2\pi i}}   \phantom{mm} \pI \nn \\
&&\nn \\ 
&=&\int_{\mathbb R^\G} f\bigl( \, \bigl\{\, e^{iudt\,-\,i\sqrt{2udt}\, \phi_j} z_j \,\bigr\}_{j\in\Gamma}\, \bigr) \;
  \pro_j   e^{i{\phi_j^2\over 2}}  \ts {d\phi_j \over \sqrt{2\pi i}}  \pI  \nn  \\ 
&&\nn \\ 
&=:&\int_{\mathbb R^\G} f( e^{-i\,dD_{\phi}} z )\;\pro_j   e^{i{\phi_j^2\over 2}}  \ts {d\phi_j \over \sqrt{2\pi i}}  \pI  \lbeq{1.24}\\  \nn
\eeqn
with the $\G\times \G$ diagonal matrix 
\beqn
dD_{\phi}&:=&{\rm diag}\bigl( \, \{\, \sqrt{2udt} \,\phi_{j}-udt\, \}_{j\in\Gamma} \,\bigr) \;\;\in\;\;\C^{\G\times \G}  \pI  \lbeq{1.25}\\ \nn
\eeqn
Thus, a single Trotter step is given by 
\beqn
(e^{-i\,dt\,h}f)(z)&=&\bigl[ e^{-i\,dt\,h_\rint}\, (e^{-i\,dt\,h_0}f)\bigr](z)  \pI  \nn\\
&&\nn \\
&=& \int_{\mathbb R^\G} (e^{-i\,dt\,h_0}f)( e^{-i \,dD_{\phi}} z ) \;\pro_j   e^{i{\phi_j^2\over 2}}  \ts {d\phi_j \over \sqrt{2\pi i}}   \pI \nn \\
&&\nn \\
&=&\int_{\mathbb R^\G} f( e^{-i dt \,\vep} e^{-i\,dD_{\phi}} z ) \,\; e^{\,i\sum_j{\phi_j^2\over 2}} 
    \pro_j\ts {d\phi_j \over \sqrt{2\pi i}}  \pI \nn \\ 
&&\nn \\
&=:&\int_{\mathbb R^\G} f( e^{-i dt \,\vep} e^{-i\,dD_{\phi}} z ) \,\; e^{\,i{\phi^2\over 2}} \ts {d^\G\phi \over (2\pi i)^{\G/2}}  \pI  \lbeq{1.26} \\  \nn
\eeqn
Iterating, 
\beqn
(e^{-i\,kdt\,h}f)(z)&=&\int_{\mathbb R^{k\G}} f\bigl(\, e^{-idt \,\vep} e^{-i\,dD_{\phi_{1}} }\cdots e^{-i dt \,\vep} e^{-i\,dD_{\phi_k} } \,z\, \bigr) 
    \,\; \pro_{\ell=1}^k e^{\,i{\phi_\ell^2\over 2}} \ts {d^\G\phi_\ell \over (2\pi i)^{\G/2}}  \pI \phantom{mmm}  \lbeq{1.27} \\  \nn
\eeqn
with the notations 
\beqn
\phi_\ell&:=&(\,\{\phi_{j,\ell}\}_{j\in\Gamma}\,)\;\;\in\;\;\mathbb R^\G  \pI \nn \\  
\phi_\ell^2&:=&\ts \sum_j\phi_{j,\ell}^2 \;\;\in\;\;\mathbb R \lbeq{1.28} \\ 
d^\G\phi_\ell&:=&\pro_j d\phi_{j,\ell } \nn \pI
\eeqn
That is, on the (real, scalar) integration variable $\phi_{j,\ell}\in\R$, the first index $j=(j_1,\cdots,j_d)$ is a lattice site index 
and the second index $\ell\in\mathbb N$ is a time index. Furthermore 
\beqn
dD_{\phi_\ell}&=&{\rm diag}\Bigl( \; \bigl\{\, \sqrt{2udt} \,\phi_{j,\ell}\,-\,udt\,\bigr\}_{j\in\Gamma} \; \Bigr) \;\;\in\;\;\mathbb R^{\G\times \G} \pI \lbeq{1.29}
\eeqn

\medskip
\noindent Let us summarize the above formula in part (a) of the following

\goodbreak
\bigskip
\bigskip
\bigskip
\noindent{\bf Theorem 1:} Let $\;h=h_0+h_\rint:\cF\to \cF\;$ be the Bose-Hubbard Hamiltonian given by (\req{1.13}) and (\req{1.14}) and let $\psi\in\cF$ be any initial state. 
Then: 

\bigskip
\begin{itemize}
\item[{\bf a)}] In the limit $dt\to 0$ with $t=t_k=kdt\,$ fixed, there is the formula 
\beqn
(e^{-ikdt\,h}\psi)(z)&=&\E\bigl[ \, \psi( U_{kdt} z )\,\bigr] \pI \lbeq{1.30}
\eeqn
with unitary evolution matrix
\beqn
U_{kdt}&:=& e^{-i dt \,\vep}\, e^{-i\,dD_{\phi_{1}} }\;\cdots\; e^{-i dt \,\vep}\, e^{-i\,dD_{\phi_k} }\;\; \in\;\; \C^{\G\times\G} \pI  \lbeq{1.31}
\eeqn
and Fresnel expectation value 
\beqn
\E\bigl[ \,\; \cdot\;\,\bigr]&:=&\int_{\mathbb R^{k\G}} 
 \;\cdot \;\; \pro_{\ell=1}^k e^{i{\phi_\ell^2\over 2} } \ts {d^{\G}\phi_\ell \over (2\pi i)^{\G/2}}  \pI  \lbeq{1.32}\\  \nn
\eeqn
\item[{\bf b)}] The evolution matrix $U_{t_k}=U_{kdt}$ is a solution of the following stochastic differential equation (SDE) 
\beqn
dU_{t_k}\;\;:=\;\;U_{t_k}-U_{t_{k-1}}&=& -\,i\,U_{t_{k-1}}\,\bigl(\,\vep\, dt\;+\;\sqrt{2u} \;dx_{t_k}\,\bigr) \phantom{mmmm}\pI \lbeq{1.33}
\eeqn
where $dx_{t_k}$ is the diagonal matrix of Fresnel Brownian motions given by (\req{1.45}) below.
\end{itemize}

\bigskip
\bigskip
\bigskip
\noindent Let's consider part (b) of the theorem. Suppose we would have Gaussian densities, let's say 1-dimensional, 
$\phantom{\bigl|} e^{-{\phi_\ell^2/ 2}} \;{d\phi_\ell / \sqrt{2\pi}}\;$ instead 
of Fresnel kernels $\; e^{\,i\,{\phi_\ell^2/ 2}} \;{d\phi_\ell / \sqrt{2\pi i}}\;$. Then the combination of integration variables 
\beqn
x_{t_k}&:=&\ts \sqrt{dt}\,\sum_{\ell=1}^k \phi_\ell  \pI \lbeq{1.34}
\eeqn
would be a standard Brownian motion, the product (with $x_{t_0}:=0$)
\medskip
\beqn
dW&:=&\pro_{\ell=1}^k e^{\,-{\phi_\ell^2\over 2} }\; {\ts {d\phi_\ell \over \sqrt{2\pi} }} \;\;=\;\; 
  \pro_{\ell=1}^k e^{\,-\,{(x_{t_\ell}-x_{t_{\ell-1}})^2\over {2dt}} }\; {\ts {dx_{t_\ell} \over \sqrt{2\pi \,dt} }} \;\;=:\;\; 
  \pro_{\ell=1}^k \,p_{dt}(x_{t_\ell},x_{t_{\ell-1}}) \;dx_{t_\ell}\pI  \phantom{mm} \lbeq{1.35}
\eeqn

\smallskip
would be standard Wiener measure and in the limit $dt\to 0$ there would be the standard Brownian motion calculation rule (appendix A.1 has a quick reminder)
\beqn
(dx_t)^2&=&dt \pI \lbeq{1.36}
\eeqn
and $\,dx_t\,dt\,=\,dt\,dt\,=\,0\,$. In the presence of Fresnel kernels, we can make the same definitons. That is, the combination of integration variables 
\beqn
x_{t_k}&:=&\ts\sqrt{dt}\,\sum_{\ell=1}^k \phi_\ell  \pI \lbeq{1.37}
\eeqn
we call a one-dimensional Fresnel Brownian motion if the $\phi_\ell$'s are to be integrated against the product of one-dimensional Fresnel kernels (again with $x_{t_0}:=0\,$)
\medskip
\beqn
dF&:=&\pro_{\ell=1}^k e^{\,i{\phi_\ell^2\over 2} } \; {\ts {d\phi_\ell \over \sqrt{2\pi i} }}  \;\;=\;\; 
  \pro_{\ell=1}^k e^{\,i\,{(x_{t_\ell}-x_{t_{\ell-1}})^2\over {2dt}} }\; {\ts {dx_{t_\ell} \over \sqrt{2\pi i \,dt} }}  \;\;=:\;\; 
  \pro_{\ell=1}^k \,q_{dt}(x_{t_\ell},x_{t_{\ell-1}}) \;dx_{t_\ell} \pI \phantom{mm}  \lbeq{1.38}
\eeqn

\smallskip
which we then refer to as Fresnel measure. Observe that for both Fresnel and Gaussian kernels we have the equations 
\beqn
\ts\int_{\mathbb R} k_t(x,y)\, k_s(y,z)\, dy &=& k_{t+s}(x,z)  \pI  \lbeq{1.39} \\ 
\ts\int_{\mathbb R} k_t(x,y)\, dy &=& 1  \lbeq{1.40}
\eeqn

\smallskip
with $k_t\in\{p_t,q_t\}$.  Now, what we can 
use is the fact that in the limit $dt\to 0$ there are analog calculation rules for Fresnel Brownian motions. That is, there are the 
following formulae (see appendix A.1 and A.4 for more background) 
\beqn
(dx_t)^2&=&i\,dt \pI  \lbeq{1.41}
\eeqn
and $\,dx_t\,dt\,=\,dt\,dt\,=\,0\,$.

\bigskip
\noindent Now let's return to the time evolution formula (\req{1.30}) of Theorem 1. Instead of 1-dimensional Fresnel kernels we have $\G$-dimensional Fresnel kernels 
and accordingly $\G$-dimensional Fresnel Brownian motions
\beqn
\phantom{mm} x_{j,t_k}\;\;=\;\; x_{j,kdt}&:=&\ts \sqrt{dt}\,\sum_{\ell=1}^k \phi_{j,\ell}  \pI \lbeq{1.42}
\eeqn
where $j\in\Gamma$ again denotes some lattice site. With 
\beqn
dx_{j,kdt}&:=&x_{j,kdt}\,-\,x_{j,(k-1)dt}\;\;=\;\;\sqrt{dt}\,\phi_{j,k}  \pI \lbeq{1.43}
\eeqn
we can write 
\beqn
dD_{\phi_k}&=&{\rm diag}\bigl( \; \{\,\sqrt{2u} \,dx_{j,kdt}\,-\,u\,dt\,\}_{j\in\Gamma} \;\bigr) \pI \nn \\ 
&=:& \sqrt{2u} \;dx_{kdt}\,\;-\,\;u\,dt\,\Id \pI  \lbeq{1.44}
\eeqn
where we introduced the diagonal matrix of Fresnel Brownian motions
\beqn
dx_{kdt}&:=&{\rm diag}\bigl(\, \{\, dx_{j,kdt} \,\}_{j\in\Gamma}\,\bigr) \;\;\in\;\; \C^{\G\times\G}  \pI \lbeq{1.45}
\eeqn
and $\,\Id\,$ is the $\G\times\G$ identity matrix. From the calculation rule (\req{1.41}) we get the matrix equation  
\beqn
(dx_{kdt})^2&=&{\rm diag}\bigl(\; \bigl\{\, (dx_{j,kdt})^2 \, \bigr\}_{j\in\Gamma} \; \bigr) \;\;=\;\; i\,dt\,\Id  \pI \lbeq{1.46}
\eeqn
Thus, up to terms $\;O\bigl(\,dt^{3/2}\,\bigr)\,$, 
\beqn
e^{-i\,dD_{\phi_k}}&=&1\;-\;i\,dD_{\phi_k}\;-\;\ts {1\over 2}\,(dD_{\phi_k})^2 \pI \nn\\ 
&=&1\;-\;i\sqrt{2u} \,dx_{kdt}\;+\;i\,u\,dt\,\Id\;-\;\ts {1\over 2}\,\bigl( \sqrt{2u} \,dx_{kdt}\;-\;u\,dt\,\Id \bigr)^2 \nn\\ 
&=&1\;-\;i\sqrt{2u} \,dx_{kdt}\;+\;i\,u\,dt\,\Id\;-\;\ts {1\over 2}\,\bigl( \sqrt{2u} \,dx_{kdt} \bigr)^2 \pI \nn\\ 
&=&1\;-\;i\sqrt{2u} \,dx_{kdt}  \lbeq{1.47}
\eeqn

\medskip
and we arrive at the following SDE for the evolution matrix $U_t$:
\smallskip
\beqn
U_{kdt}&=& e^{-i dt \,\vep} e^{-i\,dD_{\phi_{1}} }\;\times\;\cdots\;\times \;e^{-i dt \,\vep} e^{-i\,dD_{\phi_{k-1}} }\;\times\; 
    e^{-i dt \,\vep} e^{-i\,dD_{\phi_k} } \pI \nn \\ 
&=& U_{(k-1)dt} \;\times\;  e^{-i dt \,\vep} e^{-i\,dD_{\phi_k} } \nn  \\ 
&\buildrel (\req{1.47})\over=& U_{(k-1)dt} \;\bigl(\, 1\;-\;i dt\,\vep\,\bigr)\;\bigl(\,1\;-\;i\sqrt{2u} \;dx_{kdt}\,\bigr)\;  \pI \nn \\
&=& U_{(k-1)dt} \;\bigl(\, 1\;-\;i dt\,\vep\;-\;i\sqrt{2u} \;dx_{kdt}\,\bigr)\;  \lbeq{1.48}
\eeqn

\medskip
or, with $t_k=kdt$, 
\beqn
dU_{t_k}\;\;:=\;\;U_{t_k}-U_{t_{k-1}}&=& -i\,U_{t_{k-1}}\,\bigl(\, dt\,\vep\;+\;\sqrt{2u} \;dx_{t_k}\,\bigr) \phantom{mmmm}\pI \lbeq{1.49}
\eeqn
which completes the derivation of part (b) of Theorem 1. More compactly, this could be written as 
\beqn
dU_t&=& -i\,U_t\,\bigl(\,\vep\, dt\;+\;\sqrt{2u} \;dx_t\,\bigr) \pI \lbeq{1.50}
\eeqn
but we want to remind at this place that when discretizing stochastic differential equations or stochastic integrals it is usually crucial whether a particular 
time index is a $t_k$ or a $t_{k-1}$. Throughout this paper, we use the Ito definition which is as follows: If some quantity $Q_t$ satisfies the stochastic differential equation 
\beqn
dQ_t&=&A_t\,dt \;+\;B_t\,dx_t \pI  \lbeq{1.51}
\eeqn
then this is equivalent to the following discrete time update rule 
\beqn
\lefteqn{ 
Q_{t_k}(\phi_1,\cdots,\phi_k)\;\;=\;\;  \pI }  \lbeq{1.52} \\ 
&&Q_{t_{k-1}}(\phi_1,\cdots,\phi_{k-1}) \;+\;A_{t_{k-1}}(\phi_1,\cdots,\phi_{k-1})\,dt \;+\;B_{t_{k-1}}(\phi_1,\cdots,\phi_{k-1})\,\sqrt{dt}\;\phi_k \nn
\eeqn
That is, the new random number or integration variable $\phi_k$ which enters when going from time $t_{k-1}$ to $t_k$ enters in an explicitely given way, namely 
through the explicit $\phi_k$ on the very right of (\req{1.52}). There are no $\phi_k$'s in the $A$, $B$ or $Q_{t_{k-1}}$ on the right hand side of (\req{1.52}). This then 
for example has the immediate consequence that  
\beqn
\E[\,Q_{t_k}\,]&=&\E[\,Q_{t_{k-1}}\,]\;+\;\E[\,A_{t_{k-1}}\,]\,dt  \pM \lbeq{1.53}
\eeqn
since the diffusive part does not contribute to the expectation value because of $\,\E[\phi_k]=0\,$. Thus, quantities $Q_t$ which have a vanishing drift part 
$A=0$ are of special importance since their expectation value does not change over time and they are called martingales. Appendix A.2 has a quick reminder 
on Ito and Stratonovich integrals and why there are different definitions at all. 

\goodbreak

\bigskip
\bigskip
\bigskip
\noindent{\bf 1.2\; Density Matrix Elements from Stochastic Differential Equations  }

\nobreak
\bigskip
\bigskip
We consider the following normalized initial state with $\lam=\{\lam_j\}\in\mathbb C^\G$
\beqn
\psi_0(z)\;=\;\psi_0(\,\{z_j\}\,)&:=& \ts\prod_j \,e^{\lam_j z_j}\; e^{-{|\lam_j|^2\over 2}} \;\;=:\;\; e^{\lam z} \;e^{-{|\lam|^2\over 2}} \pI \lbeq{1.54}
\eeqn
This is a product of coherent states. The expected number of particles at site $j$ is 
\beqn
(\,\psi_0\,,a_j^+a^\pp_j \psi_0\,)_\cF&=&|\lam_j|^2 \pI \lbeq{1.55}
\eeqn
which means that $\;|\lam_j|=\sqrt{N_j}\;$ would be a natural choice if we want to have $N_j$ particles at site $j$. The total number 
of particles $N$ is given by
\beqn
N\;\;=\;\;\ts \sum_j N_j &=&\ts\sum_j |\lambda_j|^2\;\;=\;\;|\lambda|^2 \pI \lbeq{1.56}
\eeqn
Let's consider the time evolution of this $\psi_0$. According to Theorem 1, we have  
\beqn
\psi_t(z)\;=\;(e^{-ith}\psi_0)(z)&=&\E[\,\psi_0(U_tz)\,] \pI \lbeq{1.57}
\eeqn
where the evolution matrix $U_t$ is given by the SDE
\beqn
dU_t&=& -\,i\,U_t\,\bigl(\, dt\,\vep\;+\;\sqrt{2u} \;dx_t\,\bigr) \pI \lbeq{1.58}
\eeqn
with initial condition $\,U_0=\Id\,$ and $x_t$ being a Fresnel Brownian motion. We want to calculate the time evolution of the density matrix elements 
\beqn
(\,\psi_t\,,a_i^+a^\pp_j \,\psi_t\,)_\cF &=& (\,\psi_t\,,[a^\pp_j a_i^+ -\delta_{i,j}] \,\psi_t\,)_\cF \pI \nn \\ 
&=& (\,a_j^+\psi_t\,, a^+_i \psi_t\,)_\cF \;-\; \delta_{i,j}\,(\,\psi_t\,, \psi_t\,)_\cF \nn \\ 
&=&\ts \int_{\mathbb C^\G} z_j \bz_i\; |\psi_t(z)|^2 \,d\mu(z) \;\;-\;\;\delta_{i,j} \pI  \lbeq{1.59}
\eeqn

\smallskip
From Theorem 1 we have the representations (with \,$t=t_k=kdt$\,)
\beqn
\psi_t(z)&=&\E[\,\psi_0(U_tz)\,]\;\;=\;\;\ts \int \,\psi_0(U_{x,t}z)\; dF(\{x_t\}) \pI  \lbeq{1.60} \\ 
\overline{\psi_t(z)}&=&\overline{\E[\,\psi_0(U_tz)\,]}\;\;=\;\;\ts \int \,\overline{\psi_0(U_{y,t}z)}\; d\bar F(\{y_t\}) \nn
\eeqn

\medskip
with
\beqn
dF&=& \pro_{\ell=1}^k e^{\,+\,i{\phi_\ell^2\over 2}} {\ts {d^\G\phi_\ell \over (2\pi i)^{\G/2}} } \;\;=\;\;
   \pro_{\ell=1}^k e^{\,+\,i{(x_{t_\ell}-x_{t_{\ell-1}})^2\over 2 dt}} \ts {d^\G x_{t_\ell} \over (2\pi i\,dt)^{\G/2}} \pI \lbeq{1.61} \\ 
d\bar F&=&\pro_{\ell=1}^k e^{\,-\,i{\theta_\ell^2\over 2}} {\ts {d^\G\theta_\ell \over [2\pi (-i)]^{\G/2}} } \;\;=\;\;
   \pro_{\ell=1}^k e^{\,-\,i{(y_{t_\ell}-y_{t_{\ell-1}})^2\over 2 dt}} \ts {d^\G y_{t_\ell} \over [2\pi (-i)dt]^{\G/2}}  \nn
\eeqn

\medskip
where $\,\sqrt{-i}\,:=\,\overline{\sqrt{i}}\,=\,e^{\,-\,i{\pi\over 4}}\,$ and $x_{t_\ell}$ and $y_{t_\ell}$ are $\G$-dimensional Fresnel Brownian motions 
given by
\beqn
\ts x_{t_\ell}&=&\ts \sqrt{dt}\,\sum_{m=1}^\ell \phi_m  \lbeq{1.62} \\
\ts y_{t_\ell}&=&\ts \sqrt{dt}\,\sum_{m=1}^\ell \theta_m \pI  \nn
\eeqn
With that, we can write
\beqn
\lefteqn{
\ts \int_{\mathbb C^\G} z_j \bz_i\; |\psi_t(z)|^2 \,d\mu(z)  \pI } \lbeq{1.63}  \\ 
&=& \ts  \int_{\mathbb C^\G} z_j \bz_i\; \int \,\psi_0(U_{x,t}z)\; dF(\{x_t\})\;   \int \,\overline{\psi_0(U_{y,t}z)}\; d\bar F(\{y_t\}) \;d\mu(z)  \nn \\ 
&=&\ts \int \int \;\Bigl\{\; \int_{\mathbb C^\G} z_j \bz_i\; \psi_0(U_{x,t}z)\;\overline{\psi_0(U_{y,t}z)}\;d\mu(z)\;\Bigr\} \; dF(\{x_t\})\;  d\bar F(\{y_t\})  \pI \phantom{mm}\nn
\eeqn

\smallskip
and in the same way 
\beqn
\|\psi_t\|_\cF^2&=&\ts \int_{\mathbb C^\G}  |\psi_t(z)|^2 \,d\mu(z)  \pI \lbeq{1.64}\\
&=&\ts \int \int \;\Bigl\{\; \int_{\mathbb C^\G}  \psi_0(U_{x,t}z)\;\overline{\psi_0(U_{y,t}z)}\;d\mu(z)\;\Bigr\} \; dF(\{x_t\})\;  d\bar F(\{y_t\})  \nn
\eeqn

\medskip
The wavy brackets above are the expectations over the bosonic Fock space, written in the Bargmann-Segal representation, and can be calculated. Since 
\beqn
\psi_0(U_{x,t}z)\;\overline{\psi_0(U_{y,t}z)}&=& \exp\bigl\{\, U_{x,t}^T\lam\cdot z \,\bigr\}\;\exp\bigl\{\, \bU_{y,t}^T\blam\cdot \bz \,\bigr\}   \; e^{-|\lam|^2} \pI\lbeq{1.65}
\eeqn

with $U^T$ denoting the transpose of $U$, and because of the formulae
\medskip
\beqn
\ts\int_{\mathbb C^\G} e^{\lambda z+\bar\lambda\bar z}\; d\mu(z)&=& e^{\lambda\bar\lambda}  \nn \\ 
\ts\int_{\mathbb C^\G}  z_j\,\bz_i\; e^{\lambda z+\bar\lambda\bar z}\;d\mu(z) &=& 
\ts {\pt \over \pt \lambda_j^{\phantom{|}}} {\pt \over \pt {\bar \lambda_i}^{\phantom{|}} }\;e^{\lambda\bar\lambda}
\;\; = \;\; (\lam_i\,\blam_j+\delta_{i,j})\;e^{\lambda\bar\lambda}   \lbeq{1.66}
\eeqn

\medskip
we obtain the following representations: The norm of $\psi_t$ is given by 
\medskip
\beqn
\|\psi_t\|_\cF^2&=&  \E_x\bar\E_y\Bigl[ \;
  \exp\bigl\{\, U_{x,t}^T\lam \cdot  \bU_{y,t}^T\blam \,\bigr\} \;\Bigr] \; e^{-|\lam|^2}   \pI  \lbeq{1.67}
\eeqn

\smallskip
and density matrix elements can be written as
\smallskip
\beqn
(\,\psi_t\,,a_i^+a^\pp_j \,\psi_t\,)_\cF &=& \E_x\bar\E_y\Bigl[\, \;[U_{x,t}^T\lam]_i\;\, [\bU_{y,t}^T\blam]_j\;
  \exp\bigl\{\, U_{x,t}^T\lam \cdot  \bU_{y,t}^T\blam \,\bigr\} \;\Bigr] \;e^{-|\lam|^2} \pI \lbeq{1.68}
\eeqn

\smallskip
where we used the notation 
\beqn
\E_x\bar\E_y\bigl[\;\cdots\;\bigr]&=&\ts \int\int \;\cdots\; dF(\{x_t\})\; d\bar F(\{y_t\}) \pI \lbeq{1.69}
\eeqn
for the Fresnel expectations. 

\bigskip
\bigskip
\noindent Now recall that 
\beqn
dU_{x,t}&=& -\,i\,U_{x,t}\,\bigl(\, dt\,\vep\;+\;\sqrt{2u} \,dx_{t}\,\bigr) \pS \lbeq{1.70}
\eeqn
or, since $\vep_{ij}=\vep_{ji}\,$ or $\,\vep^T=\vep$,  
\beqn
dU_{x,t}^T&=& -\,i\,\bigl(\, dt\,\vep\;+\;\sqrt{2u} \,dx_{t}\,\bigr)\,U_{x,t}^T \pS \lbeq{1.71}
\eeqn
Thus, 
\beqn
d(U_{x,t}^T\lam)&=& -\,i\,\bigl(\, dt\,\vep\;+\;\sqrt{2u} \,dx_{t}\,\bigr)\,U_{x,t}^T\lam \pS \lbeq{1.72}
\eeqn
and in the same way
\beqn
d(\bU_{y,t}^T\blam)&=& +\,i\,\bigl(\, dt\,\vep\;+\;\sqrt{2u} \,dy_{t}\,\bigr)\,\bU_{y,t}^T\blam \pS \lbeq{1.73}
\eeqn

\medskip
Thus, if we abbreviate the quantities (where the $\bar v$ at this stage is not the complex conjugate of $v$ since it has different integration variables) 
\medskip
\beqn
v\;=\;v_{x,t}&:=&U_{x,t}^T\lam \;\;\in\;\;\mathbb C^\G  \pM \lbeq{1.74} \\ 
\bv\;=\;\bv_{y,t}&:=&\bU_{y,t}^T\blam \;\;\in\;\;\mathbb C^\G \nn
\eeqn
we obtain the SDE system 
\beqn
dv&=&\;-\;i\,\bigl(\, dt\,\vep\;+\;\sqrt{2u} \,dx_{t}\,\bigr)\,v \pM  \lbeq{1.75} \\ 
d\bv&=&\; +\; i\,\bigl(\, dt\,\vep\;+\;\sqrt{2u} \,dy_{t}\,\bigr)\,\bv \nn
\eeqn
with initial conditions 
\beqn
v_0&=&\lam   \lbeq{1.76}\\ 
\bv_0&=&\blam \nn
\eeqn
In coordinates, this reads
\beqn
dv_j&=&\;-\;i\, dt\,(\vep v)_j \;-\;i\,\sqrt{2u}\,v_j\,dx_j    \lbeq{1.77}\\ 
d\bar v_j&=&\;+\;i\, dt\,(\vep\bar v)_j \;+\;i\,\sqrt{2u}\,\bar v_j\,dy_j  \nn 
\eeqn

\medskip
or even more explicitely, making also the times and the Fresnel integration variables $x$ and $y$ explicit at the $v,\bv$,  
\beqn
dv_{j,x,t}&=&\;-\;i\, dt\,\ts\sum_i\vep_{ji}\, v_{i,x,t} \;-\;i\,\sqrt{2u}\,v_{j,x,t}\,dx_{j,t}  \pM   \lbeq{1.78} \\ 
d\bar v_{j,y,t}&=&\;+\;i\, dt\,\ts\sum_i\vep_{ji}\,\bar v_{i,y,t} \;+\;i\,\sqrt{2u}\,\bar v_{j,y,t}\,dy_{j,t}  \nn 
\eeqn

\bigskip
Then, from (\req{1.68}) we obtain the density matrix elements as 
\beqn
(\,\psi_t\,,a_i^+a^\pp_j \,\psi_t\,)_\cF &=&\E_x\bar\E_y\bigl[ \;v_i \,\bv_j\; e^{v\bv} \;\bigr] \;e^{-|\lam|^2}  \pI \lbeq{1.79}
\eeqn
if we use the notation
\beqn
v\bv&:=&\ts\sum_j\,v_j\bar v_j \;\;=\;\; v^T\bv \pI \lbeq{1.80}
\eeqn
Finally, the norm of $\psi_t$ has the representation 
\beqn
\|\psi_t\|^2_\cF &=& \E_x\bar\E_y[ \; e^{v\bv} \;] \;e^{-|\lam|^2}  \pI \lbeq{1.81}
\eeqn

\medskip
We summarize the results in the following

\bigskip
\bigskip
\bigskip
\noindent{\bf Theorem 2:} Let $\psi_0$ be the initial state 
\beqn
\psi_0(z)\;=\;\psi_0(\,\{z_j\}\,)&=& \ts\prod_j \,e^{\lam_j z_j}\; e^{-{|\lam_j|^2\over 2}} \;\;=:\;\; e^{\lam z} \;e^{-{|\lam|^2\over 2}} \pI \lbeq{1.82}
\eeqn
and let $\,\psi_t=e^{-ith}\psi_0\,$ be the time evolved state with Bose-Hubbard Hamiltonian $h=h_0+h_\rint$ given by (\req{1.13}) and (\req{1.14}). Then there are the following 
representations:
\medskip
\beqn
(\,\psi_t\,,a_i^+a^\pp_j \,\psi_t\,)_\cF &=&\E_x\bar\E_y\bigl[ \;v_{i,x,t} \;\bv_{j,y,t}\; e^{v_{x,t}\bv_{y,t}} \;\bigr] \;e^{-|\lam|^2} \phantom{mm} \lbeq{1.83} \\ 
\|\psi_t\|^2_\cF &=& \E_x\bar\E_y[ \; e^{v_{x,t}\bv_{y,t}} \;] \;e^{-|\lam|^2}   \pS  \lbeq{1.84}
\eeqn

\smallskip
with Fresnel expectations $\,\E_x\bar\E_y[ \;\cdot\;]\,$ given by (\req{1.69}) and (\req{1.61}) above, and the $v,\bv\in\mathbb C^\G\,$ are given 
by the SDE system 
\smallskip
\beqn
dv_{j}&=& \;-\;i\, dt\,(\vep v)_j \;-\;i\,\sqrt{2u}\,v_j\,dx_j  \phantom{mmm}  \nn \\ 
d\bar v_{j}&=&\;+\;i\, dt\,(\vep\bar v)_j \;+\;i\,\sqrt{2u}\,\bar v_j\,dy_j  \lbeq{1.85}
\eeqn

\smallskip
with initial conditions $\,v_{x,0}=\lam,\,\bv_{y,0}=\bar\lam\,$.

\bigskip
\bigskip
\bigskip
\noindent Now, in the next chapter we will see that the combination $v_{x,t}\bv_{y,t}$ is a martingale and as a consequence, the exponential $\,e^{v_{x,t}\bv_{y,t}}\,$ 
can be absorbed into the Fresnel integration measure. This has the effect that the density matrix elements are then simply given by 
\beq
(\,\psi_t\,,a_i^+a^\pp_j \,\psi_t\,)_\cF &=&\E_{\ti x}\bar\E_{\ti y}\bigl[ \;v_{i,\ti x,t} \;\bv_{j,\ti y,t}\;\bigr]   \pI
\eeq

\smallskip
where the $\;v_{\ti x,t},\,\bv_{\ti y,t}\;$ are given by the transformed SDE system
\medskip
\beq
dv_j &=&\;-\;i\, dt\,(\vep v)_j \;-\;i\,2u\,dt\,v_j\bv_j\,v_j \;-\;i\,\sqrt{2u}\,v_j\,d\ti x_j  \phantom{mmm}  \\ 
d\bar v_j&=& \;+\;i\, dt\,(\vep\bar v)_j\;+\;i\,2u\,dt\,v_j\bv_j\,\bv_j \;+\;i\,\sqrt{2u}\,\bar v_j\,d\ti y_j   
\eeq

\medskip
with transformed Fresnel Brownian motions
\beq
d\ti x_{j,t_\ell}&:=&dx_{j,t_\ell} \,-\, \sqrt{2u}\,dt\, (v_j\bv_j)_{t_{\ell-1}} \phantom{mmm} \\ 
d\ti y_{j,t_\ell}&:=&dy_{j,t_\ell} \,-\, \sqrt{2u}\,dt\, (v_j\bv_j)_{t_{\ell-1}} 
\eeq

\medskip
So, let's look at the details.

\goodbreak

\noindent{\large\bf 2. Martingale Property and Girsanov Transformation}
\numberwithin{equation}{section}
\renewcommand\thesection{2}
\setcounter{equation}{0}

\bigskip
\bigskip
\bigskip
\noindent{\bf 2.1\; Unitary Time Evolution as a Martingale}

\bigskip
\bigskip
Recall from Theorem 1 that the time evolution $\,\psi_t=e^{-it(h_0+h_\rint)}\psi_0\,$ of some initial state $\psi_0=\psi_0(z)$ can be written as a Fresnel expectation value  
\beqn 
\psi_t(z) &=&\ts \E\bigl[\, \psi_0(U_tz) \,\bigr]\;\;=\;\;\E_x\bigl[\, \psi_0(U_{x,t}z) \,\bigr]\;\;=\;\;\int \psi_0(U_{x,t}z) \,dF(\{x_s\}_{0<s\le t}) \pI  \lbeq{2.1}
\eeqn
where the unitary evolution matrix is given by the SDE 
\beqn
dU_{x,t}&=&-i\,U_{x,t}\,\bigl(\, dt\,\vep \;+\; \sqrt{2u}\, dx_t\,\bigr) \pI  \lbeq{2.2}
\eeqn
with initial value $U_{x,0}=\Id$. The norm of $\psi_t$ is given by 
\beqn
\|\psi_t\|_\cF^2\;\;=\;\; (\psi_t,\psi_t)_\cF&=&\ts \int_{\mathbb C^\G} \,\psi_t(z)\, \overline{\psi_t(z)}\, d\mu(z) \pI \lbeq{2.3}
\eeqn
with $d\mu(z)$ given by (\req{1.7}). For the complex conjugated $\psi_t$ we use integration variables or Fresnel Brownian motions $\{y_s\}_{0<s\le t}$ and write 
\beqn 
\overline{\psi_t(z)} &=&\ts \bar\E_y\bigl[\, \overline{\psi_0(U_{y,t}z)} \,\bigr]\;\;=\;\;\int \,\overline{\psi_0(U_{y,t}z)} \,d\bar F(\{y_s\}_{0<s\le t}) \pI \lbeq{2.4}
\eeqn
such that, as in (\req{1.64}) of the last chapter, 
\beqn
\|\psi_t\|_\cF^2&=&\ts \int_{\mathbb C^\G} \,\psi_t(z)\, \overline{\psi_t(z)}\, d\mu(z) \pI  \nn \\ 
&=&\ts \int_{\mathbb C^\G} \,\E_x\bigl[\, \psi_0(U_{x,t}z) \,\bigr]\; \bar\E_y\bigl[\, \overline{\psi_0(U_{y,t}z)} \,\bigr] \; d\mu(z) \nn \pI \\ 
&=&\ts \E_x\bar\E_y\Bigl[\; \int_{\mathbb C^\G} \, \psi_0(U_{x,t}z) \, \overline{\psi_0(U_{y,t}z)}  \; d\mu(z) \;\Bigr]   \pI \lbeq{2.5}
\eeqn
Now we have
\beqn
d\mu(Uz)&=&d\mu(z) \pM
\eeqn

\medskip
for any unitary $U$. Thus, with the substitution $z=U_{y,t}^+\,w\,$ with $U^+=\bar U^T$ being the adjoint matrix, and renaming the $w$ back to $z$, we obtain 
\beqn
\|\psi_t\|_\cF^2&=&\ts \E_x\bar\E_y\Bigl[\; \int_{\mathbb C^\G} \, \psi_0(U_{x,t}U_{y,t}^+z) \, \overline{\psi_0(z)}  \; d\mu(z) \;\Bigr]   \pI \lbeq{2.7}
\eeqn
This quantity has to be independent of $t$, we have to have 
\smallskip
\beqn
\|\psi_t\|_\cF^2&=&\ts \E_x\bar\E_y\Bigl[\; \int_{\mathbb C^\G} \, \psi_0(U_{x,t}U_{y,t}^+z) \, \overline{\psi_0(z)}  \; d\mu(z) \;\Bigr]   \pS \nn \\ 
&\buildrel ! \over =& \ts  \int_{\mathbb C^\G} \, \psi_0(z) \, \overline{\psi_0(z)}  \; d\mu(z) \;\;=\;\;  \|\psi_0\|_\cF^2  \lbeq{2.8} 
\eeqn
How can this be understood from a stochastic calculus point of view? We have 
\medskip
\beqn
dU_{x,t}&=&-\,i\,U_{x,t}\,\bigl(\, dt\,\vep \;+\; \sqrt{2u}\, dx_t\,\bigr) \nn \\ 
dU^+_{y,t}&=&+\,i\,\bigl(\, dt\,\vep \;+\; \sqrt{2u}\, dy_t\,\bigr)\, U_{y,t}^+ \pM \lbeq{2.9}
\eeqn

\medskip
Since $\,dx_t\,dy_t=0\,$, we also have $\,dU_{x,t}\,dU_{y,t}^+=0\,$ such that 
\beqn
d\bigl(\, U_{x,t}U_{y,t}^+\,\bigr) &=& dU_{x,t}\,U_{y,t}^+ \;+\; U_{x,t}\, dU_{y,t}^+ \;+\; dU_{x,t}\,dU_{y,t}^+  \pI \nn \\ 
&=&-\,i\,U_{x,t}\,\bigl(\, dt\,\vep \,+\, \sqrt{2u}\, dx_t\,\bigr)\,U_{y,t}^+ 
  \;+\;i\,U_{x,t}\,\bigl(\, dt\,\vep \,+\, \sqrt{2u}\, dy_t\,\bigr)\, U_{y,t}^+ \;\;+\;0 \nn  \\ 
&=&\,-\,i\,\sqrt{2u}\; U_{x,t}\,(dx_t-dy_t)\,U_{y,t}^+ \nn  \pI \\ 
&=&\,-\,i\,\sqrt{4u}\; U_{x,t}\,d\xi_t\,U_{y,t}^+  \lbeq{2.10}
\eeqn

\medskip
if we put (the eta's will be used later, not now)
\beqn
d\xi_t\;\;:=\;\;\ts {dx_t-dy_t\over \sqrt{2} }\;,\;\;\;\;\;\; d\eta_t\;\;:=\;\;\ts {dx_t+dy_t\over \sqrt{2} } \pS \lbeq{2.11}
\eeqn
Equation (\req{2.10}) means that the matrix $\,U_{x,t}U_{y,t}^+\,$ is a martingale, its $\,d(U_{x,t}U_{y,t}^+)\,$ has no drift part, no $dt$ part, but only 
a diffusive part, a $dx_t$ or $dy_t$ part. And since 
\beqn
\E[\,dx_t\,]\;=\;\bar\E[\,dy_t\,]\;=\;0  \pS \lbeq{2.12} 
\eeqn 
one then has 
\beqn
\E\bar\E\bigl[\,d(U_{x,t}U_{y,t}^+)\,\bigr]&=&0 \phantom{mmm}  \lbeq{2.13}
\eeqn
which results in 
\beqn
\E\bar\E\bigl[\, U_{x,t}U_{y,t}^+\,\bigr]&=&U_{x,0}\,U_{y,0}^+\;\;=\;\;\Id \pS \lbeq{2.14} 
\eeqn 

\smallskip
\noindent However, in equation (\req{2.8}) we have not directly an expectation of the matrix $\,U_{x,t}U_{y,t}^+\,$ itself, but we have an arbitrary function of it, 
so we have to consider an expectation of the form 
\beqn
\E_x\bar\E_y\Bigl[\; f\bigl( \,U_{x,t}U_{y,t}^+ \,\bigr) \;\Bigr] \pI \lbeq{2.15}
\eeqn
where $\,f:\mathbb C^{\G\times \G}\to \mathbb C\,$ is an arbitrary function. These quantities should also be time independent, how can we understand that? 
Let's abbreviate for the moment 
\beqn
M_t&:=&U_{x,t}U_{y,t}^+ \;\;=\;\; (\,M_{ij}\,)_{i,j\in\Gamma} \pI  \lbeq{2.16}
\eeqn
Then, with the Ito formula (appendix A.2 has a quick reminder),
\beqn
df(M_t)&=&\ts \sum_{i,j\in\Gamma}\, {\pt f \over \pt M_{ij}}\, dM_{ij}
  \;+\; {1\over 2} \sum_{i,j\in\Gamma} \sum_{k,\ell\in\Gamma} \,{\pt^2 f \over \pt M_{ij} \pt M_{k\ell}}\, dM_{ij}\, dM_{k\ell}  \pI \lbeq{2.17}
\eeqn
In the equation above, also the $\,k,\ell$ are temporarily used as lattice site indices, they are no time indices here. The first sum is purely diffusive since 
\beqn
dM_{ij}&=&\,-\,i\,\sqrt{4u}\, [\,U_{x,t}\,d\xi_t\,U_{y,t}^+\,]_{i,j}  \pI \lbeq{2.18}
\eeqn
has no $dt$ part. And the second sum actually vanishes since 
\beqn
dM_{ij}\, dM_{k\ell}&=&\,-\,4u\, [\,U_{x,t}\,d\xi_t\,U_{y,t}^+\,]_{i,j}\; [\,U_{x,t}\,d\xi_t\,U_{y,t}^+\,]_{k,\ell}  \pI  \nn \\ 
&=&\,-\,4u\, \sum_{m,n\in\Gamma} [U_{x,t}]_{i,m}\,d\xi_{m,t}\,[U_{y,t}^+]_{m,j}\; [U_{x,t}]_{k,n}\,d\xi_{n,t}\,[U_{y,t}^+]_{n,\ell}  \pI  \nn \\ 
&\buildrel d\xi_{m,t}\,d\xi_{n,t}\,=\,0\phantom{\bigr|} \over =& \;\;0 \pI
\eeqn
Namely, 
\beqn
d\xi_{m,t}\,d\xi_{n,t}&=&\ts {1\over 2}\, (dx_{m,t}-dy_{m,t})\,(dx_{n,t}-dy_{n,t})  \pI \nn \\ 
&=&\ts {1\over 2}\, \bigl(\, dx_{m,t}\,dx_{n,t} \, + \, dy_{m,t}\,dy_{n,t} \, - \, dx_{m,t}\,dy_{n,t} \, -\, dy_{m,t}\,dx_{n,t} \, \bigr)  \nn \\ 
&=&\ts {1\over 2}\, \bigl(\, dx_{m,t}\,dx_{n,t} \, + \, dy_{m,t}\,dy_{n,t} \, - \, 0 \, -\, 0 \, \bigr)  \pI 
\eeqn
For $m=n$, this becomes
\beqn
d\xi_{m,t}\,d\xi_{m,t}&=&\ts {1\over 2}\, \bigl(\, (dx_{m,t})^2 \, + \, (dy_{m,t})^2 \,  \bigr)  \pI \nn \\ 
&=&\ts {1\over 2}\, \bigl(\, i\,dt \, - \, i\, dt \,  \bigr) \;\;=\;\; 0  
\eeqn
And for $m\ne n$, this is simply 
\beqn
d\xi_{m,t}\,d\xi_{n,t}&=&\ts {1\over 2}\, \bigl(\, 0 \, + \, 0 \, \bigr)\;\;=\;\; 0 \pI 
\eeqn
Thus we end up with 
\beqn
df(M_t)&=&\ts \,-\,i\,\sqrt{4u}\,\sum_{i,j}\, {\pt f \over \pt M_{ij}}\,  [\,U_{x,t}\,d\xi_t\,U_{y,t}^+\,]_{i,j} \pI \\ \nn
\eeqn
which is purely diffusive and this results in 
\beqn
\E_x\bar\E_y\bigl[\,df(M_t)\,\bigr] &=&0 \pI 
\eeqn
and accordingly
\beqn
\E_x\bar\E_y\bigl[\,f(M_t)\,\bigr] &=&\ts f(M_0) \;+\; \int_0^t \, \E_x\bar\E_y\bigl[\,df(M_s)\,\bigr] \;\;=\;\; f(M_0)   \pI 
\eeqn
for arbitrary $f$. Let's summarize these observations in the following 

\bigskip
\bigskip
\bigskip
\noindent{\bf Theorem 3:} Let $U_{x,t}$ be the unitary evolution matrix of Theorem 1 such that the time evolution of an arbitrary state $\psi\in\cF$  
can be written as 
\beqn
(e^{-ikdt\,h}\psi)(z)&=&\E_x\bigl[ \, \psi( U_{x,kdt} z )\,\bigr] \pI
\eeqn
with Fresnel expectation $\E_x[\,\cdot\,]$ given by (\req{1.61}) and (\req{1.69}). Then, for arbitrary $\,f:\C^{\G\times\G}\to\C\,$, the quan\-tity $f( U_{x,t}U_{y,t}^+)$
is a martingale, its $df( U_{x,t}U_{y,t}^+)$ has no $dt$-part, and we have the following identity:
\beqn
\E_x\bar\E_y\Bigl[\; f\bigl( \,U_{x,t}U_{y,t}^+ \,\bigr) \;\Bigr]&=&f\bigl( \,U_{0}U_{0}^+ \,\bigr)\;\;=\;\;f\bigl( \,\Id \,\bigr)  \pS 
\eeqn

\medskip
In particular, for any time evolved state $\psi_t=e^{-it\,h}\psi_0\,$, 
\medskip
\beqn
\|\psi_t\|_\cF^2&=&\ts  \int_{\mathbb C^\G} \, \E_x\bar\E_y\bigl[\,\psi_0(U_{x,t}U_{y,t}^+z) \, \overline{\psi_0(z)}\,\bigr]  \; d\mu(z)    \nn \\ 
& =& \ts  \int_{\mathbb C^\G} \, \psi_0(z) \, \overline{\psi_0(z)}  \; d\mu(z) \;\;=\;\;  \|\psi_0\|_\cF^2 \;\;. \pS
\eeqn 

\goodbreak

\bigskip
\bigskip
\bigskip

\noindent{\bf 2.2\; Girsanov Transformed SDE System for the Density Matrix Elements}

\bigskip
\bigskip
\noindent Recall the representations of Theorem 2, for the initial state $\,\psi_0(z)=e^{\lam z}e^{-{|\lam|^2\over 2}}$, the density matrix elements can be written as 
\beqn
(\,\psi_t\,,a_i^+a^\pp_j \,\psi_t\,)_\cF &=&\E_x\bar\E_y\bigl[ \;v_{i,x,t} \;\bv_{j,y,t}\; e^{v_{x,t}\bv_{y,t}} \;\bigr] \;e^{-|\lam|^2}  \pI  
\eeqn
with vector valued functions $\,v=v_{x,t}\in\mathbb C^\G\,$ and $\,\bv=\bv_{y,t}\in\mathbb C^\G\,$ 
which are given by the SDE system 
\smallskip
\beqn
dv&=& \;-\;i\,\bigl(\, dt\,\vep  \;+\;\sqrt{2u}\,dx_t\,\bigr)\, v  \phantom{mm} \nn \\ 
d\bar v&=& \;+\;i\,\bigl(\, dt\,\vep  \;+\;\sqrt{2u}\,dy_t\,\bigr)\, \bv   
\eeqn

\medskip
with initial conditions $\,v_{x,0}=\lam,\,\bv_{y,0}=\bar\lam\,$. Since $\,dx_{j,t}\,dy_{j,t}=0\,$, we obtain
\smallskip
\beqn
d(v^T\bv)&=& dv^T\bv\;+\;v^Td\bv\;+\;dv^Td\bv \pS \nn \\ 
&=&\;-\;i\,v^T\bigl(\, dt\,\vep\;+\;\sqrt{2u} \,dx_{t}\,\bigr)\bv  \; +\; i\, v^T\bigl(\, dt\,\vep\;+\;\sqrt{2u} \,dy_{t}\,\bigr)\,\bv \;\;+\;0  \nn \\
&=& \;-\;i\sqrt{2u}\, v^T(dx_t-dy_t) \bv\pS \nn \\ 
&=&\ts \;-\;i\sqrt{2u}\, \sum_j v_j\bv_j\,(dx_{j,t}-dy_{j,t})  \lbeq{2.31}
\eeqn

\medskip
That is, the quantity $\,v^T\bv=\sum_{j} v_j\bv_j\,\equiv\,v\bv\,$, we omit the transpose sign in the following, is a martingale. 
Now we write 
\beqn
(v\bv)_{kdt}&=&(v\bv)_0 \;+\; \sum_{\ell=1}^k \bigl[\,(v\bv)_{\ell dt} \,-\,(v\bv)_{(\ell-1) dt}\,\bigr]   \pI  \nn \\ 
&=&(v\bv)_0 \;+\; \sum_{\ell=1}^k d(v\bv)_{\ell dt}  \pI  \nn \\ 
&=&(v\bv)_0 \;-\;i\sqrt{2u}\, \sum_{\ell=1}^k\sum_j (v_j\bv_j)_{(\ell-1)dt} \,(dx_{j,\ell dt} - dy_{j,\ell dt})  \pI \nn \\ 
&=&|\lam|^2 \;-\;i\sqrt{2udt}\; \sum_{\ell=1}^k\sum_j (v_j\bv_j)_{(\ell-1)dt} \,(\phi_{j,\ell} - \theta_{j,\ell})  \pI
\eeqn 
such that 
\beqn
e^{(v\bv)_{kdt}}\,e^{-|\lam|^2}\;\;=\;\;e^{(v\bv)_{kdt}-(v\bv)_0} 
&=&\exp\biggl\{\;-\;i\sqrt{2udt}\; \sum_{\ell=1}^k\sum_j (v_j\bv_j)_{(\ell-1)dt} \,(\phi_{j,\ell} - \theta_{j,\ell}) \;\biggr\} \pI \phantom{mm}  \lbeq{2.33}
\eeqn

\medskip
with integration variables 
\beqn 
dx_{j,\ell dt}&=&\sqrt{dt}\,\phi_{j,\ell}  \\ 
dy_{j,\ell dt}&=&\sqrt{dt}\,\theta_{j,\ell} \nn
\eeqn

\medskip
In terms of the $\,\phi_{j,\ell},\theta_{j,\ell}\,$ variables, the Fresnel measure reads 
\smallskip
\beqn
\E_x\bar\E_y\bigl[\;\cdots\;\bigr]&=& \int_{\R^{k\G}}\int_{\R^{k\G}} \;\;\cdots\;\;\; \pro_{\ell=1}^k \; e^{ {i\over 2}\sum_j (\phi_{j,\ell}^2 - \theta_{j,\ell}^2) }
 \ts \;  { d^\G\phi_\ell \,d^\G\theta_\ell \over (2\pi)^\G} \pI
\eeqn

\medskip
Consider the $\ell$'th term on the right hand side of (\req{2.33}),
\beqn
\sum_j (v_j\bv_j)_{(\ell-1)dt} \,(\phi_{j,\ell} - \theta_{j,\ell})   \pI
\eeqn
The $\phi_\ell$ and $\theta_\ell$ show up in an explicit linear form, since the quantities 
\beqn
(v_j\bv_j)_{(\ell-1)dt}&=&(v_j\bv_j)_{(\ell-1)dt}\Bigl(\, \{\phi_m,\theta_m\}_{m=1}^{\ell-1}\, \Bigr) \pI
\eeqn
depend only on $\phi$'s and $\theta$'s at earlier times $t_1,\cdots,t_{\ell-1}$. Thus, we can absorb them into the integration measure 
simply by completing the square. In the mathematics literature, the corresponding change of variables then is called a Girsanov transformation. Thus, this is a 
very elementary calculation, but, since this is a key step, let us be very explicit and proceed line by line. The result is summarized in Theorem 4 below. 

\bigskip
\noindent We have at time $\,t=t_k=kdt$
\beqn
\lefteqn{
(\,\psi_t\,,a_i^+a^\pp_j \,\psi_t\,)_\cF \;\;=\;\;\E\bar\E[ \,v_i \bv_j\, e^{v\bv} \,] \;e^{-|\lam|^2}   } \pI \\
&=&\int_{\R^{2k\G}} v_{i,kdt}\, \bv_{j,kdt}\;
 \prod_{\ell=1}^k \exp\biggl\{\;-\;i\sqrt{2udt}\;\sum_j (v_j\bv_j)_{(\ell-1)dt} \,(\phi_{j,\ell} - \theta_{j,\ell}) \;\biggr\} \;\times \pI \nn \\ 
&&\phantom{mmmmmmmmmmmm}   \prod_{\ell=1}^k \; \exp\biggl\{\, {i\over 2}\sum_j (\phi_{j,\ell}^2 - \theta_{j,\ell}^2) \,\biggr\} 
  \;  { d^\G\phi_\ell\, d^\G\theta_\ell \over (2\pi)^\G} \phantom{mmm} \pI \nn \\ \nn
\eeqn

\smallskip
Consider the $\ell$-th factor. The exponentials with the $\phi_\ell$ variables combine to 
\smallskip
\beqn
\lefteqn{
\exp\biggl\{\;-\;i\sqrt{2udt}\;\sum_j (v_j\bv_j)_{(\ell-1)dt} \,\phi_{j,\ell} \;\biggr\} \;\times\; 
  \exp\biggl\{\, {i\over 2}\sum_j \phi_{j,\ell}^2 \,\biggr\}     \pI } \lbeq{2.39}  \\ 
&=&\exp\biggl\{\; {i\over 2}\sum_j \Bigl[\, \phi_{j,\ell}^2 \;-\; 2\sqrt{2udt}\, (v_j\bv_j)_{(\ell-1)dt} \,\phi_{j,\ell} \,\Bigr] \; \biggr\}  \pI \nn\\ 
&=&\exp\biggl\{\; {i\over 2}\sum_j \bigl[\, \phi_{j,\ell} \,-\, \sqrt{2udt}\, (v_j\bv_j)_{(\ell-1)dt} \,\bigr]^2 \; \biggr\} \;\times \; 
  \exp\biggl\{\;-\, {i\over 2}\sum_j \, 2udt\, (v_j\bv_j)_{(\ell-1)dt}^2  \; \biggr\}   \pI \nn 
\eeqn

\medskip
The exponentials with the $\theta_\ell$ variables combine to 
\smallskip
\beqn
\lefteqn{
\exp\biggl\{\;+\;i\sqrt{2udt}\;\sum_j (v_j\bv_j)_{(\ell-1)dt} \,\theta_{j,\ell} \;\biggr\} \;\times\; 
  \exp\biggl\{\,-\, {i\over 2}\sum_j \theta_{j,\ell}^2 \,\biggr\}     \pI }  \lbeq{2.40} \\ 
&=&\exp\biggl\{\;-\, {i\over 2}\sum_j \Bigl[\, \theta_{j,\ell}^2 \;-\; 2\sqrt{2udt}\, (v_j\bv_j)_{(\ell-1)dt} \,\theta_{j,\ell} \,\Bigr] \; \biggr\}  \pI \nn \\ 
&=&\exp\biggl\{\;-\, {i\over 2}\sum_j \bigl[\, \theta_{j,\ell} \,-\, \sqrt{2udt}\, (v_j\bv_j)_{(\ell-1)dt} \,\bigr]^2 \; \biggr\} \;\times \; 
  \exp\biggl\{\;+\, {i\over 2}\sum_j \, 2udt\, (v_j\bv_j)_{(\ell-1)dt}^2  \; \biggr\}   \pI \nn \\  \nn
\eeqn
Observe that the last exponentials in (\req{2.39}) and (\req{2.40})
\beqn
\ts \exp\bigl\{\;-\, {i\over 2}\sum_j \, 2udt\, (v_j\bv_j)_{(\ell-1)dt}^2  \; \bigr\}\times 
  \exp\bigl\{\;+\, {i\over 2}\sum_j \, 2udt\, (v_j\bv_j)_{(\ell-1)dt}^2  \; \bigr\} &=&1  \phantom{mm} \pI 
\eeqn
cancel each other. 

\bigskip
Now we make the substitution of variables
\smallskip
\beqn
\ti\phi_{j,\ell}&:=&\phi_{j,\ell} \,-\, \sqrt{2udt}\, (v_j\bv_j)_{(\ell-1)dt} \phantom{mm} \\ 
\ti\theta_{j,\ell}&:=&\theta_{j,\ell} \,-\, \sqrt{2udt}\, (v_j\bv_j)_{(\ell-1)dt}  \nn
\eeqn
or equivalently 
\beqn
d\ti x_{j,t_\ell}&:=&dx_{j,t_\ell} \,-\, \sqrt{2u}\,dt\, (v_j\bv_j)_{(\ell-1)dt} \phantom{mm} \\ 
d\ti y_{j,t_\ell}&:=&dy_{j,t_\ell} \,-\, \sqrt{2u}\,dt\, (v_j\bv_j)_{(\ell-1)dt} \nn 
\eeqn

\smallskip
Then we can write 
\beqn
(\,\psi_t\,,a_i^+a^\pp_j \,\psi_t\,)_\cF &=&\E_x\bar\E_y[ \,v_i \bv_j\, e^{v\bv} \,] \;e^{-|\lam|^2}  \nn \phantom{mmm}  \\
&=&\E_{\ti x}\bar\E_{\ti y}[ \,v_i \bv_j\, ]  \pI  
\eeqn
where in terms of the transformed variables $\ti x,\ti y\,$ the $v_j$ and $\bv_j$ are given by the transformed SDE system 
\beqn
dv_j&=&\;-\;i\, dt\,(\vep v)_j \;-\;i\sqrt{2u}\,v_j\,dx_j  \nn \\ 
&=&\;-\;i\, dt\,(\vep v)_j  \;-\;i\sqrt{2u}\,v_j\,\bigl[ d\ti x_j \,+\, \sqrt{2u}\,dt\, v_j\bv_j \bigr] \nn \\ 
&=&\;-\;i\, dt\,(\vep v)_j  \;-\; i\,2u\,dt\, v_j\bv_j \,v_j\;-\;i\sqrt{2u}\,v_j\, d\ti x_j  \lbeq{2.45} 
\eeqn
and
\beqn
d\bar v_j&=&\;+\;i\, dt\,(\vep\bar v)_j \;+\;i\sqrt{2u}\,\bar v_j\,dy_j   \nn \\
&=&\;+\;i\, dt\,(\vep\bar v)_j  \;+\;i\sqrt{2u}\,\bar v_j\,\bigl[ d\ti y_j \,+\, \sqrt{2u}\,dt\, v_j\bv_j \bigr] \nn \\ 
&=&\;+\;i\, dt\,(\vep\bar v)_j \;+\;i\, 2u\,dt\, v_j\bv_j \,\bv_j  \;+\;i\sqrt{2u}\,\bar v_j\, d\ti y_j \lbeq{2.46} 
\eeqn 

\medskip
with initial conditions 
\beqn
v_{j,0}&=&\lam_j \nn \\ 
\bv_{j,0}&=&\blam_j  
\eeqn
Recall that 
\beqn
|\lam|^2&=&N 
\eeqn
has the meaning of total number of particles. Thus, if we devide (\req{2.45}) and (\req{2.46}) through $|\lam|=\sqrt{N}$ and put 
\beqn
w_j&:=& v_j\;/\;|\lam| \nn \\ 
\bar w_j&:=& \bv_j\;/\;|\lam| 
\eeqn
we obtain
\smallskip
\beqn
dw_j&=&\;-\;i\, dt\,(\vep w)_j \;-\; i\,2uN\,dt\, w_j\bar w_j \,w_j\;-\;i\sqrt{2u}\,w_j\, d\ti x_j  \phantom{mm} \nn \\ 
d\bar w_j&=&\;+\;i\, dt\,(\vep\bar w)_j\;+\;i\, 2uN\,dt\, w_j\bar w_j \,\bar w_j  \;+\;i\sqrt{2u}\,\bar w_j\, d\ti y_j 
\eeqn
or, with 
\beqn
g&:=&uN 
\eeqn
\beqn
dw_j&=&\;-\;i\, dt\,(\vep w)_j \;-\; i\,2g\,dt\, w_j\bar w_j \,w_j\;-\;i\sqrt{2g/N}\,w_j\, d\ti x_j  \phantom{mm} \nn \\ 
d\bar w_j&=&\;+\;i\, dt\,(\vep\bar w)_j\;+\;i\, 2g\,dt\, w_j\bar w_j \,\bar w_j  \;+\;i\sqrt{2g/N}\,\bar w_j\, d\ti y_j 
\eeqn

\bigskip
Thus, in the limit $\,N\to\infty\,$ with $\,g=uN\,$ fixed, the diffusive part vanishes, the SDE reduces to a deterministic ODE system and the exact density matrix elements are 
given by 
\smallskip
\beqn
(\,\psi_t\,,a_i^+a^\pp_j \,\psi_t\,)_\cF &=&\E_{\ti x}\bar\E_{\ti y}[ \,v_i \bv_j\, ] \;\;=\;\; 
N \;\times \; w_i(t) \,\bar w_j(t)  \pI
\eeqn

\smallskip
with the $w,\bar w$ given by the ODE system
\beqn
\dot w_j&=&\;-\;i\,(\vep w)_j \;-\; i\,2g\, w_j\bar w_j \,w_j  \nn \\ 
\dot{\bar w}_j&=&\;+\;i\,(\vep\bar w)_j\;+\;i\, 2g\, w_j\bar w_j \,\bar w_j \phantom{mm}
\eeqn

\smallskip
with  initial conditions 
\beqn
w_j(0)&=& \lam_j\;/\;|\lam| \nn \\ 
\bar w_j(0)&=& \blam_j\;/\;|\lam| 
\eeqn

\smallskip
Observe that now the $\bar w_j$ are the true complex conjugates of the $w_j$. Before we proceed to the analog calculation for number states 
in the next section, let's summarize the results in the following 

\goodbreak 

\medskip
\bigskip
\bigskip
\noindent{\bf Theorem 4:}  Recall the SDE representation of Theorem 2 above,
\smallskip
\beqn
(\,\psi_t\,,a_i^+a^\pp_j \,\psi_t\,)_\cF &=&\E_x\bar\E_y\bigl[ \;v_{i,x,t} \;\bv_{j,y,t}\; e^{v_{x,t}\bv_{y,t}} \;\bigr] \;e^{-|\lam|^2}  
\eeqn
with
\beqn
dv_j&=& \;-\;i\, dt\,(\vep v)_j \;-\;i\,\sqrt{2u}\,v_j\,dx_j  \nn  \\ 
d\bar v_j&=& \;+\;i\, dt\,(\vep\bar v)_j \;+\;i\,\sqrt{2u}\,\bar v_j\,dy_j 
\eeqn

\smallskip
Then, with the Girsanov transformation 
\beqn
d\ti x_{j,t_\ell}&:=&dx_{j,t_\ell} \,-\, \sqrt{2u}\,dt\, (v_j\bv_j)_{(\ell-1)dt} \phantom{mm} \nn \\ 
d\ti y_{j,t_\ell}&:=&dy_{j,t_\ell} \,-\, \sqrt{2u}\,dt\, (v_j\bv_j)_{(\ell-1)dt} 
\eeqn

\smallskip
the exponential $\,e^{(v\bv)_t-|\lam|^2}=e^{(v\bv)_t-(v\bv)_0}\,$ can be absorbed into the Fresnel integration measure, there is the identity 
\beqn
(\,\psi_t\,,a_i^+a^\pp_j \,\psi_t\,)_\cF &=&\E_{\ti x}\bar\E_{\ti y}\bigl[ \;v_{i,\ti x,t} \;\bv_{j,\ti y,t}\; \bigr] \lbeq{2.59}
\eeqn
with
\beqn
dv_j&=&\;-\;i\, dt\,(\vep v)_j \;-\; i\,2u\,dt\, v_j\bv_j \,v_j\;-\;i\sqrt{2u}\,v_j\, d\ti x_j  \phantom{mmm}\nn \\
d\bar v_j&=&\;+\;i\, dt\,(\vep\bar v)_j\;+\;i\, 2u\,dt\, v_j\bv_j \,\bv_j  \;+\;i\sqrt{2u}\,\bar v_j\, d\ti y_j  \lbeq{2.60}
\eeqn 

\medskip
In the limit $\,N\to\infty\,$ with $\,g=uN\,$ fixed, we have the exact representation 
\beqn
(\,\psi_t\,,a_i^+a^\pp_j \,\psi_t\,)_\cF &=&N \;\times \; w_i(t) \,\bar w_j(t)  \pS
\eeqn
with $\bar w_j$ now being the true complex conjugate of $w_j$ and the $w_j$ are given by the ODE system
\beqn
\dot w_j&=&\;-\;i\,(\vep w)_j \;-\; i\,2g\, w_j\bar w_j \,w_j  \pS  \lbeq{2.62}
\eeqn
with initial conditions $\,w_j(0)= \lam_j\,/\,|\lam|\,$. Equation (\req{2.62}) is the time dependent discrete Gross Pitaevskii equation, 
here with a general hopping matrix $\vep=(\vep_{ij})_{i,j\in\Gamma}$ which may also include some on-diagonal trapping potentials $\eps_j=\vep_{jj}$ 
from the Hamiltonian (\req{1.1}).    

\bigskip
\bigskip
\bigskip

\noindent{\bf 2.3\; Number States}

\bigskip
\bigskip
\noindent Let's consider the dynamics of number states which are used to describe the dynamics of Bose-Einstein condensates. They are given 
by the following initial state 
\beqn
\psi_0(z)\;=\;\psi_0(\,\{z_j\}\,)&=& \ts {1 \over \sqrt{N!N^N}} \;\bigl(\, \sum_j \lam_j z_j \,\bigr)^N\;\;=\;\; {(\lam z)^N \over \sqrt{N!N^N}} \pI
\eeqn
with 
\beqn
|\lam|^2&=&\ts \sum_j |\lam_j|^2\;\;=\;\; N \lbeq{2.64}
\eeqn
There are the following formulae which are standard expectations over the bosonic Fock space, here written again in the Bargmann-Segal 
representation:
\beqn
\ts {1\over N! }\;\int_{\C^\G}\, (\lam z)^N\, (\blam \bz)^N \; d\mu(z) &=&\ts (\,|\lam|^2\,)^N \pI \\ 
\ts {1\over N!}\;\int_{\C^\G}\, z_i\,\bar z_j\;(\lam z)^N\, (\blam \bz)^N \; d\mu(z) &=&\ts \delta_{i,j}\;(\,|\lam|^2\,)^N \;+\;N\,\blam_i\lam_j\;(\,|\lam|^2\,)^{N-1}
\eeqn

\medskip
Thus, with the condition (\req{2.64}), we have the following time zero expectations:

\goodbreak

\smallskip
\beqn
(\psi_0,\psi_0)_\cF&=&\ts \int_{\mathbb C^\G}  |\psi_0(z)|^2 \,d\mu(z) \;\;=\;\; { |\lam|^{2N} \over N^N} \;\;=\;\;1 \pS \\
(\,\psi_0\,,a_i^+a^\pp_j \,\psi_0\,)_\cF&=&\ts \int_{\mathbb C^\G} z_j \bz_i\; |\psi_0(z)|^2 \,d\mu(z) \;\;-\;\; \delta_{i,j} \;\;=\;\;
  \ts N\;{\lam_i\blam_j \over |\lam|^2}\;\;=\;\;\lam_i\blam_j  \pM
\eeqn

\medskip
Now, let's consider the time $t$ formulae. From equations (\req{1.59}) and (\req{1.63}) we have for an arbitrary initial state $\psi_0$
\beqn
\lefteqn{ 
(\,\psi_t\,,a_i^+a^\pp_j \,\psi_t\,)_\cF \;\;=\;\;\ts \int_{\mathbb C^\G} z_j \bz_i\; |\psi_t(z)|^2 \,d\mu(z) \;\;-\;\; \delta_{i,j} \pI   }\\ 
&=&\ts \int \int \;\Bigl\{\; \int_{\mathbb C^\G} z_j \bz_i\; \psi_0(U_{x,t}z)\;\overline{\psi_0(U_{y,t}z)}\;d\mu(z)\;\Bigr\} \; dF(\{x_t\})\;  d\bar F(\{y_t\}) 
    \;\;-\;\; \delta_{i,j} \nn
\eeqn

\medskip
The wavy bracket we can evaluate with the formulae from above. We obtain
\medskip
\beqn
\lefteqn{ 
\ts  \int_{\mathbb C^\G} z_j \bz_i\; \psi_0(U_{x,t}z)\;\overline{\psi_0(U_{y,t}z)}\;d\mu(z)  }  \\ 
&=&\ts {1\over N!N^N}\;\int_{\C^\G}\, z_j\,\bar z_i\;\bigl(\,U_{x,t}^T\lam\cdot z\,\bigr)^N\, \bigl(\,\bar U_{y,t}^T\lam\cdot \bz\,\bigr)^N \; d\mu(z) \pI \nn \\ 
&=&\ts {1\over N^N}\;\Bigl\{\; \delta_{i,j}\;(\,\lam\cdot U_{x,t}U_{y,t}^+\blam \,)^N 
  \;+\;N\,[U_{x,t}^T\lam]_i\; [\bar U_{y,t}^T\blam]_j\;(\,\lam\cdot U_{x,t}U_{y,t}^+\blam \,)^{N-1} \;\Bigr\}  \nn
\eeqn

Thus, 
\beqn
\lefteqn{ 
(\,\psi_0\,,a_i^+a^\pp_j \,\psi_0\,)_\cF\;\;=\;\;   } \\ 
&&\ts {1\over N^N}\;\E_x\bar\E_y\Bigl[\; \delta_{i,j}\;(\,\lam\cdot U_{x,t}U_{y,t}^+\blam \,)^N 
  \;+\;N\,[U_{x,t}^T\lam]_i\; [\bar U_{y,t}^T\blam]_j\;(\,\lam\cdot U_{x,t}U_{y,t}^+\blam \,)^{N-1} \;\Bigr] \;-\;\delta_{i,j} \pI \nn \\ 
&=&\ts \E_x\bar\E_y\Bigl[\; [U_{x,t}^T\lam]_i\; [\bar U_{y,t}^T\blam]_j
  \;\bigl(\, U_{x,t}^T\lam\,\cdot\, U_{y,t}^+\blam\,\bigr)^{N-1} \;\Bigr] \;\bigr/\;(\lam\cdot\blam)^{N-1}   \nn
\eeqn

\medskip
since 
\beqn
\ts {1\over N^N}\;\E_x\bar\E_y\Bigl[\; \delta_{i,j}\;(\,\lam\cdot U_{x,t}U_{y,t}^+\blam \,)^N \;\Bigr] &=& \delta_{i,j} \pM
\eeqn

\medskip
because of Theorem 3 of section 2.1. Now we can proceed as in the preceeding section 2.2. We introduce the variables
\beqn
v_j\;=\;v_{x,t,j}&:=&[\,U_{x,t}^T\lam\,]_{j} \nn \\ 
\bv_j\;=\;\bv_{y,t,j}&:=&[\,\bar U_{y,t}^T\blam\,]_{j}
\eeqn

\smallskip
and from the matrix equations
\smallskip
\beqn
dU_{x,t}^T&=& -i\,\bigl(\, dt\,\vep\;+\;\sqrt{2u} \,dx_{t}\,\bigr)\,U_{x,t}^T  \nn \\
d\bU_{y,t}^T&=& +i\,\bigl(\, dt\,\vep\;+\;\sqrt{2u} \,dy_{t}\,\bigr)\,\bU_{y,t}^T 
\eeqn
we obtain 
\beqn
dv_j&=& -i\, dt\,(\vep v)_j\;+\;\sqrt{2u} \,dx_{j,t}\,v_j  \nn \\
d\bv_j&=& +i\, dt\,(\vep\bv)_j\;+\;\sqrt{2u} \,dy_{j,t}\,\bv_j 
\eeqn
Furthermore, 
\beqn
\bigl(\, U_{x,t}^T\lam\,\cdot\, U_{y,t}^+\blam\,\bigr)^{N-1}&=&\ts \bigl(\, \sum_jv_j\bv_j\,\bigr)^{N-1}\;\;=:\;\; (v\bv)^{N-1} \pI 
\eeqn
Thus, for the density matrix elements we obtain the following representation: 
\beqn
(\,\psi_t\,,a_i^+a^\pp_j \,\psi_t\,)_\cF &=&\E\bar\E\bigl[ \, v_i\,\bv_j\;   (v\bv)^{N-1} \,\bigr] \;\bigr/\;(\lam\blam)^{N-1} \pI
\eeqn
As in the last section, we can make a Girsanov transformation and absorb the quantity $\,(v\bv)^{N-1}\,$ into the Fresnel measure. 
Since the $v,\bv$ obey exactly the same SDEs as in the last section with the same initial conditions 
\beqn
v_{0}&=&\lam \nn \\
\bv_{0}&=&\blam 
\eeqn
equation (\req{2.31}) remains unchanged: 
\beqn
d(v\bv)&=&\ts \;-\;i\sqrt{2u}\, \sum_j v_j\bv_j\,(dx_{j,t}-dy_{j,t})  \pS
\eeqn
Now we abbreviate 
\beqn
P(v\bv)&:=&(v\bv)^{N-1} \pM  \\
p(v\bv)&:=&P'(v\bv)/P(v\bv)\;=\;[\log P]'(v\bv)\;=\; (N-1)/(v\bv) \pI
\eeqn
Then, since $(v\bv)_0=\lam\blam$ and, more importantly, $[d(v\bv)]^2=0$, we can write 
\beqn
\ts {(v_t\bv_t)^{N-1}\over (\lam\blam)^{N-1}} &=&e^{\log P(v_t\bv_t)-\log P(v_0\bv_0)}\;\;=\;\;e^{ \int_0^t p(v_s\bv_s)\,d(v\bv)_s }  \pI 
\eeqn
With the discrete time variables of section 2.2, the exponent reads as follows:
\beqn
\ts\int_0^{t_k} p(v_s\bv_s)\,d(v\bv)_s&=& \ts\sum_{\ell=1}^k p[(v\bv)_{(\ell-1)dt}] \, d(v\bv)_{\ell dt} \pI \nn \\ 
&=&\;-\;i\sqrt{2u}\, \ts \sum_{\ell=1}^k p[(v\bv)_{(\ell-1)dt}]\, \sum_j (v_j\bv_j)_{(\ell-1)dt} \,(dx_{j,\ell dt} - dy_{j,\ell dt})  \pI \nn \\ 
&=& \;-\;i\sqrt{2udt}(N-1)\;\ts \sum_{\ell=1}^k\sum_j \bigl({v_j\bv_j\over v\bv}\bigr)_{(\ell-1)dt} \,(\phi_{j,\ell} - \theta_{j,\ell})  \pI
\eeqn 
Thus, now we have to make the following substitution of variables: 
\smallskip
\beqn
\ti\phi_{j,\ell}&:=&\phi_{j,\ell} \,-\, \sqrt{2udt}(N-1)\, (v_j\bv_j/v\bv)_{(\ell-1)dt} \phantom{mm} \nn \\ 
\ti\theta_{j,\ell}&:=&\theta_{j,\ell} \,-\, \sqrt{2udt}(N-1)\, (v_j\bv_j/v\bv)_{(\ell-1)dt}  
\eeqn
or equivalently 
\beqn
d\ti x_{j,t_\ell}&:=&dx_{j,t_\ell} \,-\, \sqrt{2u}(N-1)\,dt\, (v_j\bv_j/v\bv)_{(\ell-1)dt} \phantom{mm} \nn \\ 
d\ti y_{j,t_\ell}&:=&dy_{j,t_\ell} \,-\, \sqrt{2u}(N-1)\,dt\, (v_j\bv_j/v\bv)_{(\ell-1)dt} 
\eeqn

\smallskip
Then we can write 
\beqn
(\,\psi_t\,,a_i^+a^\pp_j \,\psi_t\,)_\cF &=&\E_x\bar\E_y[ \,v_i \bv_j\, (v\bv)^{N-1} \,]\,/\,(\lam\blam)^{N-1} \;\;=\;\;
 \E_{\ti x}\bar\E_{\ti y}[ \,v_i \bv_j\, ]  \pI  
\eeqn
where in terms of the transformed variables $\ti x,\ti y\,$ the $v_j$ and $\bv_j$ are given by the transformed SDE system 
\beqn
dv_j&=&\;-\;i\, dt\,(\vep v)_j \;-\;i\sqrt{2u}\,v_j\,dx_j  \nn \\ 
&=&\ts \;-\;i\, dt\,(\vep v)_j  \;-\; i\,2u(N-1)\,dt\, {v_j\bv_j\over v\bv} \,v_j\;-\;i\sqrt{2u}\,v_j\, d\ti x_j  
\eeqn
and
\beqn
d\bar v_j&=&\;+\;i\, dt\,(\vep\bar v)_j \;+\;i\sqrt{2u}\,\bar v_j\,dy_j   \nn \\
&=&\ts \;+\;i\, dt\,(\vep\bar v)_j \;+\;i\, 2u(N-1)\,dt\, {v_j\bv_j\over v\bv} \,\bv_j  \;+\;i\sqrt{2u}\,\bar v_j\, d\ti y_j 
\eeqn 

\medskip
with initial conditions $(v,\bv)_0=(\lam,\blam)\,$. Dividing the system through $|\lam|=\sqrt{N}$ and introdu\-cing again the normalized quantities 
\beqn
w&:=&v\,/\,|\lam| \nn \\ 
\bar w&:=&\bv\,/\,|\lam| 
\eeqn
we obtain the following ODE system in the limit $N\to \infty\,$ with $g=UN$ fixed: 
\medskip
\beqn
\dot w_j&=&\ts \;-\;i\,(\vep w)_j  \;-\; i\,2g\, {w_j\bar w_j\over w\bar w} \,w_j \nn \\ 
{\dot{\bar w}}_j&=&\ts \;+\;i\,(\vep\bar w)_j \;+\;i\,2g\, {w_j\bar w_j\over w\bar w} \,\bar w_j  \pS  \lbeq{2.90}
\eeqn 

\smallskip
In particular, for a symmetric hopping matrix $\,\vep=\vep^T$
\medskip
\beqn
\ts {d\over dt}(w\bw)&=&\ts\sum_j\bigl\{\, \dot w_j\,\bw_j\,+\,w_j\,{\dot\bw}_j\,\bigr\}  \nn  \\ 
&=&\ts \;-\;i\,\sum_j\bigl\{\,(\vep w)_j\,\bw_j\,-\,w_j\,(\vep\bar w)_j\,\bigr\} 
   \ts \;-\;i\,2g\,\sum_j\bigl\{\,  {w_j\bw_j\over w\bw} \,w_j\,\bw_j\,- \,w_j\, {w_j\bw_j\over w\bw} \,\bw_j \,\bigr\}  \pS \nn \\ 
&=& \;\;0 
\eeqn

\medskip
Thus we have $(w\bw)_t=1$ for all $t$ and the ODE system (\req{2.90}) reduces again to the time dependent discrete 
Gross-Pitaevskii equation. We summarize in the following 

\medskip
\bigskip
\bigskip
\noindent{\bf Theorem 5:} Consider the following normalized initial number state, 
\beqn
\psi_0(z)\;=\;\psi_0(\,\{z_j\}\,)&=& \ts {1 \over \sqrt{N!N^N}} \;\bigl(\, \sum_j \lam_j z_j \,\bigr)^N\;\;=\;\; {(\lam z)^N \over \sqrt{N!N^N}} \pI
\eeqn
with $|\lam|^2=N$. Then the time $t$ density matrix elements can be written as 
\beqn
(\,\psi_t\,,a_i^+a^\pp_j \,\psi_t\,)_\cF &=&  \E\bar\E[ \,v_i \bv_j\, ]  \pS
\eeqn
where the $v_j,\bv_j$ are given by the SDE system 
\smallskip
\beqn
dv_j&=&\ts \;-\;i\,(\vep v)_j\, dt  \;-\; i\,2u(N-1)\, {v_j\bv_j\over v\bv} \,v_j\,dt\;-\;i\sqrt{2u}\,v_j\, dx_j  \nn \\
d\bar v_j&=&\ts \;+\;i\,(\vep\bar v)_j\, dt \;+\;i\, 2u(N-1)\, {v_j\bv_j\over v\bv} \,\bv_j\,dt  \;+\;i\sqrt{2u}\,\bar v_j\, dy_j \pS
\eeqn 
with initial conditions $(v,\bv)_0=(\lam,\blam)\,$. In the lage $N$ limit with $g=uN$ fixed, this reduces again, as in Theorem 4 where the initial state 
was a coherent state, to the time dependent discrete Gross-Pitaevskii equation 
\beqn
\dot w_j&=&\ts \;-\;i\,(\vep w)_j  \;-\; i\,2g\, {w_j\bw_j} \,w_j \nn \\ 
{\dot\bw}_j&=&\ts \;+\;i\,(\vep\bar w)_j \;+\;i\,2g\, {w_j\bw_j} \,\bw_j  
\eeqn 
with normalized quantities $(w,\bw):=(v,\bv)/|\lam|$, initial conditions $(w,\bw)_0=(\lam,\blam)/|\lam|\,$ and density matrix elements given by  
$\,(\psi_t\,,a_i^+a^\pp_j \psi_t)_\cF\,=\, N\,w_{i,t} \bw_{j,t}\;$.

\bigskip
\bigskip
\bigskip
The results of Theorems 4 and 5 are in line with rigorous results in the continuous case in the large $N$ limit. In [2], Benedikter, Porta 
and Schlein give an overview on rigorous derivations of effective evolution equations and results concerning the continuous time dependent Gross-Pitaevskii 
equation are summarized in chapter 5. The article [3] focusses solely on the GP equation. Pickl [4,5] and more recently Jeblick, Leopold and Pickl [6] also 
gave rigorous derivations of the continuous time dependent GP equation. The issue has a longer history with more people involved, more background can be found in [2]. 
The fact that the coherent states and the number states of Theorem 4 and 5 show similar dynamics in the large $N$ limit has also been observed by Schachenmayer, 
Daley and Zoller in [7]. 

\goodbreak

\bigskip
\bigskip
\bigskip
\bigskip

\noindent{\large\bf 3. PDE Representations}
\numberwithin{equation}{section}
\renewcommand\thesection{3}
\setcounter{equation}{0}

\bigskip
\bigskip
\noindent{\bf 3.1\; Untransformed Case, before Girsanov Transformation} 

\bigskip
In the untransformed case, the SDE representation for the density matrix elements is given by Theorem 2 of section 1.2. We have 
\beqn
(\,\psi_t\,,a_i^+a^\pp_j \,\psi_t\,)_\cF &=&\E\bar\E[ \,v_i \bv_j\, e^{v\bv} \,] \;e^{-|\lam|^2}\pI \lbeq{3.1}
\eeqn
with 
\beqn
dv_j&=& \;-\;i\, dt\,(\vep v)_j \;-\;i\,\sqrt{2u}\,v_j\,dx_j   \lbeq{3.2}  \\ 
d\bar v_j&=& \;+\;i\, dt\,(\vep\bar v)_j \;+\;i\,\sqrt{2u}\,\bar v_j\,dy_j   \nn 
\eeqn

\smallskip
\medskip
According to the Fresnel version of Kolmogorov's backward equation, a one dimensional version is given in formula (\req{A.83}) with Fresnel expectation (\req{A.75}) in appendix A.4, 
the quantity (\req{3.1}) has a PDE representation. In order to write it down, we need the operator $A$ which is associated with the SDE system (\req{3.2}). 
To this end, we consider some arbitrary complex-valued function $f$ of $\,2\G$ arguments, 
\beqn
f\;=\;f(\,\{v_j\},\{\bar v_j\} \,)&:& \C^{2\G} \; \to \;\mathbb C  \pI
\eeqn

\goodbreak

Because of 
\beqn
(dv_j)^2&=&\bigl\{\,-\;i\, dt(\vep v)_j \;-\;i\sqrt{2u}\,v_j\,dx_j \,\bigr\}^2  \pI \nn \\
&=& \bigl\{\,-\;i\sqrt{2u}\,v_j\,dx_j \,\bigr\}^2 \nn \\ 
&=&\;-\;2u\,v_j^2\,(dx_j)^2  \pI \nn \\ 
&=&\;-\;i\,2u\,dt\, v_j^2
\eeqn
and
\beqn
(d\bv_j)^2&=&\bigl\{\,+\;i\, dt(\vep\bv)_j \;+\;i\sqrt{2u}\,\bv_j\,dy_j \,\bigr\}^2  \pI \nn \\
&=& \bigl\{\,+\;i\sqrt{2u}\,\bv_j\,dy_j \,\bigr\}^2 \nn \\ 
&=&\;-\;2u\,\bv_j^2\,(dy_j)^2  \pI \nn \\ 
&=&\;+\;i\,2u\,dt\,\bv_j^2
\eeqn
and, for $i\ne j$, 
\beqn
dv_i \,d\bv_i\;=\; dv_i\, d\bv_j\;=\;dv_i \,dv_j\;=\;d\bv_i \,d\bv_j\;=\; 0 \pI
\eeqn
we obtain with the Fresnel version of the Ito lemma 
\medskip
\beqn
df&=&\ts\sum_j {\ts \Bigl\{\, {\pt f\over \pt v_j}\, dv_j \,+\, {\pt f\over \pt \bar v_j}\, d\bar v_j \,\Bigr\} }
  \;+\;{\ts {1\over 2}}\,\sum_j \ts \Bigl\{\, {\pt^2 f\over \pt v_j^2}\, (dv_j)^2 \,+\, {\pt^2 f\over \pt \bar v_j^2}\, (d\bar v_j)^2 \,\Bigr\} \pI \nn \\
&=&\ts\;-\;i \,dt\;\sum_j {\ts \Bigl\{\,(\vep v)_j {\pt f\over \pt v_j}\,-\,(\vep\bar v)_j {\pt f\over \pt \bar v_j} \,\Bigr\} }\pI  \nn  \\ 
&&\ts  \;-\;iu\,dt\;\sum_j \ts \Bigl\{\,v_j^2\, {\pt^2 f\over \pt v_j^2} \,-\,\bar v_j^2 \, {\pt^2 f\over \pt \bar v_j^2}\,\Bigr\}
  \;\;\;+\;\;{\rm diffusive} \pI  \nn \\ 
&=:& Af \;\;\;+\;\;{\rm diffusive} \pI
\eeqn
Thus, the expectation
\beqn
F_t&:=&\E\bar\E\Bigl[ \,f(\,\{v_{j,x,t}\},\{\bar v_{j,y,t}\} \,)\;\Bigr] \pI 
\eeqn
considered as a function of its initial values $\,(v_0,\bv_0)=(\,\{v_{j,0}\},\{\bar v_{j,0}\} \,)=(\lam,\bar\lam)\,$, 
\beqn
F_t&=&F_t\bigl(\,\{v_{j,0}\},\{\bar v_{j,0}\} \,\bigr)  \pI 
\eeqn

\smallskip
can be obtained as the solution of the parabolic second order PDE (the zero subscripts on the $v$'s are then usually omitted in the notation, 
$v_{j,0}\to v_j\,$ in the following PDE)
\medskip
\beqn
\ts {\pt F\over \pt t}\;=\;AF&=&\ts -i\,\sum_j {\ts \Bigl\{\,(\vep v)_j\, {\pt F\over \pt v_j}\,-\, 
  (\vep\bar v)_j \,{\pt F\over \pt \bar v_j}\,\Bigr\} } 
  \;-\;iu\,\sum_j \ts \Bigl\{\; v_j^2 \,{\pt^2 F\over \pt v_j^2}\;-\; \bar v_j^2 \,{\pt^2 F\over \pt \bar v_j^2}\;\Bigr\}  \pI \lbeq{3.9}
\eeqn

\smallskip
with initial condition 
\beqn
F_0&=&f(\,\{v_j\},\{\bar v_j\}\,)  \pI
\eeqn
If we introduce the differential operators
\beqn
\cL&=& \cL_0\;+\; \cL_\rint \pI
\eeqn
with
\beqn
\cL_0&:=&\ts \sum_j \Bigl\{\; {\ts (\vep v)_j\, {\pt \over \pt v_j} 
  \,-\,   (\vep\bar v)_j\, {\pt \over \pt \bar v_j}  } \;\Bigr\}  \pI \\ 
\cL_\rint&:=&\ts u\;\sum_j \Bigl\{\;\ts  v_j^2\, {\pt^2 \over \pt v_j^2} \,-\,\bar v_j^2\, {\pt^2 \over \pt \bar v_j^2} \;\Bigr\}  \pI
\eeqn
the solution may be written as 
\beqn
F_t&=&e^{-it\cL}\,F_0   \pI
\eeqn
which one could try to evaluate through small or approximate large Trotter steps according to 
\beqn
F_{t}\;\;=\;\;e^{-it(\cL_0+\cL_\rint)}\,F_0 &\approx& \bigl( e^{-i{t\over k}t\cL_0} \, e^{-i{t\over k}\cL_\rint} \bigr)^k \,F_0 \phantom{mmm}  
\eeqn

\medskip
For the density matrix elements, we have to calculate the expectation (\req{3.1}) and hence the initial condition $F_0=f$ is given by 
\beqn
F_0(\,\{v_j\},\{\bar v_j\}\,)&=&v_i\,\bv_j\; e^{\,v\bv\,-\,|\lam|^2} \pI
\eeqn

\smallskip
The time evolution is then obtained through 
\smallskip
\beqn
(\,\psi_t\,,a_i^+a^\pp_j \,\psi_t\,)_\cF &=&e^{-it(\cL_0+\cL_\rint)}\,\bigl\{\, v_i\,\bv_j\; e^{\,v\bv\,-\,|\lam|^2} \,\bigr\}\, \bigr|_{v=\lam,\bv=\blam}  \pI  \lbeq{3.17}
\eeqn

\medskip
In particular, the solution of the PDE (\req{3.9}) is not needed in the whole $(v,\bv)$\,-\,space, but only at one specific point 
\beqn
(\,\{v_j\},\{\bar v_j\}\,)&=&(\,\{\lam_j\},\{\bar\lam_j\}\,)\;\;.
\eeqn

\bigskip
\bigskip
\noindent{\bf 3.2\; Transformed Case, after Girsanov Transformation} 

\bigskip
After Girsanov transformation, we have the SDE representation which is given by Theorem 4 \\ of section 2.2, 
\beqn
(\,\psi_t\,,a_i^+a^\pp_j \,\psi_t\,)_\cF &=&\E\bar\E[ \,v_i \bv_j\, ]\pI \lbeq{3.19}
\eeqn
with 
\beqn
dv_j&=&\;-\;i\, dt\,(\vep v)_j \;-\; i\,2u\,dt\, v_j\bv_j \,v_j \;-\;i\,\sqrt{2u}\,v_j\,dx_j   \phantom{mmm} \\
d\bar v_j&=&\;+\;i\, dt\,(\vep\bar v)_j  \;+\;i\, 2u\,dt\, v_j\bv_j \,\bv_j  \;+\;i\,\sqrt{2u}\,\bar v_j\,dy_j   \nn
\eeqn 

\medskip
We still have 
\beqn
(dv_j)^2&=&\;-\;i\,2u\,dt\, v_j^2  \\ 
(d\bv_j)^2&=&\;+\;i\,2u\,dt\,\bv_j^2  \nn
\eeqn
and, for $i\ne j$, 
\beqn
dv_i \,d\bv_i\;=\; dv_i\, d\bv_j\;=\;dv_i \,dv_j\;=\;d\bv_i \,d\bv_j\;=\; 0 \pI
\eeqn

\smallskip
Thus, again with the Ito lemma, for some arbitrary $\,f=f(\,\{v_j\},\{\bar v_j\} \,)\,$,
\medskip
\beqn
df&=&\ts \sum_j {\ts \Bigl\{\, {\pt f\over \pt v_j}\, dv_j \,+\, {\pt f\over \pt \bar v_j}\, d\bar v_j \,\Bigr\} }
  \;+\;{\ts {1\over 2}}\,\sum_j \ts \Bigl\{\, {\pt^2 f\over \pt v_j^2}\, (dv_j)^2 \,+\, {\pt^2 f\over \pt \bar v_j^2}\, (d\bar v_j)^2 \,\Bigr\} \pI \nn \\
&=&\ts \;-\;i \,dt\;\sum_j {\ts \Bigl\{\,(\vep v)_j {\pt f\over \pt v_j}\,-\,(\vep\bar v)_j {\pt f\over \pt \bar v_j} \,\Bigr\} } 
  \ts \;-\;i\,2u \,dt\;\sum_j {\ts \Bigl\{\, v_{j}\bv_{j}\,v_j \, {\pt f\over \pt v_j}\,-\,v_{j}\bv_{j}\,\bv_j \, {\pt f\over \pt \bar v_j} \,\Bigr\} }\pI \nn \\ 
&&\ts  \;-\;iu\,dt\;\sum_j \ts \Bigl\{\,v_j^2\, {\pt^2 f\over \pt v_j^2} \,-\,\bar v_j^2 \, {\pt^2 f\over \pt \bar v_j^2}\,\Bigr\}
  \;\;\;+\;\;{\rm diffusive} \pI
\eeqn

\medskip
such that the expectation
\beqn
F_t&:=&\E\bar\E\Bigl[ \,f(\,\{v_{j,x,t}\},\{\bar v_{j,y,t}\} \,)\;\Bigr] \pI 
\eeqn
considered again as a function of its initial values $\,(v_0,\bv_0)=(\,\{v_{j,0}\},\{\bar v_{j,0}\} \,)=(\lam,\bar\lam)\,$, 
\beqn
F_t&=&F_t\bigl(\,\{v_{j,0}\},\{\bar v_{j,0}\} \,\bigr)  \pI 
\eeqn

\smallskip
and omitting the subscripts 0 on the $v,\bv$ in the following, can be obtained as the solution of the second order PDE 
\smallskip
\beqn
\ts {\pt F\over \pt t}&=&\ts -i\,\sum_j {\ts \Bigl\{\,(\vep v)_j\, {\pt F\over \pt v_j}\,-\,  (\vep\bar v)_j \,{\pt F\over \pt \bar v_j}\,\Bigr\} } 
   \;-\;i\,2u \,\sum_j v_{j}\bv_{j}\, {\ts \Bigl\{\, v_j \, {\pt F\over \pt v_j}\,-\,\bv_j \, {\pt F\over \pt \bar v_j} \,\Bigr\} } \pI \nn \\ 
&&\ts \;-\;i\,u\,\sum_j \ts \Bigl\{\; v_j^2 \,{\pt^2 F\over \pt v_j^2}\;-\; \bar v_j^2 \,{\pt^2 F\over \pt \bar v_j^2}\;\Bigr\}  \pI \lbeq{3.26}
\eeqn
with initial condition 
\beqn
F_0&=&f(\,\{v_j\},\{\bar v_j\}\,)  \pI
\eeqn
We write $\,\cL\,=\, \cL_0+ \cL_\rint \,$ as in the untransformed case and abbreviate the additional term, which is due to the exponential $e^{v\bv}$ which 
in turn comes from the initial state $\psi_0$ which was chosen to be a product of coherent states, as  
\beqn
\cL_{\psi_0}&:=&\ts 2u \,\sum_j v_{j}\bv_{j}\, {\ts \Bigl\{\, v_j \, {\pt \over \pt v_j}\,-\,\bv_j \, {\pt \over \pt \bar v_j} \,\Bigr\} }   \pI
\eeqn
Then the solution of the PDE (\req{3.26}) can be written as 
\beqn
F_t&=&e^{-it(\cL+\cL_{\psi_0})}\,F_0   \pI
\eeqn
For the density matrix elements, we now have to calculate the expectation (\req{3.19}) instead of (\req{3.1}) and hence the initial condition $F_0$ is simply given by 
\beqn
F_0(\,\{v_j\},\{\bar v_j\}\,)&=&v_i\,\bv_j \pI
\eeqn
instead of the $\,v_i\,\bv_j\; e^{\,v\bv\,-\,|\lam|^2}\,$ which we had in the untransformed case. 
Thus, the time evolved density matrix elements are obtained through 
\beqn
(\,\psi_t\,,a_i^+a^\pp_j \,\psi_t\,)_\cF &=&e^{-it(\cL_0+ \cL_\rint+\cL_{\psi_0})}\,\{\, v_i\,\bv_j \,\}\, \bigr|_{v=\lam,\bv=\blam}  \pI  \lbeq{3.31}
\eeqn
Again, the solution of the PDE (\req{3.26}) is not needed in the whole $(v,\bv)$\,-\,space, but only 
at one specific point $(v,\bv)=(\lam,\bar\lam)$\;.

\goodbreak

\bigskip
\bigskip
\bigskip
\noindent{\bf 3.3\; The PDE Version of the Girsanov Transformation Formula} 

\bigskip
\bigskip
By comparison of (\req{3.17}) and (\req{3.31}), apparently there has to be the identity 
\smallskip
\beqn
e^{-it(\cL_0+\cL_\rint)}\,\bigl\{\, f(v,\bv)\; e^{\,v\bv\,-\,|\lam|^2} \,\bigr\}\, \bigr|_{v=\lam,\bv=\blam} &=& 
  e^{-it(\cL_0+\cL_\rint+\cL_{\psi_0})}\,\{\, f(v,\bv) \,\}\, \bigr|_{v=\lam,\bv=\blam} \phantom{mm} \pI
\eeqn
for arbitrary functions $f(v,\bv)$. This identity has the following generalization which can be obtained by redoing the calculations of chapter 1 and 2 for 
an arbitrary initial state, not necessarily for a coherent state: Let $P=P(v\bv)$ be some (usually positive when restricted to real values, since it come from 
a $\|\psi_t\|^2$\,) arbitrary function of one real or complex variable. Then 
\smallskip
\beqn
e^{-it(\cL_0+\cL_\rint)}\,\bigl\{\, f(v,\bv)\; \ts {P(v\bv)\over P(\lam\blam)} \,\bigr\}\, \bigr|_{v=\lam,\bv=\blam} &=& 
  e^{-it(\cL_0+\cL_\rint+\cL_{P})}\,\{\, f(v,\bv) \,\}\, \bigr|_{v=\lam,\bv=\blam}  \phantom{mm} \pI
\eeqn
or equivalently 
\beqn
e^{-it(\cL_0+\cL_\rint)}\,\bigl\{\,  P(v\bv)\,f(v,\bv) \,\bigr\} &=& 
  P(v\bv)\;\times\; e^{-it(\cL_0+\cL_\rint+\cL_{P})}\,\{\, f(v,\bv) \,\} \pI  \lbeq{3.34}
\eeqn
with
\beqn
\cL_{P}&:=&\ts 2u \,{\ts { P'(v\bv)\over P(v\bv)}} \;\sum_j v_{j}\bv_{j}\, {\ts \Bigl\{\, v_j \, {\pt \over \pt v_j}\,-\,\bv_j \, {\pt \over \pt \bar v_j} \,\Bigr\} }   \pI
\eeqn
and $P'(x)=dP/dx$. 

\bigskip
\noindent Since (\req{3.34}) is completely independent of any stochastics, just some algebraic statement concerning derivatives, let us also give an 
independent proof, thereby confirming the validity of the stochastic formalism\footnote{recall that we are using the stochastic calculus formalism with 
respect to Fresnel measure, not with Wiener measure, which is not part of standard rigorous textbook mathematics} which has been used so far: First, one calculates 
that for any $\,P=P(v\bv)=P\bigl(\Sigma_j v_j\bv_j)\,$, one has, using $\vep_{ij}=\vep_{ji}$ for the $\cL_0$ equation, 
\beqn
\cL_0P&=& 0 \\ 
\cL_\rint P&=& 0 
\eeqn
Since $\cL_0$ is a first order operator, one has for arbitrary $f=f(\,\{v_j\},\{\bar v_j\} \,)$
\beqn
\cL_0(Pf)&=&\cL_0(P) \times f\;+\; P\times \cL_0(f) \;\;=\;\; P\times \cL_0(f) \pI
\eeqn
Furthermore, 
\beqn
\cL_\rint(Pf)&=&\ts u\;\sum_j \Bigl\{\;\ts  v_j^2\, {\pt^2 \over \pt v_j^2} \,-\,\bar v_j^2\, {\pt^2 \over \pt \bar v_j^2} \;\Bigr\} (Pf) \pI \nn \\ 
&=&\ts \cL_\rint(P)\times f\;+\; P\times\cL_\rint(f)\;+\; 
  2u\;\sum_j \Bigl\{\;\ts  v_j^2\, {\pt P \over \pt v_j}\,{\pt f \over \pt v_j} \,-\,\bar v_j^2\, {\pt P \over \pt \bar v_j}\,{\pt f \over \pt \bar v_j} \;\Bigr\} \pI\nn \\ 
&=&\ts P\times\cL_\rint(f)\;+\; 
  2u\;\sum_j \Bigl\{\;\ts  v_j^2\, P'\,\bv_j\,{\pt f \over \pt v_j} \,-\,\bar v_j^2\, P'\,v_j\,{\pt f \over \pt \bar v_j} \;\Bigr\} \pI \nn\\ 
&=&\ts P\times\cL_\rint(f)\;+\; 
  2uP'\;\sum_j v_j\bv_j\,\Bigl\{\;\ts  v_j\,{\pt f \over \pt v_j} \,-\,\bar v_j\,{\pt f \over \pt \bar v_j} \;\Bigr\} \pI \nn\\ 
&=&\ts P\times\bigl[\,\cL_\rint(f) \,+\,\cL_P(f) \, \bigr]  \pI
\eeqn
Thus, 
\beqn
(\cL_0+\cL_\rint)(Pf)&=&P\;\times \;(\cL_0+\cL_\rint+\cL_P)(f)  \pI \lbeq{3.41n}
\eeqn
and by induction 
\beqn
(\cL_0+\cL_\rint)^{n+1}(Pf)&=&(\cL_0+\cL_\rint)\,\bigl\{\, (\cL_0+\cL_\rint)^{n}(Pf) \,\bigr\} \pI \nn \\ 
&=&(\cL_0+\cL_\rint)\,\bigl\{\, P\,\times\, (\cL_0+\cL_\rint+\cL_P)^{n}(f) \,\bigr\} \pI \nn \\ 
&\buildrel (\req{3.41n})\over =& P\;\times \; (\cL_0+\cL_\rint+\cL_P)\,\Bigl\{\, (\cL_0+\cL_\rint+\cL_P)^{n}(f) \, \Bigr\} \pI \nn \\ 
&=&P\;\times \; (\cL_0+\cL_\rint+\cL_P)^{n+1}(f) \pI
\eeqn
This proves (\req{3.34}). Let us summarize the results of this chapter in the following 

\bigskip
\bigskip
\bigskip
\noindent{\bf Theorem 6:} For an arbitrary real symmetric hopping matrix $\vep=(\vep_{ij})=(\vep_{ji})$, define the differential operators 
\beqn
\cL_0&:=&\ts\sum_j \Bigl\{\; {\ts (\vep v)_j\, {\pt \over \pt v_j}  \,-\,   (\vep\bar v)_j\, {\pt \over \pt \bar v_j}  } \;\Bigr\} 
\;\;=\;\;\sum_{i,j} {\ts \,\vep_{ij}\Bigl\{\; v_i\, {\pt \over \pt v_j} \,-\, \bar v_i\, {\pt \over \pt \bar v_j}  } \;\Bigr\} \pI \\ 
\cL_\rint&:=&\ts u\;\sum_j \Bigl\{\;\ts  v_j^2\, {\pt^2 \over \pt v_j^2} \,-\,\bar v_j^2\, {\pt^2 \over \pt \bar v_j^2} \;\Bigr\}  \pI
\eeqn
Furthermore, for some arbitrary function of one variable $\,P=P(v\bv)=P\bigl(\Sigma_j v_j\bv_j\bigr)$,\, put 
\beqn
\cL_{P}&:=&\ts 2u \,{\ts { P'(v\bv)\over P(v\bv)}} \;\sum_j v_{j}\bv_{j}\, {\ts \Bigl\{\, v_j \, {\pt \over \pt v_j}\,-\,\bv_j \, {\pt \over \pt \bar v_j} \,\Bigr\} }   \pI
\eeqn
Then:
\begin{itemize}
\item[{\bf a)}] The density matrix elements of Theorem 2 can be calculated through the following formula, this is the untransformed case before Girsanov transformation: 
\beqn
(\,\psi_t\,,a_i^+a^\pp_j \,\psi_t\,)_\cF &=&e^{-it(\cL_0+\cL_\rint)}\,\bigl\{\, v_i\,\bv_j\; e^{\,v\bv\,-\,|\lam|^2} \,\bigr\}\, \bigr|_{v=\lam,\bv=\blam}  \pI
\eeqn
\item[{\bf b)}] The density matrix elements of Theorem 2 can be calculated through the following equivalent formula, obtained from the SDE representation 
(\req{2.59},\req{2.60}) after Girsanov transformation: 
\beqn
(\,\psi_t\,,a_i^+a^\pp_j \,\psi_t\,)_\cF &=&e^{-it(\cL_0+\cL_\rint+\cL_P)}\,\bigl\{\, v_i\,\bv_j \,\bigr\}\, \bigr|_{v=\lam,\bv=\blam}  \pS
\eeqn
with $P(x)=e^x$\;.
\item[{\bf c)}] For an arbitrary function $P=P(x)$ of one variable, there is the following general identity:
\beqn
e^{-it(\cL_0+\cL_\rint)}\,\bigl\{\,  P(v\bv)\,f(v,\bv) \,\bigr\} &=& P(v\bv)\;\times\; e^{-it(\cL_0+\cL_\rint+\cL_{P})}\,\{\, f(v,\bv) \,\} \phantom{mm}  \pI
\eeqn
with $f=f(v,\bv)=f(\,\{v_j\},\{\bar v_j\} \,)\,$ being an arbitrary function of $2\G$ variables. 
\end{itemize}

\bigskip
\bigskip
\bigskip
\bigskip

\goodbreak

\noindent{\large\bf 4. An Explicit Solvable Test Case: The 0D Bose-Hubbard Model}
\numberwithin{equation}{section}
\renewcommand\thesection{4}
\setcounter{equation}{0}

\bigskip
\bigskip
\noindent Since stochastic calculus with Fresnel Brownian motions instead of standard Brownian motions is not part of rigorous textbook mathematics, 
let's make an additional check of the formalism by applying it to an explicit solvable test case, the 0D Bose-Hubbard model. The term 0D Bose-Hubbard model 
we borrowed from the paper [8] of Ray, Ostmann, Simon, Grossmann and Strunz, where the model also had to serve as a test example. Besides of just being 
a test case, the purpose of this chapter is also to give some intuition for the appro\-ximation which will be used in section 5.3 to take the diffusive part 
of the SDE system into account. 

\bigskip
In the Bargmann-Segal representation (\req{1.4}), the Hamiltonian of the 0D Bose-Hubbard model is simply 
\beqn
h\;\;=\;\; h_0\,+\,h_\rint&:=&\ts \vep\,z {d\over dz} \;+\; u\, z^2 {d^2\over dz^2}  \pS
\eeqn
We choose the initial state 
\beqn
\psi_0(z)&:=&e^{\lambda z} \;e^{-|\lambda|^2/2} \pS
\eeqn
with $\lambda\in\mathbb C$ and consider the time evolution 
\beqn
\psi_t&:=&e^{-ith}\psi_0  \pS
\eeqn
We want to calculate the quantity, with $\,a=d/dz$\,, $\,a^+=z\,$,
\beqn
(\psi_t,a\,\psi_t)_\cF&=& (a^+\psi_t,\psi_t)_\cF  \;\;=\;\; \ts \int_{\mathbb C} d\mu(z)\;z\;|\psi_t(z)|^2 \pI  \nn \\ 
&=&\ts \int_{\mathbb R^2} {d\re z\,d\im z\over \pi}\; e^{-|z|^2} \; z\;|\psi_t(z)|^2 
\eeqn
Since 
\beqn
h\, z^n&=&\bigl\{\vep n +un(n-1)\bigr\}\,z^n \;\;=:\;\; h_n\,z^n 
\eeqn
we have 
\beqn
\psi_t(z)&=&\sum_{n=0}^\infty \ts{\lambda^n\over n!}\; e^{-ith_n}\,z^n \;e^{-|\lambda|^2/2} \pI
\eeqn
such that 
\beqn
(\psi_t,a\,\psi_t)_\cF &=&\sum_{n,m=0}^\infty e^{-it(h_n-h_m)}\; \ts{\lambda^n\bar\lambda^m\over n!\,m!}\; \ts \int_{\mathbb R^2} {d^2z\over \pi}\; e^{-|z|^2} \;
     {z}^{n+1} {\bar z}^{m}\;e^{-|\lambda|^2} \pI \nn \\ 
&=&\sum_{n=0}^\infty e^{-it(h_{n}-h_{n+1})}\;\ts {\lambda^{n}{\bar\lambda}^{n+1}\over n!\,(n+1)!}\; 
  \int_{\mathbb R^2} {d^2z\over \pi}\; e^{-|z|^2} \; |z|^{2(n+1)}\;e^{-|\lambda|^2} \pI \nn \\ 
&=&\sum_{n=0}^\infty e^{+it(h_{n+1}-h_{n})}\;\ts {\lambda^{n}{\bar\lambda}^{n+1}\over n!}\;\;e^{-|\lambda|^2}  \pI 
\eeqn
Since 
\beqn
h_{n+1}-h_n&=&\vep-u+u(2n+1)\;\;=\;\;\vep +2u\,n 
\eeqn
we end up with 
\beqn
\phantom{mmmm}(\psi_t,a\,\psi_t)_\cF &=& \bar\lam\, \sum_{n=0}^\infty e^{+it(\vep +2u\,n )}\; {\ts {(\lambda{\bar\lambda})^n\over n!} }\;e^{-|\lambda|^2}   \pI  \nn \\
&=&\bar\lambda\; e^{i\vep t}\; \exp\bigl\{\, -\,(1-e^{2iut})\,|\lambda|^2\,\bigr\} \pI  \lbeq{4.9}
\eeqn

\smallskip
This quantity has already collapse and revivals, if we plot $\,\re(\psi_t,a\,\psi_t)_\cF\,$ for $\,\vep=2\,$, $\,u=g/N\,$ with $\,g=0.5\,$ and $\,N=|\lam|^2=20\,$, we get 

\medskip
\bigskip
\bigskip
\centerline{\includegraphics[width=11cm]{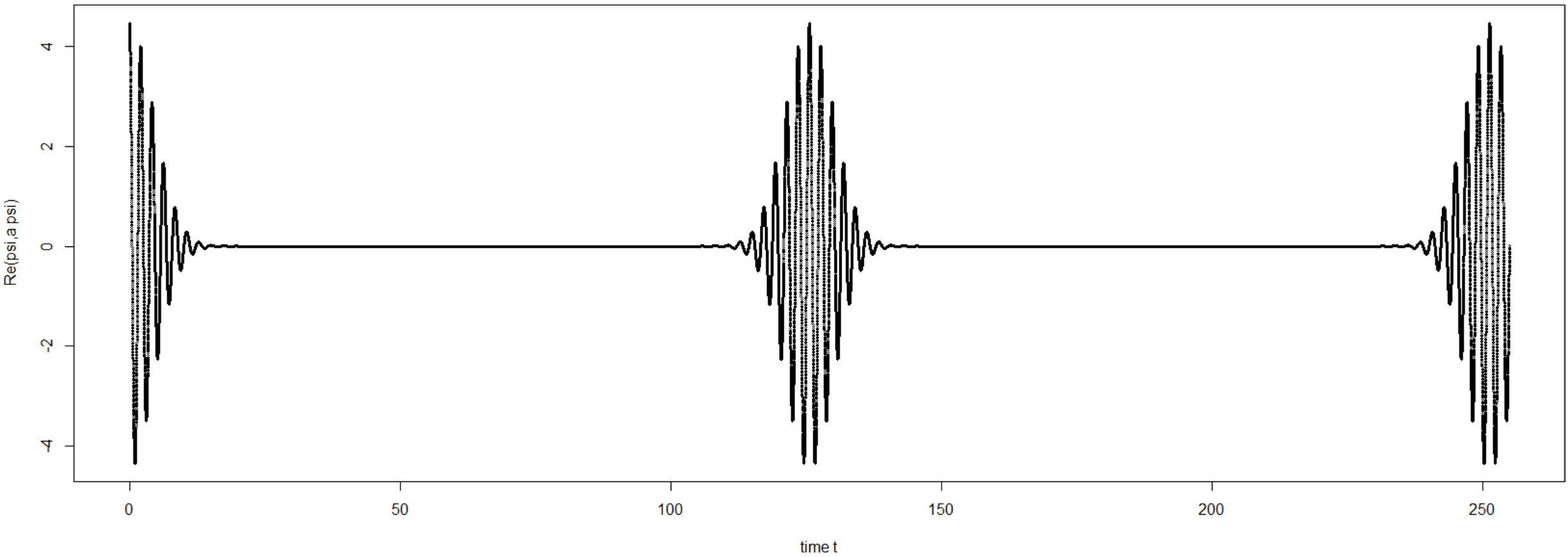}}

\bigskip
\bigskip 
Let's apply Theorem 2. We get the following representation: 
\medskip
\beqn
(\,\psi_t\,,a \,\psi_t\,)_\cF &=&\E_x\bar\E_y\bigl[ \;\bv_{y,t}\; e^{v_{x,t}\bv_{y,t}} \;\bigr] \;e^{-|\lam|^2}   \phantom{mmm} \lbeq{4.10}
\eeqn

\medskip
where the $v_t,\bv_t\in\C$ are given by the SDEs 
\medskip
\beqn
dv_t&=& \;-\;i\, dt\,\vep\, v_t \;-\;i\,\sqrt{2u}\,v_t\,dx_t  \phantom{mmm} \nn  \\ 
d\bar v_t&=&\;+\;i\, dt\,\vep\,\bar v_t \;+\;i\,\sqrt{2u}\,\bar v_t\,dy_t  
\eeqn

\medskip
with initial conditions $\,v_{0}=\lam,\,\bv_{0}=\bar\lam\,$. If we would have Wiener measure instead of Fresnel measure, 
this would be a geometric Brownian motion. Here we have Fresnel measure with calculation rules
\beqn
(dx_t)^2&=&+i\,dt  \nn \\ 
(dy_t)^2&=&-i\,dt  \lbeq{4.12}
\eeqn
and obtain the solutions
\beqn
v_t&=&v_0\, e^{-i(\vep-u)t\,-\,i\sqrt{2u}\,x_t } \nn \\
\bv_t&=&\bv_0\, e^{+i(\vep-u)t\,+\,i\sqrt{2u}\,y_t }
\eeqn
Namely, the Ito formula applied to $\,v_t=v_0\, e^{-i(\vep-u)t\,-\,i\sqrt{2u}\,x_t }= v(x_t,t)\,$ gives
\beqn
dv_t&=&\ts {\pt v \over \pt x} \, dx_t \;+\; {1\over 2}\,{\pt^2 v \over \pt x^2}\,(dx_t)^2 \;+\; {\pt v \over \pt t} \, dt  \pI \nn \\ 
&=&\ts \,-\,i\sqrt{2u}\,v_t\,dx_t \;+\; {1\over 2}\,(-i\sqrt{2u})^2\,v_t\,(dx_t)^2 \;-\;i\,(\vep-u)\,v_t\, dt  \nn  \\ 
&\buildrel (\req{4.12})\over=& \ts \,-\,i\sqrt{2u}\,v_t\,dx_t \;+\; {1\over 2}\,(-i\sqrt{2u})^2\,v_t\;(+i)\,dt \;-\;i\,(\vep-u)\,v_t\, dt  \nn  \pI \\ 
&=&\ts \,-\,i\sqrt{2u}\,v_t\,dx_t \;-\;i\,\vep\,dt\, v_t
\eeqn

\medskip
For some integrand which depends on $x_t$ only and not on earlier $\{x_{t'}\}_{t'<t}\,$, Fresnel or Wiener expectations reduce to 1-dimensional integrals 
(appendix A4 has a general formula in (\req{A.86})). We have 
\beqn
\E_x\bar\E_y\bigl[ \;f(x_t,y_t)\;\bigr]&=&\ts \int_{\R^2} f(x_t,y_t)\; e^{\,i{x_t^2-y_t^2\over 2t}} \,\ts{dx_t\,dy_t\over 2\pi t} \pI \nn \\ 
&=&\ts \int_{\R^2} \,f(\,\sqrt{t}\,x\,,\,\sqrt{t}\,y\,)\; e^{\,i{x^2-y^2\over 2}} \,\ts{dx\,dy\over 2\pi} \pI \lbeq{4.15}
\eeqn
Now we substitute 
\beqn
\ts \xi\;\;=\;\;{x-y\over \sqrt{2}}\;,\;\;\;\; \eta\;\;=\;\;{x+y\over \sqrt{2}} 
\eeqn
which gives 
\beqn
 e^{i{x^2-y^2\over 2}} \,\ts{dx\,dy\over 2\pi} &=&  e^{i\xi\eta} \;\ts{d\xi\,d\eta\over 2\pi}  \pI 
\eeqn
The quantities in the integrand in (\req{4.15}) are given by (\req{4.10}),
\beqn
\;e^{v_{x,t}\bv_{y,t}}&=&\exp\bigl\{\, |\lam|^2\,e^{\,-\,i\sqrt{2u}\,(x_t-y_t)\,} \,\bigr\} \;\;=\;\;\exp\bigl\{\, |\lam|^2\,e^{\,-\,i\sqrt{4ut}\,\xi\,} \,\bigr\} \pI
\eeqn
and
\beqn
\;\bv_{y,t}\; e^{v_{x,t}\bv_{y,t}}&=& \bar\lam\; e^{+i(\vep-u)t\,+\,i\sqrt{2u}\,y_t } \;\times\; \exp\bigl\{\, |\lam|^2\,e^{\,-\,i\sqrt{4ut}\,\xi\,} \,\bigr\} \pI \nn \\ 
&=&\bar\lam\; e^{+i(\vep-u)t} \; e^{\,+\,i\sqrt{ut}\,(\eta-\xi) } \;\times\; \exp\bigl\{\, |\lam|^2\,e^{\,-\,i\sqrt{4ut}\,\xi\,} \,\bigr\} 
\eeqn
Thus we have to evaluate 
\beqn
(\,\psi_t\,,a \,\psi_t\,)_\cF &=&\E_x\bar\E_y\bigl[ \;\bv_{y,t}\; e^{v_{x,t}\bv_{y,t}} \;\bigr] \;e^{-|\lam|^2}   \pI \nn \\ 
&=&\ts\int_{\R^2} \bar\lam\, e^{+i(\vep-u)t} \; e^{\,+\,i\sqrt{ut}\,(\eta-\xi) } \,\times\, \exp\bigl\{\, |\lam|^2\,(e^{\,-\,i\sqrt{4ut}\,\xi\,}-1) \,\bigr\} 
  \;  e^{i\xi\eta} \;\ts{d\xi\,d\eta\over 2\pi}  \pI   \nn \\ 
&=& \ts \bar\lam\; e^{+i(\vep-u)t} \int_{\R} d\xi\; e^{\,-\,i\sqrt{ut}\,\xi } \,\ts \int_{\R}  e^{\,+\,i(\xi+\sqrt{ut})\,\eta } \;\ts{d\eta\over 2\pi} \; 
   \times\; \exp\bigl\{\, |\lam|^2\,(e^{\,-\,i\sqrt{4ut}\,\xi\,}-1) \,\bigr\}  \pI   \nn \\ 
&=&\bar\lam\;\ts  e^{+i(\vep-u)t} \int_{\R} d\xi\; e^{\,-\,i\sqrt{ut}\,\xi } \;\delta(\xi+\sqrt{ut}) \; 
   \times\; \exp\bigl\{\, |\lam|^2\,(e^{\,-\,i\sqrt{4ut}\,\xi\,}-1) \,\bigr\}  \pI   \nn \\ 
&=&\bar\lam\;\ts  e^{+i(\vep-u)t} \; e^{\,+\,iut }  \;  \times\; \exp\bigl\{\, |\lam|^2\,(e^{\,+\,i\,2ut\,}-1) \,\bigr\}  \pI   \nn \\ 
&=&\bar\lam\;\ts  e^{+i\vep t}  \;  \times\; \exp\bigl\{\, |\lam|^2\,(e^{\,+\,i\,2ut\,}-1) \,\bigr\}  \pI 
\eeqn
and this coincides with the exact result (\req{4.9}). Now let's apply Theorem 4. After Girsanov transformation, we have the following representation (we omit the 
tilde on the transformed Fresnel BMs in the following):
\beqn
(\,\psi_t\,,a \,\psi_t\,)_\cF &=&\E_x\bar\E_y\bigl[ \;\bv_{t} \;\bigr]    \pI
\eeqn
with $v,\bv$ given by 
\beqn
dv_t&=& \;-\;i\, dt\,\vep\, v_t \;-\;i\,2u\,dt\,v_t\bv_t\,v_t \;-\;i\,\sqrt{2u}\,v_t\,dx_t  \phantom{mmm} \nn  \\ 
d\bar v_t&=&\;+\;i\, dt\,\vep\,\bar v_t \;+\;i\,2u\,dt\,v_t\bv_t\,\bv_t \;+\;i\,\sqrt{2u}\,\bar v_t\,dy_t  
\eeqn

\smallskip
This SDE system can still be solved in closed form. First we calculate 
\beqn
d(v\bv)&=&dv\,\bv \;+\;v\,d\bv\;+\; dv\,d\bv   \nn  \pI\\ 
&=& \;-\;i\, dt\,\vep\, v\bv \;-\;i\,2u\,dt\,v\bv\,v\bv \;-\;i\,\sqrt{2u}\,v\bv\,dx_t \nn \\ 
&& \;+\;i\, dt\,\vep\,v\bar v \;+\;i\,2u\,dt\,v\bv\,v\bv \;+\;i\,\sqrt{2u}\,v\bar v\,dy_t  \;\;+\;\;0   \nn \\ 
&=& \;-\;i\,\sqrt{2u}\,v\bv\,(dx_t-dy_t) \;\;=\;\; \;-\;i\,\sqrt{4u}\,v\bv\, d\xi_t \pI \lbeq{4.23}
\eeqn
Since 
\beqn
(d\xi_t)^2\;\;=\;\;\ts \bigl(\,{dx_t-dy_t\over \sqrt{2}}\,\bigr)^2 \;\;=\;\;\ts { (dx_t)^2 \,-\,2dx_tdy_t \,+\,(dy_t)^2 \over 2 } \;\;=\;\;
 \ts  { +i\,dt \,-\,0\,-\,i\,dt \over 2} \;\;=\;\; 0  \pI
\eeqn
the solution to (\req{4.23}) is 
\beqn
(v\bv)_t&=&(v\bv)_0\; e^{-i\sqrt{4u}\,\xi_t} \;\;=\;\;|\lam|^2\; e^{-i\sqrt{4u}\,\xi_t}  \pI 
\eeqn
Thus, the equation for $\bv$ becomes 
\beqn
d\bv&=&\Bigl[\;+\;i\,\bigl(\,\vep\,+\,2u|\lam|^2\, e^{-i\sqrt{4u}\,\xi_t} \,\bigr)\, dt \;+\;i\,\sqrt{2u}\,dy_t \;\Bigr]\, \bv   \pI 
\eeqn
This is a geometric Fresnel BM with a time dependent and stochastic drift. To solve it, we have to take into account that $\,(dy_t)^2=-i\,dt\,$ and obtain
\beqn
\bv_t&=&\bv_0\; e^{\,+\,i\int_0^t\bigl(\,\vep\,+\,2u|\lam|^2\, e^{-i\sqrt{4u}\,\xi_s} \,\bigr)\, ds \,+\,i\sqrt{2u}\,y_t\,-\,iut } \pI \nn \\ 
&=&\bar\lam\;e^{+i(\vep-u)t}\;e^{\,+\,i\,2u|\lam|^2\int_0^t e^{-i\sqrt{4u}\,\xi_s}\, ds \;+\;i\sqrt{2u}\,y_t\, } \nn \\ 
&=&\bar\lam\;e^{+i(\vep-u)t}\;e^{\,+\,i\,2u|\lam|^2\int_0^t e^{-i\sqrt{4u}\,\xi_s}\, ds \;+\;i\sqrt{u}\,(\eta_t-\xi_t)\, } \pI
\eeqn
Since $\bv_t$ is not just a function of the Fresnel BMs at time $t$, but it depends also through the $ds$-integral in the exponent 
on the Fresnel BMs $\xi_s$ at earlier times $s<t$, we can't no longer make a large step evaluation of the expectation value through a one or two dimensional 
integral, but we have to write down the full small step path integral. Recall the notations 
\beqn
x_{t_k}&=&\ts\sqrt{dt}\,\sum_{\ell=1}^k \phi_\ell \pI \nn\\ 
y_{t_k}&=&\ts\sqrt{dt}\,\sum_{\ell=1}^k \theta_\ell
\eeqn
and the Fresnel measure 
\beqn
dF\,d\bar F&=& \pro_{\ell=1}^k e^{\,i\,{\phi_\ell^2-\theta_\ell^2\over 2}}\;\ts {d\phi_\ell \,d\theta_\ell\over 2\pi} \pS 
\eeqn
Let us write
\beqn
\xi_{t_k}\;\;=\;\;\ts {x_{t_k}-y_{t_k}\over \sqrt{2}} \;\;=\;\;\sqrt{dt}\,\sum_{\ell=1}^k {\phi_\ell-\theta_\ell \over \sqrt{2}} &=:&\ts\sqrt{dt}\,\sum_{\ell=1}^k \alpha_\ell \pI \nn\\ 
\eta_{t_k}\;\;=\;\;\ts {x_{t_k}+y_{t_k}\over \sqrt{2}} \;\;=\;\;\sqrt{dt}\,\sum_{\ell=1}^k {\phi_\ell+\theta_\ell \over \sqrt{2}} &=:&\ts\sqrt{dt}\,\sum_{\ell=1}^k \beta_\ell
\eeqn

\medskip
such that the Fresnel measure becomes
\beqn
dF\,d\bar F&=& \pro_{\ell=1}^k e^{\,i\,\alpha_\ell\beta_\ell}\;\ts {d\alpha_\ell \,d\beta_\ell\over 2\pi} \pI 
\eeqn
Then in discrete time $t=t_k=kdt\,$ the $\bv$ is given by  
\beqn
\bv_{t_k}&=&\bar\lam\;e^{+i(\vep-u)t_k}\;\ts \exp\Bigl\{\,+\,i\,2u|\lam|^2\sum_{\ell=1}^k e^{-i\sqrt{4u}\,\xi_{t_\ell}}\, dt
    \;+\;i\sqrt{udt}\,\sum_{\ell=1}^k(\beta_\ell-\alpha_\ell)\, \Bigr\} \pI  \lbeq{4.32}
\eeqn
We have to calculate
\beqn
(\,\psi_t\,,a \,\psi_t\,)_\cF &=&\E_x\bar\E_y\bigl[ \;\bv_{t_k} \;\bigr]   \;\;=\;\; 
   {\ts \int_{\R^{2k}}  \;\bv_{t_k}(\alpha,\beta) }\; \pro_{\ell=1}^k e^{\,i\,\alpha_\ell\beta_\ell}\;\ts {d\alpha_\ell \,d\beta_\ell\over 2\pi} \pS 
\eeqn
The $\beta$-integrals produce $\delta$-functions,
\beqn
{\ts \int_{\R^{k}} \exp\bigl\{\;+\;i\sqrt{udt}\,\sum_{\ell=1}^k \beta_\ell\, \bigr\} }\; 
  \pro_{\ell=1}^k e^{\,i\,\alpha_\ell\beta_\ell}\;\ts {d\beta_\ell\over 2\pi} &=& \pro_{\ell=1}^k \,\delta(\sqrt{udt}+\alpha_\ell)  \pI
\eeqn
Thus we have $\,\alpha_\ell=-\sqrt{udt}\,$ for all $\ell$ and the $\xi_{t_\ell}$ in (\req{4.32}) becomes  
\beqn
\xi_{t_\ell}&=&\ts\sqrt{dt}\,\sum_{m=1}^\ell \alpha_m \;\;=\;\; -\,\sqrt{u}\,dt\,\ell \;\;=\;\; -\, \sqrt{u}\,t_\ell  \pI 
\eeqn
We end up with 
\beqn
\E_x\bar\E_y\bigl[ \;\bv_{t_k} \;\bigr] &=& \bar\lam\;e^{+i(\vep-u)t_k}\;\ts \exp\Bigl\{\,+\,i\,2u|\lam|^2\sum_{\ell=1}^k e^{+i\sqrt{4u}\,\sqrt{u}\,t_\ell}\, dt
    \;+\;i\sqrt{udt}\,\sum_{\ell=1}^k\,\sqrt{udt} \; \Bigr\} \pI  \nn \\ 
&=& \bar\lam\;e^{+i(\vep-u)t_k}\;\ts \exp\Bigl\{\,+\,i\,2u|\lam|^2\sum_{\ell=1}^k e^{+i\,2u\,t_\ell}\, dt \;+\;i\,udt\, k \; \Bigr\} \pI  \nn \\ 
&\buildrel dt\to 0 \over =& \bar\lam\;e^{+i \vep t}\;\ts \exp\Bigl\{\,+\,i\,2u|\lam|^2\int_0^t e^{+i\,2u\,s}\, ds \; \Bigr\} \pI  \nn  \\ 
&=& \bar\lam\;e^{+i \vep t}\;\ts \exp\Bigl\{\;|\lam|^2( e^{+i\,2u\,t}-1) \; \Bigr\} \pI
\eeqn
and this again coincides with the original result (\req{4.9}). 

\bigskip
\bigskip
Finally, let us check the PDE representations of chapter 3 which are summarized in Theorem\,6. There we have the operators $\cL_0$, $\cL_\rint$ and $\cL_P$ with $P(x)=e^x$ 
which for the current 0D case reduce to 
\beqn
\cL_0&=&\ts  \vep\, \bigl\{\, v\, {\pt \over \pt v} \,-\, \bar v\, {\pt \over \pt \bar v}   \,\bigr\}  \nn \\ 
\cL_\rint&=&\ts u\,\bigl\{\,  v^2\, {\pt^2 \over \pt v^2} \,-\,\bar v^2\, {\pt^2 \over \pt \bar v^2} \,\bigr\} \pS\phantom{mm} \\
\cL_{P}&=&\ts 2u \,v\bv\, \bigl\{\, v \, {\pt \over \pt v}\,-\,\bv \, {\pt \over \pt \bar v} \,\bigr\}    \nn
\eeqn
Theorem 6 makes the following statements: 

\medskip
\bigskip 
\noindent{a)} Untransformed Representation: 
\beqn
(\,\psi_t\,,a \,\psi_t\,)_\cF &\buildrel !\over=&e^{-it(\cL_0+\cL_\rint)}\,\bigl\{\,\bv\; e^{\,v\bv\,-\,|\lam|^2} \,\bigr\}\, \bigr|_{v=\lam,\bv=\blam}  \pI  \lbeq{4.38}
\eeqn
\noindent{b)} Transformed Representation: 
\beqn
(\,\psi_t\,,a \,\psi_t\,)_\cF &\buildrel !\over=& e^{-it(\cL_0+\cL_\rint+\cL_P)}\,\bigl\{\,\bv \,\bigr\}\, \bigr|_{v=\lam,\bv=\blam}  \pI  \lbeq{4.39}
\eeqn

\bigskip 
In case (b), we have to show that the function 
\beqn
F_t(v,\bv)&:=& \bv\;e^{+i \vep t}\,\ts \exp\bigl\{\,v\bv( e^{+i\,2u\,t}-1) \, \bigr\}\pI
\eeqn
this is the original result (\req{4.9}) with the $(\lam,\blam)$ replaced by $(v_0,\bv_0)\to(v,\bv)$, is a solution of the PDE 
\beqn
\ts i\,{\pt \over \pt t}\, F_t&=&(\cL_0+\cL_\rint+\cL_P)\,F_t  \pI \lbeq{4.41}
\eeqn
with initial condition $F_0=\bv\,$. The initial condition is obvious, so let's check the derivatives. We have 
\beqn
\ts i\,{\pt \over \pt t}\, F_t&=&\,-\,\vep\,F_t\;-\;2u\,e^{+i2ut}v\bv\, F_t 
\eeqn
and one calculates
\beqn
\cL_0\,F_t&=&\,-\,\vep\,F_t  \nn\\ 
\cL_\rint\,F_t&=&\,-\,2u\,(e^{+i2ut}-1)v\bv\, F_t \\ 
\cL_P\,F_t&=&\,-\,2u\,v\bv\, F_t \nn
\eeqn
which validates equation (\req{4.41}) or (\req{4.39}). 

\bigskip
\noindent In case (a), we have to show that the function 
\beqn
G_t(v,\bv)&:=&F_t(v,\bv)\;e^{\,+\,v\bv}\;\;=\;\; \bv\;e^{+i \vep t}\,\ts \exp\bigl\{\,v\bv\, e^{+i\,2u\,t} \, \bigr\}\pI 
\eeqn
is a solution of the PDE 
\beqn
\ts i\,{\pt \over \pt t}\, G_t&=&(\cL_0+\cL_\rint)\,G_t  \pI 
\eeqn
with initial condition $G_0=\bv\,e^{v\bv}\,$. The initial condition is obvious, so let's check the derivatives. We have 
\beqn
\ts i\,{\pt \over \pt t}\, G_t&=&\,-\,\vep\,G_t\;-\;2u\,e^{+i2ut}v\bv\, G_t 
\eeqn
and one calculates
\beqn
\cL_0\,G_t&=&\,-\,\vep\,G_t \nn \\ 
\cL_\rint\,G_t&=&\,-\,2u\,e^{+i2ut}v\bv\, G_t
\eeqn
which verifies equation (\req{4.38}) of the untransformed representation (a).

\bigskip
\bigskip
\bigskip
\bigskip

\goodbreak

\noindent{\large\bf 5. The Two Site Bose-Hubbard Model}
\numberwithin{equation}{section}
\renewcommand\thesection{5}
\setcounter{equation}{0}

\bigskip
\bigskip
In this chapter we consider Bose-Hubbard model with just two lattice sites, 1 and 2. Because of its numerical simplicity, 
everything can be calculated easily with exact diagonalization, but its highly nontrivial physics, the model exhibits collapse and revivals 
and has a phase transition between an oscillatory and a self trapping regime, it provides an extremely beautiful test case for any calculation 
scheme which aims at an efficient and reasonable description of quantum many body systems. In fact, about 80 to 90 percent of the research time 
has been spent within that model, the generalization to arbitrary dimensions in the end then being more or less straightforward. 

\bigskip
The Hamiltonian in the Bargmann-Segal representation is
\beqn
h&=&\ts \sum_{i,j=1}^2 \vep_{ij}\,a^+_i a_j\;+\;u\,\sum_{j=1}^2 a_j^+a_j^+a_ja_j \pI \nn \\ 
&=&\ts \vep\,\bigl(\,z_1{\pt\over \pt z_2} \,+\, z_2{\pt\over \pt z_1}\,\bigr) \;+\: u\,\bigl(\; z_1^2\,{\pt^2\over \pt z_1^2}\,+\, z_2^2\,{\pt^2\over \pt z_2^2}\;\bigr) \pI 
\eeqn
Thus, the connection to the standard notation with hopping $J$ and interaction strength $U$ is made through 
\beqn
\vep&=&-J \\ 
u&=&U/2 
\eeqn
and we ignore any on-diagonal trapping potentials, that is, we put $\vep_{jj}=\eps_j=0$. Then Theorem 4 of section 2.2 for coherent states and 
Theorem 5 of section 2.3 for number states can be summarized by the following SDE system 
\medskip
\beqn
dv_1&=&\ts \;-\;i\,\vep\, dt\,v_2 \;-\; i\,2u\,dt\, p(v\bv)\,v_1\bv_1 \,v_1\;-\;i\sqrt{2u}\,v_1\, dx_1 \nn \\ 
dv_2&=&\ts \;-\;i\,\vep\, dt\,v_1 \;-\; i\,2u\,dt\, p(v\bv)\,v_2\bv_2  \,v_2\;-\;i\sqrt{2u}\,v_2\, dx_2 \nn \\  
d\bar v_1&=&\ts \;+\;i\,\vep\, dt\,\bar v_2\;+\;i\, 2u\,dt\, p(v\bv)\,v_1\bv_1  \,\bv_1  \;+\;i\sqrt{2u}\,\bar v_1\, dy_1 \nn \\
d\bar v_2&=&\ts \;+\;i\,\vep\, dt\,\bar v_1\;+\;i\, 2u\,dt\, p(v\bv)\,v_2\bv_2  \,\bv_2  \;+\;i\sqrt{2u}\,\bar v_2\, dy_2 \lbeq{5.4}
\eeqn

\medskip
with $\,v\bv=v_1\bv_1+v_2\bv_2\,$ and 
\beqn
p(x)&=&P'(x)/P(x)\;=\;[\log P]'(x)
\eeqn
with 
\beqn
P(x)&=&\begin{cases} e^x &{\rm \;if\;the\;initial\;state\;is\;a\;coherent\;state} \\ 
  x^{N-1}&{\rm  \;if\;the\;initial\;state\;is\;a\;number\;state}  \end{cases} 
\eeqn
such that 
\beqn
p(x)&=&\begin{cases} 1 &{\rm \;if\;the\;initial\;state\;is\;a\;coherent\;state} \\ 
  (N-1)/x&{\rm  \;if\;the\;initial\;state\;is\;a\;number\;state}  \end{cases} 
\eeqn

\medskip
Recall that the initial coherent and number states were given by
\beqn
\psi_0(z_1,z_2)&=&\begin{cases}\; e^{\lam_1z_1+\lam_2z_2}\,e^{-{|\lam_1|^2+|\lam_2|^2\over 2}} \;=\;e^{\lam z}\,e^{-{|\lam|^2\over 2}} \pI&\;\;{\rm coherent\;state} \\ 
  \; \ts {1 \over \sqrt{N!N^N}} \;(\lam_1z_1+\lam_2z_2)^N\;=\; {(\lam z)^N \over \sqrt{N!N^N}} &\;\;{\rm number\;state}   \end{cases}
\eeqn
with 
\beqn
|\lam|^2\;\;=\;\;|\lam_1|^2+|\lam_2|^2&=&N 
\eeqn
being the total number of particles. The density matrix elements are then given by 
\beqn
(\psi_t,a_i^+a_j\,\psi_t)_\cF&=& \E\bar\E[\,v_i\bv_j\,]  \pI 
\eeqn
with the $v_1,v_2$ and $\bv_1,\bv_2$ given by (\req{5.4}) with initial conditions
\beqn
v_0\;=\;\bmat v_{1,0}\\ v_{2,0} \emat \;=\;\bmat \lam_1\\ \lam_2 \emat ,\;\;\;\; \bv_0\;=\;\bmat \bv_{1,0}\\ \bv_{2,0} \emat\;=\;\bmat \blam_1\\ \blam_2 \emat   \pI 
\eeqn

\medskip
Let's introduce the quadratic quantities 
\beqn
n_1&:=&v_1\bv_1 \nn \\ 
n_2&:=&v_2\bv_2 \nn \\ 
q&:=&v_1\bv_2 \nn \\ 
\bq&:=&\bv_1 v_2 
\eeqn

\medskip
where the $\bq$, as long as no expectation values are taken, is not necessarily the complex conjugate of the $q$. Then, 
because of $\,dv_id\bv_i=0\,$, these quantities satisfy the following SDE system. The on-diagonal elements are given by
\medskip
\beqn
dn_1&=&\;+\;i\,\vep\,dt\,(q-\bq) \;-\;i\sqrt{2u}\,n_1\, (dx_1-dy_1)  \nn\\ 
dn_2&=&\;-\;i\,\vep\,dt\,(q-\bq) \;-\;i\sqrt{2u}\,n_2\, (dx_2-dy_2) \pM \lbeq{5.13}
\eeqn

\medskip
and for the off-diagonal elements one obtains, with $n=n_1+n_2=v_1\bv_1+v_2\bv_2=v\bv$, 
\medskip
\beqn
dq&=&\ts \;+\;i\,\vep\, dt\,(n_1-n_2)\;-\; i\,2u\,dt\;p(n)\,(n_1-n_2)\,q \;-\;i\sqrt{2u}\,q\, (dx_1-dy_2)   \nn \\ 
d\bq&=&\ts \;-\;i\,\vep\, dt\,(n_1-n_2)\;+\; i\,2u\,dt\;p(n)\,(n_1-n_2)\,\bq \;-\;i\sqrt{2u}\,\bq\, (dx_2-dy_1) \pM \lbeq{5.14}
\eeqn

\medskip
since for example, 
\medskip
\beqn
dq&=&dv_1\,\bv_2\;+\;v_1\,d\bv_2\;+\;dv_1\,d\bv_2   \\ 
&=&\ts\bigl\{\;-\;i\,\vep\, dt\,v_2 \;-\; i\,2u\,dt\; p(v\bv)\,v_1\bv_1 \,v_1\;-\;i\sqrt{2u}\,v_1\, dx_1\bigr\}\, \bv_2\nn \\ 
&&\ts\;+\; v_1\,\bigl\{\;i\,\vep\, dt\,\bar v_1\;+\;i\, 2u\,dt\; p(v\bv)\,v_2\bv_2 \,\bv_2  \;+\;i\sqrt{2u}\,\bar v_2\, dy_2 \; \bigr\} \;\;+\;0 \nn \\ 
&=&\ts\;+\;i\,\vep\, dt\,(v_1\bv_1-v_2\bv_2)\;-\; i\,2u\,dt\;p(v\bv)\,(v_1\bv_1-v_2\bv_2)\,v_1\bv_2 \;-\;i\sqrt{2u}\,v_1\bv_2\, (dx_1-dy_2) \nn
\eeqn

\medskip
Let's introduce again, now for $j\in\{1,2\}$, 
\beqn
d\xi_j&:=&\ts { dx_j-dy_j \over \sqrt{2}} \nn \\ 
d\eta_j&:=&\ts { dx_j+dy_j \over \sqrt{2}}
\eeqn
and let's also put
\beqn
d\nu_{12}\;\;:=\;\; \sqrt{2}\,(dx_1-dy_2)&=&d\xi_1+d\xi_2 +d\eta_1-d\eta_2  \pM \nn \\ 
d\nu_{21}\;\;:=\;\; \sqrt{2}\,(dx_2-dy_1)&=&d\xi_1+d\xi_2 -d\eta_1+d\eta_2  
\eeqn

\medskip
Then, with the abbreviations 
\beqn
n_{12}&:=&n_1-n_2  \nn \\
n&:=&n_1+n_2
\eeqn

\medskip
the system (\req{5.13}) and (\req{5.14}) looks as follows: 
\medskip
\beqn
dn&=&\;-\;i\sqrt{4u}\,( n_1\, d\xi_1 + n_2\, d\xi_2) \nn \\
dn_{12}&=&\;+\;i\,2\vep\,dt\,(q-\bq) \;-\;i\sqrt{4u}\, (n_1\, d\xi_{1}-n_2\, d\xi_{2}) \phantom{\sum} \nn  \\ 
dq&=&\ts \;+\;i\,\vep\, dt\,n_{12}\;-\; i\,2u\,dt\,p(n)\,n_{12}\,q \;-\;i\sqrt{u}\,q\, d\nu_{12}  \nn  \\ 
d\bq&=&\ts \;-\;i\,\vep\, dt\,n_{12}\;+\; i\,2u\,dt\,p(n)\,n_{12}\,\bq \;-\;i\sqrt{u}\,\bq\, d\nu_{21}  \lbeq{5.19}
\eeqn

\medskip
from which we immediately get $\,\E\bar\E[\,dn_t\,]=0\,$ or
\beqn
\la n_t\ra\;\;:=\;\;\E\bar\E[\,n_t\,]&=& n_0\;\;=\;\;n_{1,0}+n_{2,0}\;\;=\;\;N \;\;\;\; \forall t \pI  \lbeq{5.20}
\eeqn
Furthermore we have the following exact equations:
\beqn
\ts {d\over dt}\la n_{12,t}\ra&=&\;+\;i\,2\vep\,\bigl(\,\la q_t\ra\,-\,\la\bq_t\ra\,\bigr)  \nn  \\ 
\ts {d\over dt}\la q_t\ra&=&\ts \;+\;i\,\vep\,\la n_{12,t}\ra\;-\; i\,2u\,\bigl\la \,p(n_t)\,n_{12,t}\,q_t\,\bigr\ra  \pS \nn \\ 
\ts {d\over dt}\la\bq_t\ra&=&\ts \;-\;i\,\vep\,\la n_{12,t}\ra\;+\; i\,2u\,\bigl\la \,p(n_t)\,n_{12,t}\,\bq_t\,\bigr\ra  \lbeq{5.21}
\eeqn

\medskip
Here we reencounter the typical feature of quantum many body systems, namely, the system is non closed and when trying to close it by deriving 
SDEs for quantities like $n_{12,t}\,q_t$ or $\,p(n_t)\,n_{12,t}\,q_t\,$, we generate higher and higher products which basically corresponds to an expansion 
of the exponential which generates the collapse and revivals.

\bigskip
\bigskip
\noindent{\bf 5.1\; Large $N$ Limit} 

\medskip
\bigskip
If we introduce the normalized quantities (so the $w$'s used in this section are different from the $w$'s used in chapter 2)
\beqn
\rho_1&:=& n_1/N\;, \phantom{mm} \rho_2\;\;:=\;\; n_2/N  \nn \\ 
w &:=&q/N\;, \phantom{miim}    \bw\;\;:=\;\;\bq/N   \\
&& \phantom{im} g\;\;:=\;\;uN \nn
\eeqn
and
\beqn
\rho_{12}&:=&\rho_1-\rho_2 \nn \\ 
\rho&:=&\rho_1+\rho_2
\eeqn
the system (\req{5.19}) is equivalent to \;(with $\,p(n)=p(N\rho)\,$, we write $p(n)$ for brevity)
\medskip
\beqn
d\rho&=&\;-\;i\,\sqrt{4g/N}\,( \rho_1\, d\xi_1 + \rho_2\, d\xi_2) \pS  \nn \\
d\rho_{12}&=&\;+\;i\,2\vep\,dt\,(w-\bw) \;-\;i\,\sqrt{4g/N}\, (\rho_1\, d\xi_{1}-\rho_2\, d\xi_{2})   \nn \pS \\ 
dw&=&\ts\;+\;i\,\vep\, dt\,\rho_{12}\;-\; i\,2g\,dt\,p(n)\,\rho_{12}\,w \;-\;i\,\sqrt{g/N}\,w\, d\nu_{12}  \nn \\ 
d\bw&=&\ts\;-\;i\,\vep\, dt\,\rho_{12}\;+\;i\,2g\,dt\,p(n)\,\rho_{12}\,\bw \;-\;i\,\sqrt{g/N}\,\bw\, d\nu_{21} \pS
\eeqn
In the limit $\,N\to\infty\,$ with $\,g=uN\,$ held fixed, the diffusive part vanishes and we obtain the ODE system 
\beqn
\dot\rho&=&0 \nn \\ 
\dot\rho_{12}&=&\;+\;i\,2\vep\,(w-\bw) \nn \\ 
\dot w &=& \;+\;i\,\vep\,\rho_{12}\;-\; i\,2g\,\rho_{12}\,w \nn \\
\dot{\bw}&=& \;-\;i\,\vep\,\rho_{12}\;+\; i\,2g\,\rho_{12}\,\bw
\eeqn

\medskip
Here we can ignore the $\,p(n_t)=p(N\rho_t)\,$ also for number states since we have $\,p(N\rho_t)=(N-1)/(N\rho_t)=(N-1)/N\to 1\,$ 
because of $\,\rho_t=1\,$ for all $t$. Now let's put all $N$ particles on lattice site 1, that is, we choose the initial conditions 
\beqn
(\rho_1,\rho_2,w,\bw)_0=(1,0,0,0) 
\eeqn
The quantity $\bw_t$ is now the true complex conjugate of $w_t$. We are left with the two equations
\smallskip
\beqn
\dot\rho_{12}&=&\;-\;4\vep\;\im\,w   \lbeq{5.27}\\ 
\dot w &=& \;+\;i\,\vep\,\rho_{12}\;-\; i\,2g\,\rho_{12}\,w \lbeq{5.28}
\eeqn

\medskip
Equation (\req{5.28}) is solved by 
\beqn
w_t&=&\ts {\vep \over 2g}\;+\; \bigl( w_0 \,-\,{\vep \over 2g}\bigr)\; e^{\,-2ig\int_0^t \rho_{12,s}ds} \pM  \nn \\ 
&\buildrel w_0=0\over =&\ts {\vep \over 2g}\;-\; {\vep \over 2g}\; e^{\,-2ig\int_0^t \rho_{12,s}ds}  
\eeqn
We obtain
\beqn
\im\,w_t&=&\ts \,+\,{\vep \over 2g}\,\sin\bigl[\, 2g\int_0^t \rho_{12,s} ds\,\bigr] \pS
\eeqn
Putting this into (\req{5.27}) gives 
\beqn
\dot\rho_{12,t}&=&\;-\;4\vep\,\im\,w_t\;\;=\;\; \ts \,-\,{2\vep^2 \over g}\,\sin\bigl[\, 2g\int_0^t \rho_{12,s} ds\,\bigr]  \pS
\eeqn
Thus, the quantity 
\beqn
\vp_t&:=&\ts 2g\int_0^t \rho_{12,s} ds 
\eeqn
is the solution of the second order equation 
\beqn
\ddot \vp_t \;+\; \ts 4\vep^2 \sin \vp_t &=& 0 \pS  \lbeq{5.33}
\eeqn
which is just the equation of motion for the mathematical pendulum. We have the following initial conditions: 
\beqn
\vp_0&=& 0 \nn \\ 
\dot\vp_0&=&2g\,\rho_{12,0}\;=\;2g\,(1-0)\;\;=\;\;2g 
\eeqn
with total energy 
\beqn
E&=&\ts {\dot\vp_t^2\over 2}\,-\,4\vep^2 \cos \vp_t\;\;=\;\;{\dot\vp_0^2\over 2}\,-\,4\vep^2 \cos \vp_0 \;\;=\;\;2g^2\,-\,4\vep^2 \pS
\eeqn
The potential energy at $\,\vp=\pi\,$ is $\,E_{\rm pot}=+4\vep^2\,$. We have rollovers if the total energy is bigger than that, 
that is, if $\,2g^2\,-\,4\vep^2 > +4\vep^2\,$ or  
\beqn
g^2 &>& (2\vep)^2 \;.
\eeqn

\medskip
\bigskip
\noindent{\bf Numerical Test} 

\bigskip
\noindent Let's make a numerical check. We choose the following values: $\,\vep=1\,$ and
\medskip
\beqn
N&\in&\{\,2500\,,\,5000\,,\,10000\,,\,20000\,\}  \lbeq{5.37} \\  
g&\in& \{\,0.5\,,\,1.0\,,\,1.8\,,\,2.2\,,\,3.0\,,\,6.0\,\} \lbeq{5.38}
\eeqn

\medskip
and calculate the quantity 
\beqn
\rho_{1,t}\;\;=\;\;n_{1,t}/N&=&(\psi_t,a_1^+a_1\,\psi_t)_\cF\;/\;N \pS
\eeqn
in two different ways: First, by exact diagonalization. There we have to use the different values for $N$ given by (\req{5.37}). Second, by simulating the ODE system (\req{5.33}) 
for the mathematical pendulum and calculating $\,\rho_{1,t}\,$ through 
\beqn
\rho_{1,t}&=&\ts {1\over 2}\bigl(\, 1\,+\, {\dot\vp_t \over 2g}\,\bigr) \pM  \lbeq{5.40}
\eeqn
This is the large $N$ limit and accordingly no $N$ enters the calculation, but only a value for $g$. We obtain the following results as displayed in 
figure 5.1.1 and figure 5.1.2 below. The red line is the ODE solution and the dots come from exact diagonalization.

\nobreak

\bigskip
\bigskip
\nobreak
\bigskip
\nobreak
\centerline{\includegraphics[width=16cm]{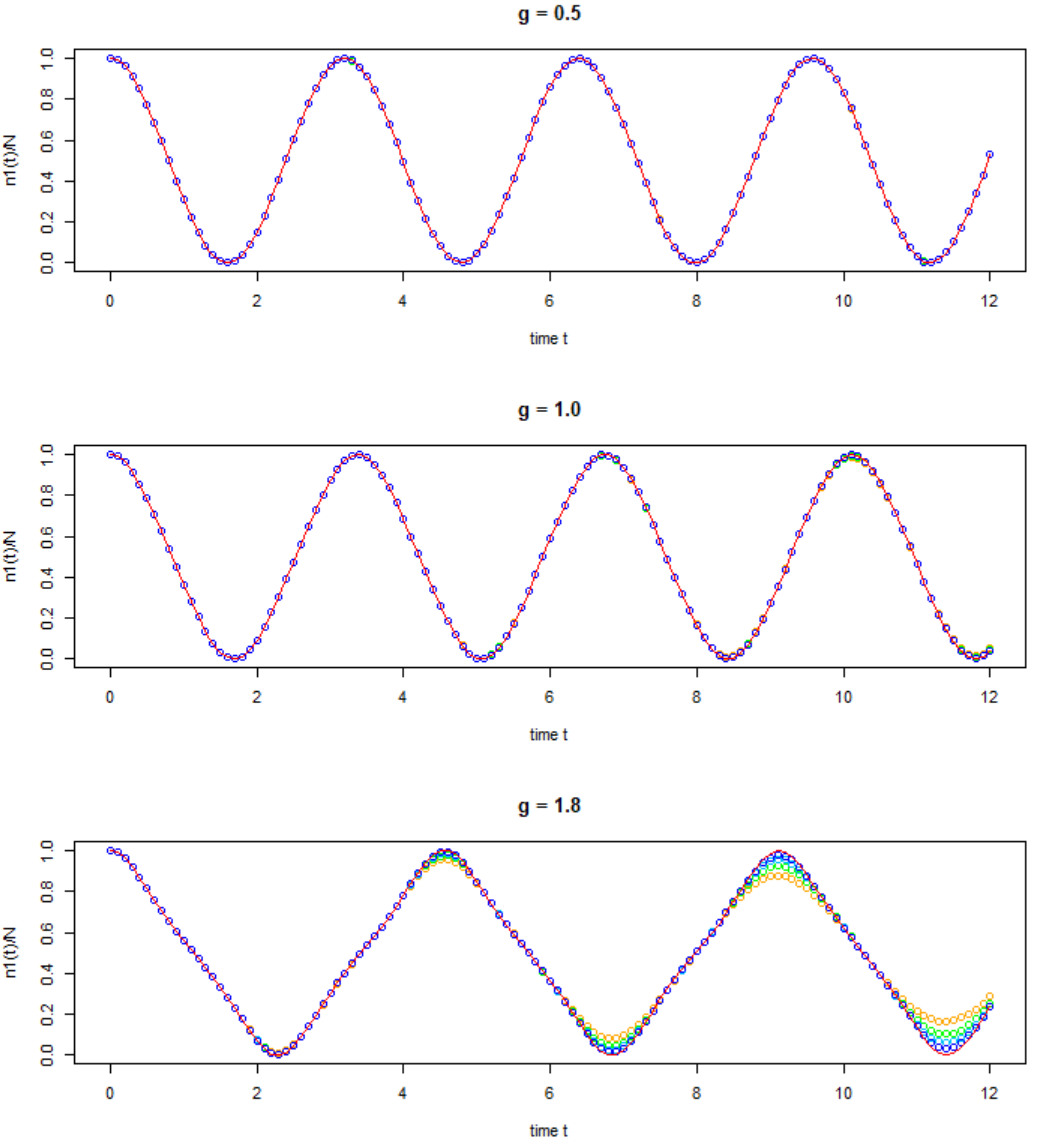}}

\bigskip
\centerline{\footnotesize Figure 5.1.1: Exact diagonalization results vs.~ODE solution for the quantity $\la n_{1,t}\ra/N$ }

\bigskip 

\bigskip
\centerline{\includegraphics[width=16cm]{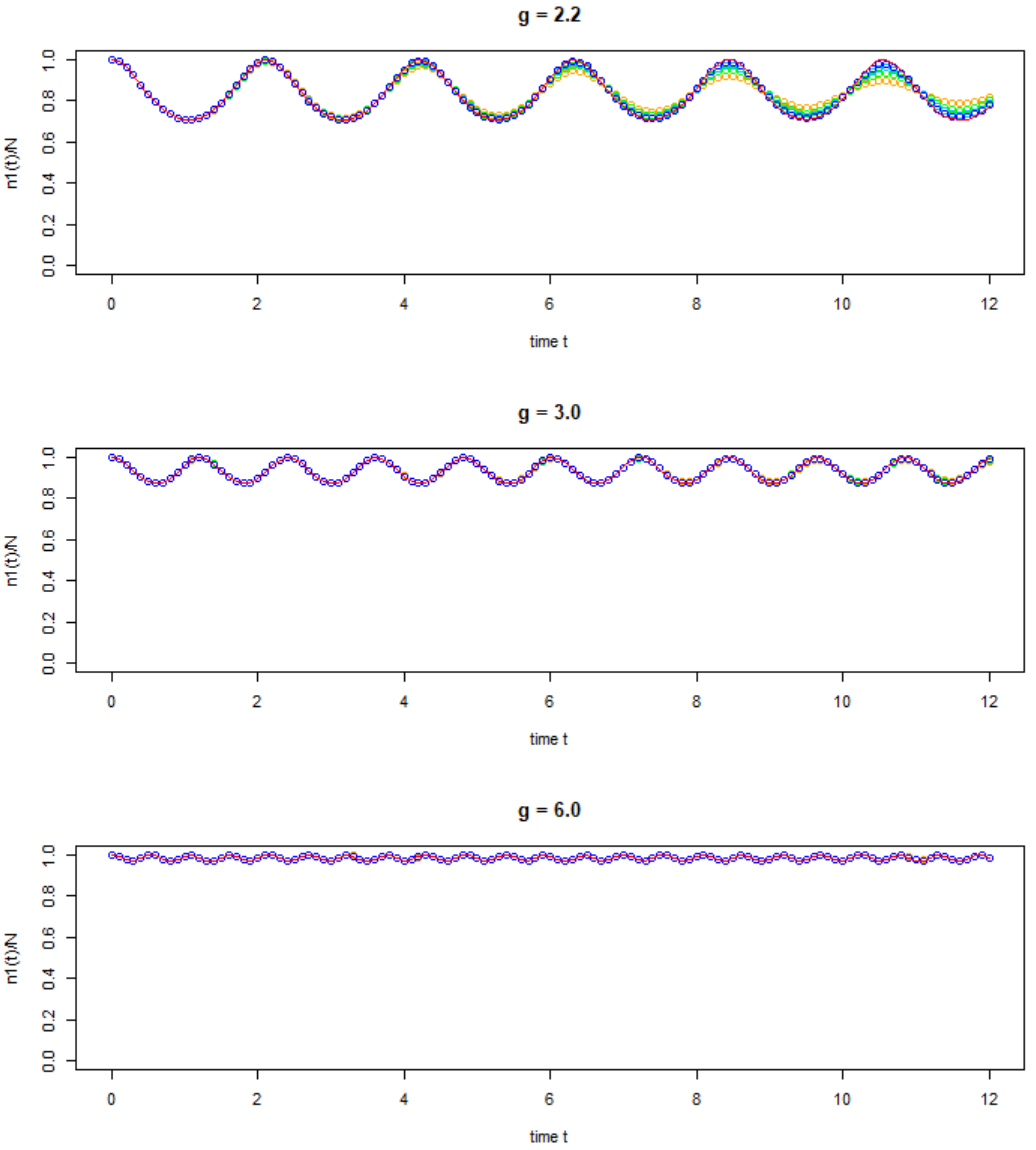}}

\medskip
\bigskip
\centerline{\footnotesize Figure 5.1.2: Exact diagonalization results vs.~ODE solution for the quantity $\la n_{1,t}\ra/N$ }

\bigskip
\bigskip
\bigskip
\noindent The closer the value of $g$ approaches 2, the larger $N$ has to be chosen in order to numerically reach the true $N=+\infty$ limit. 
The different colors of the dots represent the different values of $N$, with the obvious ordering of orange, green, light blue and dark blue for 
increasing values of $N$. The fact that $\,g=2\vep=2\,$ marks the transition point is also visualized through the following picture,

\smallskip
\bigskip
\centerline{\includegraphics[width=9cm]{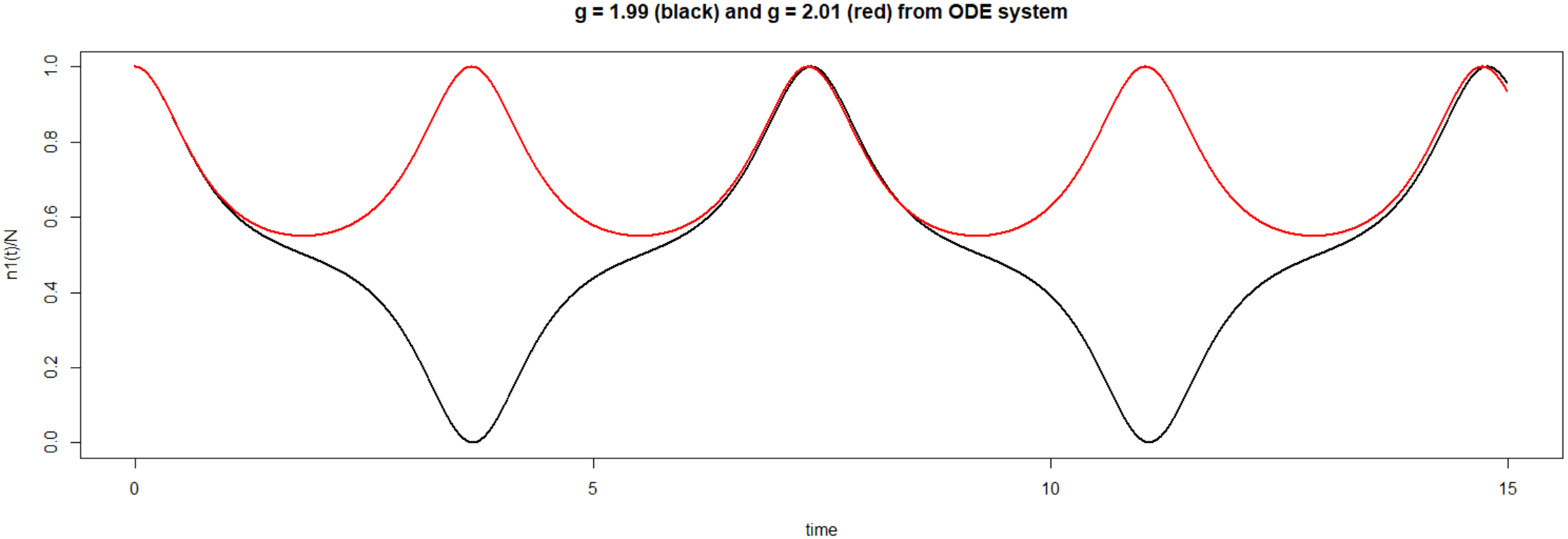}}

\bigskip 
\bigskip

\noindent which shows the quantity (\req{5.40}) for $g=1.99$ in black and for $g=2.01$ in red. 
The fact that the mathematical pendulum shows up in the dynamics of the two site Bose-Hubbard model has been observed by several authors, for example 
in refs [9-12]. The very beautiful thesis of Lena Simon [13] also provides a detailed discussion of the dynamics. 

\bigskip
\bigskip
{\bf Equivalence to Quartic Double Well Potential}

\bigskip
Actually we can obtain $n_{12,t}$ or $\rho_{12,t}$ also as a solution of a classical particle moving in a quartic double well potential. 
Recall the ODE system (\req{5.27},\req{5.28}),
\beqn
\dot\rho_{12}&=&\;-\;4\vep\;\im\,w  \nn \\ 
\dot w &=& \;+\;i\,\vep\,\rho_{12}\;-\; i\,2g\,\rho_{12}\,w 
\eeqn
We have 
\beqn
\ddot\rho_{12}&=&\;-\;4\vep\;\im\,\dot w \nn \\ 
&=&\; -\,4\vep^2\rho_{12}\;+\;  8g\vep\,\rho_{12}\,\re\,w  
\eeqn
and
\beqn
\re\,\dot w&=&\;+\; 2g\,\rho_{12}\,\im\,w 
\eeqn
or
\beqn
4\vep\,\re\,\dot w&=&\;-\; 2g\,\rho_{12}\,\dot\rho_{12} 
\eeqn

\smallskip
which gives, with initial conditions $w_0=0$ and $\rho_{12,0}=1$, 
\beqn
4\vep\,\re\, w_t&=&g(\rho_{12,0}^2-\rho_{12,t}^2)\;\;=\;\;g(1-\rho_{12,t}^2)
\eeqn
Thus, 
\beqn
\ddot\rho_{12,t}\; +\,4\vep^2\rho_{12,t}&=&\;+\;  2g\,\rho_{12,t}\,4\vep\,\re\,w_t\, \nn \\ 
&=&\;+\;  2g^2\,\rho_{12,t}\,(1-\rho_{12,t}^2)
\eeqn  
or
\beqn
\ddot\rho_{12,t}\; +\;(4\vep^2-2g^2)\rho_{12,t}\;+\;2g^2\,\rho_{12,t}^3&=& 0 
\eeqn
with initial conditions $\rho_{12,0}=1$ and $\dot\rho_{12,0}=0\,$. Let's summarize in the following

\bigskip
\bigskip
\noindent{\bf Theorem 7:} The mathematical pendulum 
\beqn
\ddot\vp_t\;+\;4\vep^2\,\sin\vp_t&=&0 
\eeqn
with $\vp_0=0$ and $\dot\vp_0=2g$ is equivalent to the cubic equation  
\beqn
\ddot\rho_{12,t}\; +\;(4\vep^2-2g^2)\rho_{12,t}\;+\;2g^2\,\rho_{12,t}^3&=& 0 \pS  \lbeq{5.49}
\eeqn
with $\rho_{12,0}=1$ and $\dot\rho_{12,0}=0$ through the following transformation 
\beqn
\ts \vp_t\;\;=\;\;\ts 2g\int_0^t \rho_{12,s}\ds \;\;\;\;\;\; \Leftrightarrow\;\;\;\;\;\; \ts\rho_{12,t}\;\;=\;\;{1 \over 2g}\; \dot\vp_t \pI
\eeqn

\bigskip
\bigskip
The transition between the oscillatory and the self trapping regime is intuitive for the mathe\-matical pendulum, let's try to understand this also 
by using the cubic equation. The total energy for equation (\req{5.49}) is 
\beqn
E&=&\ts {\dot\rho_{12,t}^2\over 2} \;+\;(2\vep^2-g^2)\rho_{12,t}^2\;+\;{g^2\over 2}\,\rho_{12,t}^4\;\;=\;\;\ts {\,4\vep^2 \,-\, g^2\,\over 2} \pI
\eeqn
The potential energy is
\beqn
E_{\rm pot}(\rho_{12})&=&\ts (2\vep^2-g^2)\rho_{12,t}^2\;+\;{g^2\over 2}\,\rho_{12,t}^4 
\eeqn
and has stationary points at $\rho_{12}=0$ and 
\beqn
4\vep^2\,-\,2g^2\;+\;2\,g^2\,\rho_{12,t}^2&=&0 
\eeqn
or
\beqn 
\rho_{12,t}^2&=&\ts 1\;-\;{2\vep^2\over g^2}\;\;<\;\; 1
\eeqn
Thus, for $g^2>2\vep^2$ we have a double well potential which, for $E<0$ or $g^2>4\vep^2$, is sufficiently deep such that $\rho_{12,t}$ cannot escape the right 
well when starting at $\rho_{12,0}=1$ with $\dot\rho_{12,0}=0\,$. Since, $\rho_{12,t}$ starts at 1 with a negative energy and the potential energy at $\rho_{12}=0$ is 
zero such that the energy required to cross this point is $\,{\dot\rho_{12,t}^2/ 2} >0\,$ which is not available.

\bigskip
Obviously, the mathematical pendulum has no collapse and revivals, so in order to see these, we have to take the diffusive part of the 
SDE system into account. Before we do this, let's have a look at PDE representations.

\bigskip
\bigskip
\bigskip
\noindent{\bf  5.2\; PDE Representations } 

\bigskip
\bigskip
In chapter 3, we used the $v_j,\bv_j$ as the basic variables to obtain PDE representations for the density matrix elements. 
Now, it is very instructive to see the corresponding PDE representations if the quadratic quantities
\beqn
(n_1,n_2,q,\bq)&=&(v_1\bv_1,v_2\bv_2,v_1\bv_2,v_2\bv_1) 
\eeqn
are used directly as variables. To this end recall the SDE system (\req{5.13}) and (\req{5.14}) for the quadratic quantities from the last section, 
\medskip
\beqn
dn_1&=&\;+\;i\,\vep\,(q-\bq)\,dt \;-\;i\sqrt{4u}\,n_1\, d\xi_1 \nn \\ 
dn_2&=&\;-\;i\,\vep\,(q-\bq)\,dt \;-\;i\sqrt{4u}\,n_2\, d\xi_2 \nn \\ 
dq&=&\;+\;i\,\vep\,n_{12}\,dt\;-\; i\,2u\, p(n)\,n_{12}\,q\,dt \;-\;i\sqrt{u}\,q\, d\nu_{12}   \nn \\ 
d\bq&=&\;-\;i\,\vep\,n_{12}\,dt\;+\; i\,2u\, p(n)\,n_{12}\,\bq\,dt \;-\;i\sqrt{u}\,\bq\, d\nu_{21}  \lbeq{5.56}
\eeqn

\medskip
with $n=n_1+n_2=v\bv$, $\,n_{12}=n_1-n_2\,$ and 
\beqn
p(x)&=&P'(x)/P(x)\;=\;[\log P]'(x)
\eeqn
with 
\beqn
P(x)&=&\begin{cases} e^x &{\rm \;if\;the\;initial\;state\;is\;a\;coherent\;state} \\ 
  x^{N-1}&{\rm  \;if\;the\;initial\;state\;is\;a\;number\;state}  \end{cases} 
\eeqn
such that 
\beqn
p(x)&=&\begin{cases} 1 &{\rm \;if\;the\;initial\;state\;is\;a\;coherent\;state} \\ 
  (N-1)/x&{\rm  \;if\;the\;initial\;state\;is\;a\;number\;state}  \end{cases} 
\eeqn

\medskip
Recall the abbreviations
\beqn
d\xi_1&=& \ts {(dx_1-dy_1)/ \sqrt{2}} \nn \\
d\xi_2&=& \ts {(dx_2-dy_2)/ \sqrt{2}} \nn \\
d\nu_{12}&=&\ts \sqrt{2}\;(dx_1-dy_2) \nn \\
d\nu_{21}&=&\ts \sqrt{2}\;(dx_2-dy_1) 
\eeqn

\medskip
To write down a PDE representation which is again obtained as a Kolmogorov backward equation, recall the logic of appendix A.3 and A.4, we need to determine the 
differential operator $A$ which is associated to the SDE system (\req{5.56}). To do this, we need the following identities which determine the 
second order part of $A$:
\medskip
\beqn
(d\xi_1)^2\;=\;(d\xi_2)^2\;=\;(d\nu_{12})^2\;=\;(d\nu_{21})^2\;=\;d\xi_1d\xi_2\;=\;d\nu_{12}d\nu_{21}\;=\;0 
\eeqn
and
\beqn
d\xi_1d\nu_{12}&=&(dx_1)^2\;=\;+\,i\,dt \nn\\
d\xi_1d\nu_{21}&=&(dy_1)^2\;=\;-\,i\,dt \nn\\
d\xi_2d\nu_{12}&=&(dy_2)^2\;=\;-\,i\,dt \nn \\
d\xi_2d\nu_{21}&=&(dx_2)^2\;=\;+\,i\,dt \lbeq{5.62}
\eeqn

\medskip
\noindent Now let 
\beq
F&=&F(n_1,n_2,q,\bq)
\eeq

\medskip
by any function. We plug in the $n_{1t},n_{2,t}$ and $q_t,\bq_t$ from the SDE system above and calculate the $dF$ with the Ito lemma: 
\beqn
dF&=&\ts {\pt F\over \pt n_1}\,dn_1\;+\;{\pt F\over \pt n_2}\,dn_2\;+\;{\pt F\over \pt q}\,dq\;+\;{\pt F\over \pt \bq}\,d\bq \pI \\ 
&&\ts \;+\;  {\pt^2 F\over \pt n_1\pt q}\,dn_1dq \;+\;{\pt^2 F\over \pt n_1\pt \bq}\,dn_1d\bq 
  \;+\;{\pt^2 F\over \pt n_2\pt q}\,dn_2dq \;+\;{\pt^2 F\over \pt n_2\pt \bq}\,dn_2d\bq \nn \pS\\
&\buildrel\phantom{{{I\over I}\over I}}\over=&\ts\;+\;i\vep dt\,(q-\bq )\, {\pt F\over \pt n_1}\;-\;i\vep dt\,(q-\bq ) \,{\pt F\over \pt n_2}  \pI  \nn \\
&&\ts \;+\;\bigl\{\,i dt\,\vep\,n_{12,t} \,-\,idt\,2u\,p(n)\,n_{12,t}\,q_t \,\bigr\}\, {\pt F\over \pt q}  
  \ts \;-\;\bigl\{\,i dt\,\vep\,n_{12,t} \;-\;idt\,2u\,p(n)\,n_{12,t}\,\bq_t \,\bigr\}\,{\pt F\over \pt \bq} \phantom{\Bigr)} \nn \\ 
&&\ts \;-\;2u\,\Bigl\{\; {\pt^2 F\over \pt n_1\pt q}\,n_1q \,d\xi_1d\nu_{12}  \;+\;{\pt^2 F\over \pt n_1\pt \bq}\,n_1\bq \,d\xi_1d\nu_{21} \pI \nn \\
&&\ts \phantom{ \;-\;2u\,\Bigl\{\;}   \;+\;{\pt^2 F\over \pt n_2\pt q}\,n_2q\, d\xi_2d\nu_{12} \;+\;{\pt^2 F\over \pt n_2\pt \bq}\,n_2\bq\, d\xi_2d\nu_{21} \;\Bigr\}
  \;\;+\;\; {\rm diffusive} \pS \nn
\eeqn

\medskip
or, using (\req{5.62}), 
\medskip
\beqn
dF&=&\ts \;+\;i\vep \,(q-\bq )\, \bigl( {\pt F\over \pt n_1}\,- \,{\pt F\over \pt n_2} \bigr)\,dt
   \;+\;i \,\vep\,(n_1-n_2) \, \bigl( {\pt F\over \pt q} \,- \,{\pt F\over \pt \bq} \bigr) \,dt  \\ 
&&\ts \;-\;i\,2u\,p(n)\,(n_1-n_2)\,\bigl( q \, {\pt F\over \pt q} \,-\,\bq \,{\pt F\over \pt \bq} \bigr)\,dt  \nn \pI \\ 
&&\ts  \;-\;i\,2u\,\bigl(\, {\pt^2 F\over \pt n_1\pt q}\,n_1q   \,-\,{\pt^2 F\over \pt n_1\pt \bq}\,n_1\bq 
    \,-\,{\pt^2 F\over \pt n_2\pt q}\,n_2q \,+\,{\pt^2 F\over \pt n_2\pt \bq}\,n_2\bq\,\bigr)\,dt \;\;+\;\; {\rm diffusive}   \nn \\ 
&=:& AF\,dt \;\;+\;\; {\rm diffusive} \pI
\eeqn
Thus, any expectation 
\beqn
F&:=&\E\bar\E_0^t[\,f(n_{1,t},n_{2,t},q_t,\bq_t)\,\bigr]\;\;=\;\;F_t(n_{1,0},n_{2,0},q_0,\bq_0)  \pM  \lbeq{5.66}
\eeqn

\medskip
considered as a function of its initial values, has to be a solution of (again, we drop the zero subscripts on the right hand side of (\req{5.66}))
\medskip
\beqn
\ts {\pt F \over \pt t}\;=\;AF&=&\ts \;+\;i\vep \,(q-\bq )\, \bigl( {\pt F\over \pt n_1}\,- \,{\pt F\over \pt n_2} \bigr)
   \;+\;i \,\vep\,(n_1-n_2) \, \bigl( {\pt F\over \pt q} \,- \,{\pt F\over \pt \bq} \bigr) \nn \\ 
&&\ts \;-\;i\,2u\,p(n)\,(n_1-n_2)\,\bigl( q \, {\pt F\over \pt q} \,-\,\bq \,{\pt F\over \pt \bq} \bigr)  \nn \pI \\ 
&&\ts  \;-\;i\,2u\,\bigl(\, {\pt^2 F\over \pt n_1\pt q}\,n_1q   \,-\,{\pt^2 F\over \pt n_1\pt \bq}\,n_1\bq 
    \,-\,{\pt^2 F\over \pt n_2\pt q}\,n_2q \,+\,{\pt^2 F\over \pt n_2\pt \bq}\,n_2\bq\,\bigr) \pM \nn \\ 
&=:&\;-\;i\,(\cL_\vep\,+\,\cL_u\,)F \pI
\eeqn
with initial condition $\,F_0=f\,$ and differential operators
\beqn
\cL_\vep&:=&\ts \;-\; \vep \,\Bigl\{\; (q-\bq )\, \bigl({\pt \over \pt n_1}-{\pt \over \pt n_2}\bigr)  
   \;+\; (n_1-n_2) \,\bigl( {\pt \over \pt q} - {\pt \over \pt \bq}\bigr)\;\Bigr\}   \pI \lbeq{5.68}  \\ 
\cL_u&:=&\ts \;+\;2u \,\Bigl\{\; \bigl(\, n_1 \, {\pt \over \pt n_1} \,-\,n_2 \,{\pt \over \pt n_2} \,\bigr) 
   \bigl(\, q \, {\pt \over \pt q} \,-\,\bq \,{\pt \over \pt \bq} \,\bigr) 
  \;+\; p(n)\,(n_1-n_2)\bigl(\, q \, {\pt \over \pt q} \,-\,\bq \,{\pt \over \pt \bq} \,\bigr) \;\Bigr\}  \pI \phantom{mm}  \lbeq{5.69}
\eeqn
This proves part (a) of the following 

\medskip
\bigskip
\bigskip
\noindent{\bf Theorem 8:} Consider the two site Bose-Hubbard model with Hamiltonian 
\beqn
h&=&\ts \vep\,(\,a^+_1 a_2\,+\,a_2^+a_1)\;+\;u\,(\, a_1^+a_1^+a_1a_1\,+\, a_2^+a_2^+a_2a_2\, )  \pI \nn \\ 
&=&\ts \vep\,\bigl(\,z_1{\pt\over \pt z_2} \,+\, z_2{\pt\over \pt z_1}\,\bigr) \;+\: u\,\bigl(\, z_1^2\,{\pt^2\over \pt z_1^2}\,+\, z_2^2\,{\pt^2\over \pt z_2^2}\;\bigr) \pM 
\eeqn
and initial state 
\beqn
\psi_0(z_1,z_2)&=&\begin{cases} e^{\,\lam_1z_1\,+\,\lam_2z_2}\; e^{\,-\,{|\lam_1|^2\,+\,|\lam_2|^2\over 2}} &{\rm\;if\;coherent\;is\;chosen} \\ 
  \ts {1 \over \sqrt{N!N^N}} \;(\lam_1z_1 +\lam_2z_2 )^N &{\rm\;if\;number\;is\;chosen} \end{cases}
\eeqn

\medskip
with a total number of $N=|\lam_1|^2+|\lam_2|^2$ particles. Let $P$ be the function of one variable given by 
\beqn
P(x)&=&\begin{cases} e^x &{\rm\;if\;coherent\;is\;chosen} \\ 
  x^{N-1}&{\rm\;if\;number\;is\;chosen}  \end{cases} 
\eeqn
and let $p(x)=P'(x)/P(x)$. Then the following statements hold:

\bigskip
\begin{itemize}
\item[{\bf a)}] The expected number of particles $\,\la n_{1,t}\ra=\E\bar\E[n_{1,t}]=(\psi_t,a_1^+a_1\psi_t)_{\cF}\,$ at lattice site 1 can be written as 
\beqn
\phantom{mmmmm}\la n_{1,t}\ra&=&e^{-it(\cL_\vep\,+\,\cL_u\,)}\,n_1\;{\bigr|_{\,(n_1,n_2,q,\bq)\,=\,(\,|\lam_1|^2,\,|\lam_2|^2,\,\lam_1\blam_2\,,\,\blam_1\lam_2\,)} }  \pI 
\eeqn
with the differential operators $\cL_\vep$ and $\cL_u$ given by (\req{5.68}) and (\req{5.69}) above. 

\bigskip
\item[{\bf b)}] The actions of $e^{-it\cL_\vep}$ and $e^{-it\cL_u}$ are as follows: For $e^{-it\cL_\vep}$ we obtain 
\beqn
(e^{-it\cL_\vep}F)(n_1,n_2,q,\bq)&=&F\bigl(\, R_t\,(n_1,n_2,q,\bq)^T \,\bigr) \pI
\eeqn
where $F=F(n_1,n_2,q,\bq)$ is an arbitrary function and $R_t$ is the $4\times 4$ matrix 
\medskip
\beqn
R_t&=&\bmat \cos^2\vep t & \sin^2\vep t & +i\sin\vep t\cos\vep t & -i\sin\vep t\cos\vep t  \\
     \sin^2\vep t &\cos^2\vep t & -i\sin\vep t\cos\vep t & +i\sin\vep t\cos\vep t  \\
      +i\sin\vep t\cos\vep t & -i\sin\vep t\cos\vep t &\cos^2\vep t & \sin^2\vep t  \\
       -i\sin\vep t\cos\vep t & +i\sin\vep t\cos\vep t & \sin^2\vep t &\cos^2\vep t \emat \pM
\eeqn

\medskip
For $e^{-it\cL_u}$ we find: 
\beqn
\lefteqn{
e^{-it\cL_u}\,\bigl\{\, G(n_1,n_2)\,q^b\,\bq^{\,\bb}\,\bigr\} \;\;=\;\;  \pI  } \\
&& \phantom{} G\bigl(\,e^{-i2u t\,(b-\bb)}n_1 \,, \, e^{+i2u t\,(b-\bb)}n_2\,\bigr) \;\times\;
   { P(\,e^{-i2u t\,(b-\bb)}n_1 \, + \, e^{+i2u t\,(b-\bb)}n_2\,) \over P(n_1+n_2)}\, \times \,q^b\,\bq^{\,\bb}  \phantom{mm} \nn 
\eeqn 
where $G=G(n_1,n_2)$ is an arbitrary function and $b,\bb$ are arbitrary natural numbers.  
\end{itemize}

\medskip
\bigskip
\noindent{\bf Proof:} It remains to prove part (b). If we make the Ansatz 
\beqn
(e^{-it\cL_\vep}F)(n_1,n_2,q,\bq)&=&F\bigl(\, e^{-itA}\, (n_1,n_2,q,\bq)^T \,\bigr) \pI
\eeqn
and take the time derivative, we find
\beqn
A&=&-\vep\bmat 0 & \sigma \\ \sigma & 0 \emat \;\; \in\;\;\R^{4\times 4} 
\eeqn
with 
\beqn
\sigma&:=&\bmat +1 & -1 \\ -1 & + 1 \emat \;\;\in\;\; \R^{2\times 2} 
\eeqn
Since
\beqn
A^{2k}\;\;=\;\;(-\vep)^{2k}\,\bmat \sigma^{2k} & 0 \\ 0 & \sigma^{2k} \emat,\;\;\;\;\;\; 
A^{2k+1}\;\;=\;\;(-\vep)^{2k+1}\,\bmat 0& \sigma^{2k+1} \\   \sigma^{2k+1} & 0 \emat    \pI 
\eeqn
and because of $\,\sigma^n=2^{n-1}\sigma=2^n\,{\sigma\over 2}\,$ for $n\ge 1$, we obtain 
\medskip
\beqn
e^{-itA}&=& Id \;\;+\;\;\sum_{k=1}^\infty { (+2it\vep)^{2k}\over (2k)!} \;{1\over 2}\bmat \sigma & 0 \\ 0 & \sigma \emat \;\;+\;\;
  \sum_{k=0}^\infty { (+2it\vep)^{2k+1}\over (2k+1)!} \;{1\over 2}\bmat 0& \sigma  \\ \sigma & 0 \emat   \nn \\ 
&&\phantom{-}\nn \\ 
&=&\bmat Id & 0 \\ 0 & Id \emat \;\;+\;\;{\cos(2\vep t)-1\over 2}\bmat \sigma & 0 \\ 0 & \sigma \emat \;\;+\;\;i\;{\sin(2\vep t)\over 2}\bmat 0& \sigma  \\ \sigma & 0 \emat  
\eeqn

\medskip
which coincides with $\,R_t\,$ since $\,\cos^2\vep t=(1+\cos2\vep t)/2\,$ and $\,\sin^2\vep t=(1-\cos2\vep t)/2\,$. 

\bigskip
\noindent The action of $e^{-it\cL_u}$ we calculate by evaluating the Fresnel expectation directly. That is, we write down the $\,\vep=0\,$ SDE system
\beqn
dn_1&=& \;-\;i\sqrt{4u}\,n_1\, d\xi_1 \nn \\ 
dn_2&=& \;-\;i\sqrt{4u}\,n_2\, d\xi_2 \nn \\ 
dq&=&\;-\; i\,2u\,dt\, p(n)\,n_{12}\,q \;-\;i\sqrt{u}\,q\, d\nu_{12}   \nn \\ 
d\bq&=&\;+\; i\,2u\,dt\, p(n)\,n_{12}\,\bq \;-\;i\sqrt{u}\,\bq\, d\nu_{21}  
\eeqn
which is solved by 
\beqn
n_{1,t}&=&n_1\,e^{\,-\,i\sqrt{4u}\int_0^t d\xi_{1,s} } \nn \\ 
n_{2,t}&=&n_2\,e^{\,-\,i\sqrt{4u}\int_0^t d\xi_{2,s} } 
\eeqn
and
\beqn
q_{t}&=&q\,e^{\,-\,i2u\int_0^t p(n_s)\,n_{12,s}\,ds\,-\,i\sqrt{u}\int_0^t d\nu_{12,s} } \nn \\ 
\bq_{t}&=&\bq\,e^{\,+\,i2u\int_0^t p(n_s)\,n_{12,s}\,ds\,-\,i\sqrt{u}\int_0^t d\nu_{21,s} } \pI
\eeqn
We have to evaluate
\beqn
\lefteqn{ 
e^{-it\cL_u}\,\bigl\{\, G(n_1,n_2)\,q^b\,\bq^{\,\bb}\,\bigr\}\;\;=\;\;\E\bar\E\bigl[\; G(n_{1,t},n_{2,t})\,q_t^b\,\bq_t^{\,\bb}\;\bigr]   \pI } \nn \\ 
&=&\E\bar\E\Bigl[\, G(n_{1,t},n_{2,t})\,e^{\,-\,i2u(b-\bb)\int_0^t p(n_s)\,n_{12,s}\,ds }
  \,e^{ \,-\,i\sqrt{u}\,\int_0^t[\,b\,d\nu_{12,s}\,+ \,\bb\,d\nu_{21,s}\,]  } \,\Bigr]  \,q^b\,\bq^{\,\bb} \pI\pM \nn \\ 
&=:&\E\bar\E\Bigl[\, \tilde G(\xi_1,\xi_2)\, \,e^{\,-\,i\sqrt{u}\,\int_0^t[\,b\,d\nu_{12,s}\,+ \,\bb\,d\nu_{21,s}\,] }\,\Bigr]  \,q^b\,\bq^{\,\bb} \pI\pM \lbeq{5.85}
\eeqn
where we abbreviated the quantity 
\beqn
\tilde G(\xi_1,\xi_2)&:=&\bigl[\,G(n_{1,t},n_{2,t})\,e^{\,-\,i2u(b-\bb)\int_0^t p(n_s)\,n_{12,s}\,ds }\,\bigr](\xi_1,\xi_2) \pI 
\eeqn
which depends only on the $\xi$-variables, but is independent of the $\eta$-variables which only show up in the last exponential in (\req{5.85}). This means 
that as in chapter 4 the $\eta$-integrals can be performed and give $\delta$-functions for the $\xi$-variables. With Fresnel BMs given by, for $j\in\{1,2\}$ and $t=t_k=kdt$,
\beqn
x_{j,t_k}&=&\ts\sqrt{dt}\,\sum_{\ell=1}^k \phi_{j,\ell} \nn \\ 
y_{j,t_k}&=&\ts\sqrt{dt}\,\sum_{\ell=1}^k \theta_{j,\ell}
\eeqn
the Fresnel measure is
\beqn
dFd\bar F&=& \pro_{\ell=1}^k e^{\,i\,\sum_{j=1}^2{\phi_{j,\ell}^2-\theta_{j,\ell}^2\over 2}}\;\ts {d^2\phi_\ell \,d^2\theta_\ell\over (2\pi)^2} \pI 
\eeqn
We write again
\beqn
\xi_{j,t_k}\;\;=\;\;\ts {x_{j,t_k}-y_{j,t_k}\over \sqrt{2}} \;\;=\;\;\sqrt{dt}\,\sum_{\ell=1}^k {\phi_{j,\ell}-\theta_{j,\ell} \over \sqrt{2}} 
&=:&\ts\sqrt{dt}\,\sum_{\ell=1}^k \alpha_{j,\ell} \pI \nn \\ 
\eta_{j,t_k}\;\;=\;\;\ts {x_{j,t_k}+y_{j,t_k}\over \sqrt{2}} \;\;=\;\;\sqrt{dt}\,\sum_{\ell=1}^k {\phi_{j,\ell}+\theta_{j,\ell} \over \sqrt{2}} 
&=:&\ts\sqrt{dt}\,\sum_{\ell=1}^k \beta_{j,\ell}
\eeqn

\medskip
such that the Fresnel measure becomes
\medskip
\beqn
dFd\bar F(\alpha,\beta)&=& \pro_{\ell=1}^k e^{\,i\,(\alpha_{1,\ell}\beta_{1,\ell}+\alpha_{2,\ell}\beta_{2,\ell})  }\;\ts {d^2\alpha_\ell \,d^2\beta_\ell\over (2\pi)^2} \pI 
\eeqn
We have to calculate
\beqn
\lefteqn{ 
\E\bar\E\Bigl[\, \tilde G(\xi_1,\xi_2)\, \,e^{\,-\,i\sqrt{u}\,\int_0^t[\,b\,d\nu_{12,s}\,+ \,\bb\,d\nu_{21,s}\,] }\,\Bigr]  \;\;=\;\; \pI \phantom{mm} } \\ 
&&\ts  \int_{\R^{4k}}\, \tilde G(\xi_1,\xi_2) \; 
   \exp\bigl\{\,-\,i\sqrt{u}\,b\int_0^t d\nu_{12} \,-\,i\sqrt{u}\,\bb\int_0^t d\nu_{21}\, \bigr\} \;  dFd\bar F(\alpha,\beta) \phantom{m}   \nn
\eeqn

\medskip
with the discrete time expressions
\smallskip
\beqn
\ts \int_0^t d\nu_{12}&=&\ts\int_0^t (\, d\xi_1+d\xi_2 +d\eta_1-d\eta_2\, ) \;\;=\;\;
  \sqrt{dt}\,\ts \sum_{\ell=1}^k \bigl(\, \alpha_{1,\ell}+\alpha_{2,\ell}+\beta_{1,\ell}-\beta_{2,\ell}\,\bigr)  \nn  \\ 
\ts \int_0^t d\nu_{21}&=&\ts\int_0^t (\, d\xi_1+d\xi_2 -d\eta_1+d\eta_2\, ) \;\;=\;\;
  \sqrt{dt}\,\ts \sum_{\ell=1}^k \bigl(\, \alpha_{1,\ell}+\alpha_{2,\ell}-\beta_{1,\ell}+\beta_{2,\ell}\,\bigr)  \pM\phantom{m}
\eeqn

\medskip
The $\beta$-integrals can be performed and produce $\delta$-functions: 
\beqn
{\ts \int_{\R^{2k}} \exp\Bigl\{\,-\,i\sqrt{udt}\,\ts \sum_{\ell=1}^k (b-\bb) [\beta_{1,m}-\beta_{2,m}]\,\Bigr\} }
     \pro_{\ell=1}^k e^{\,i\,(\alpha_{1,\ell}\beta_{1,\ell}+\alpha_{2,\ell}\beta_{2,\ell})  }\;\ts { d\beta_{1,\ell}\,d\beta_{2,\ell}\over (2\pi)^2}
  \phantom{mm} \pI\nn \\
\;\;=\;\;\pro_{\ell=1}^k \delta\Bigl(\,\alpha_{1,\ell}-[b-\bb]\sqrt{udt}\,\Bigr)\;\delta\Bigl(\,\alpha_{2,\ell}+[b-\bb]\sqrt{udt}\,\Bigr)\,  \pI \phantom{mmm}
\eeqn
Thus, 
\beqn
\alpha_{1,\ell}&=&\,+\,(b-\bb)\sqrt{udt}  \nn \\ 
\alpha_{2,\ell}&=&\,-\,(b-\bb)\sqrt{udt} 
\eeqn

\medskip
such that $n_{1,t}$ and $n_{2,t}$ become, with $t=t_k=kdt$,  
\beqn
n_{1,t}&=&n_1\,e^{\,-\,i\sqrt{4udt}\sum_{\ell=1}^k \alpha_{1,\ell} } \;\;=\;\; n_1\,e^{\,-\,i\,2u(b-\bb)\,t} \nn \\ 
n_{2,t}&=&n_2\,e^{\,-\,i\sqrt{4udt}\sum_{\ell=1}^k \alpha_{2,\ell} } \;\;=\;\; n_2\,e^{\,+\,i\,2u(b-\bb)\,t} 
\eeqn
and therefore
\beqn
\lefteqn{ 
\ts\,-\,i2u(b-\bb)\int_0^t p(n_s)\,n_{12,s}\,ds  \pI } \nn \\ 
&=&\ts \,-\,i2u(b-\bb)\int_0^t p(n_1\,e^{\,-\,i\,2u(b-\bb)\,s}+n_2\,e^{\,+\,i\,2u(b-\bb)\,s})\,(n_1\,e^{\,-\,i\,2u(b-\bb)\,s}-n_2\,e^{\,+\,i\,2u(b-\bb)\,s})\,ds  \nn \\ 
&=&\ts \,+\,\int_0^t p(n_1\,e^{\,-\,i\,2u(b-\bb)\,s}+n_2\,e^{\,+\,i\,2u(b-\bb)\,s})\,{d\over ds}(n_1\,e^{\,-\,i\,2u(b-\bb)\,s}+n_2\,e^{\,+\,i\,2u(b-\bb)\,s})\,ds \pI  \nn \\ 
&=&\ts \,+\,\int_0^t {d\over ds}\bigl[\,\log P(n_1\,e^{\,-\,i\,2u(b-\bb)\,s}+n_2\,e^{\,+\,i\,2u(b-\bb)\,s})\,\bigr]\,ds   \nn \\ 
&=&\log P(n_1\,e^{\,-\,i\,2u(b-\bb)\,t}+n_2\,e^{\,+\,i\,2u(b-\bb)\,t}) \;-\; \log P(n_1+n_2) \pI
\eeqn
Hence we arrive at 
\beqn
\lefteqn{ 
\E\bar\E\Bigl[\, G(n_{1,t},n_{2,t})\,e^{\,-\,i2u(b-\bb)\int_0^t p(n_s)\,n_{12,s}\,ds }
  \,e^{ \,-\,i\sqrt{u}\,\int_0^t[\,b\,d\nu_{12,s}\,+ \,\bb\,d\nu_{21,s}\,]  } \,\Bigr]   \;\;=\;\; \phantom{mmmmmm} \pI }  \\ 
&&\ts \phantom{mmmm}  G\bigl(\,e^{-i2u t\,(b-\bb)}n_1 \,, \, e^{+i2u t\,(b-\bb)}n_2\,\bigr) \;\times \;
  {P(n_1\,e^{\,-\,i\,2u(b-\bb)\,t}\,+\,n_2\,e^{\,+\,i\,2u(b-\bb)\,t}) \over  P(n_1+n_2) }  \pI \nn
\eeqn 

\smallskip
and this proves part (b) of the theorem. \;$\blacksquare$

\goodbreak

\bigskip
\bigskip
\bigskip
\noindent{\bf  5.3\; Collapse and Revivals } 

\bigskip
\bigskip
For small $u$, the two site Bose-Hubbard model shows the intriguing phenomenom of collapse and revivals. In two very beautiful papers, 
Fishman and Veksler [14] and Bakman, Fishman and Veksler [15] gave a very precise quantitative description of this phenomenom not only 
for the two site Bose-Hubbard model, but also, to emphasize the general mechanism, for a quantum mechanical oscillator with a small 
anharmonic perturbation. The main technical tool there was a careful semiclassical analysis of the energy spectrum. Lena Simon and 
Walter Strunz also used semiclassical methods in their article [16]. 

\bigskip
Here in our setting we have SDEs and ODEs and of course we want to use them in order to demonstrate the phenomenom. In this paper, we do not aim at the most 
sophisticated version of doing that, solving this problem would basically mean to solve the quantum mechanical many body problem, but here we just want to give a `proof of concept', 
namely, to show that the formalism is able to do that at all. To this end, recall the exact equations (\req{5.20}) and (\req{5.21}) 
which in the coherent state case read as follows: 
\beqn
\la n_t\ra&=& N \;\;\;\; \forall t  \nn \\
\ts {d\over dt}\la n_{12,t}\ra&=&\;+\;i\,2\vep\,\bigl(\,\la q_t\ra\,-\,\la\bq_t\ra\,\bigr)  \nn \pS \\ 
\ts {d\over dt}\la q_t\ra&=&\ts \;+\;i\,\vep\,\la n_{12,t}\ra\;-\; i\,2u\,\la \,n_{12,t}\,q_t\,\ra   \nn \\ 
\ts {d\over dt}\la\bq_t\ra&=&\ts \;-\;i\,\vep\,\la n_{12,t}\ra\;+\; i\,2u\,\la \,n_{12,t}\,\bq_t\,\ra \pS  \lbeq{5.98}
\eeqn
If we would simply factorize $\,\la n_{12}\,q\ra\approx \la n_{12}\ra\la q\ra\,$, we would recover the mathematical pendulum which does not have collapse and revivals. 
From part (b) of Theorem 8 of the previous section, we have 
\medskip
\beqn
\la \,n_{1,t}\,q_t\,\ra_{u}\;\;:=\;\;e^{-it\cL_u}\{\, n_{1}q\,\}&=&\,e^{-i2u t\,}n_1\;\times\;
  { P(\,e^{-i2u t}n_1 \, + \, e^{+i2u t}n_2\,) \over P(n_1+n_2)}\;q \pI \nn \\
\la \,n_{2,t}\,q_t\,\ra_{u}\;\;:=\;\;e^{-it\cL_u}\{\, n_{2}q\,\}&=&\,e^{+i2u t\,}n_2\;\times\;
  { P(\,e^{-i2u t}n_1 \, + \, e^{+i2u t}n_2\,) \over P(n_1+n_2)}\;q \pI
\eeqn

\medskip
or, since we are considering the coherent state case with $P(x)=e^x$, 
\beqn
\la \,n_{1,t}\,q_t\,\ra_{u}&=&\,e^{-i2u t\,}n_1\;\times\;
  \exp\bigl\{\,(e^{-i2u t}-1)n_1 \, + \, (e^{+i2u t}-1)n_2\,\bigr\} \;\times\;q \pI \nn \\
\la \,n_{2,t}\,q_t\,\ra_{u}&=&\,e^{+i2u t\,}n_2\;\times\;
  \exp\bigl\{\,(e^{-i2u t}-1)n_1 \, + \, (e^{+i2u t}-1)n_2\,\bigr\} \;\times\;q 
\eeqn

\medskip
We also have 
\beqn
\la n_{1,t}\ra_{u}\;\;=\;\;e^{-it\cL_u} \,n_{1} &=&\, n_1   \nn \\ 
\la n_{2,t}\ra_{u}\;\;=\;\;e^{-it\cL_u} \,n_{2} &=&\, n_2
\eeqn 
and
\beqn
\la q_t\ra_{u}\;\;=\;\;e^{-it\cL_u}\,q&=&\,\exp\bigl\{\,(e^{-i2u t}-1)n_1 \, + \, (e^{+i2u t}-1)n_2\,\bigr\} \;\times\;q   \pI 
\eeqn
Thus, for the dynamics under $\vep=0$, we can write 
\beqn
 {\la \,n_{1,t}\,q_t\,\ra_{u} \over \la n_{1,t}\ra_{u}\,\la q_t\ra_{u}} &=& \,e^{\,-\,i2u t\,}  \pI  \nn \\ 
 {\la \,n_{2,t}\,q_t\,\ra_{u} \over \la n_{2,t}\ra_{u}\,\la q_t\ra_{u}} &=& \,e^{\,+\,i2u t\,} 
\eeqn

\medskip
\noindent Now consider the dynamics under $u=0$. Since $e^{-it\cL_\vep}$ simply rotates the argument when applied to 
an arbitrary function $F$, 
\beqn
\bigl\la\,F(n_{1,t},n_{2,t},q_t,\bq_t)\,\bigr\ra_\vep \;\;:=\;\;(e^{-it\cL_\vep}F)(n_1,n_2,q,\bq)&=&F\bigl(\, R_t\,(n_1,n_2,q,\bq)^T \,\bigr) \pI
\eeqn
the action of $e^{-it\cL_\vep}$ factorizes when applied to an arbitrary product, 
\medskip
\beqn
 e^{-it\cL_\vep}\,(FG)&=& e^{-it\cL_\vep}\,F \;\times\; e^{-it\cL_\vep}\,G   \pM
\eeqn
Thus, under $u=0$, 
\beqn
{\la \,n_{1,t}\,q_t\,\ra_{\vep} \over \la n_{1,t}\ra_{\vep}\,\la q_t\ra_{\vep}} &=& \;1   \pM  \nn \\ 
{\la \,n_{2,t}\,q_t\,\ra_{\vep} \over \la n_{2,t}\ra_{\vep}\,\la q_t\ra_{\vep}} &=& \;1 
\eeqn

\medskip
Then, for the full dynamics with both $\vep$ and $u$ being nonzero, one may try the approximation 
\beqn
{\la \,n_{1,t}\,q_t\,\ra \over \la n_{1,t}\ra\,\la q_t\ra} &\approx& \;{1\over 2}\,\Bigl(\,1\,+\,e^{\,-\,i2u t\,}\,\Bigr)    \pI  \nn \\ 
{\la \,n_{2,t}\,q_t\,\ra \over \la n_{2,t}\ra\,\la q_t\ra} &\approx& \;{1\over 2}\,\Bigl(\,1\,+\,e^{\,+\,i2u t\,}\,\Bigr) \lbeq{5.107}
\eeqn

\medskip
and this in fact generates collapse and revivals. That is, we modify the exact system (\req{5.98}) to the following approximate system \;(recall the abbreviation $\,n_{12}:=n_1-n_2\,$)
\beqn
\la n_t\ra&=& N \;\;\;\; \forall t   \\
\ts {d\over dt}\la n_{12,t}\ra&=&\;+\;i\,2\vep\,\bigl(\,\la q_t\ra\,-\,\la\bq_t\ra\,\bigr)  \nn \pS \\ 
\ts {d\over dt}\la q_t\ra&\approx&\ts \;+\;i\,\vep\,\la n_{12,t}\ra\;-\; i\,u\,\bigl\{\,(1+e^{-i2ut})\,\la n_{1,t}\ra\la q_t\ra 
     \,-\,(1+e^{+i2ut})\,\la n_{2,t}\ra\la q_t\ra \,\bigr\} \nn \\ 
\ts {d\over dt}\la\bq_t\ra&\approx&\ts \;-\;i\,\vep\,\la n_{12,t}\ra\;+\; i\,u\,\bigl\{\,(1+e^{+i2ut})\,\la n_{1,t}\ra\la \bq_t\ra 
     \,-\,(1+e^{-i2ut})\,\la n_{2,t}\ra\la \bq_t\ra \,\bigr\} \pS   \nn
\eeqn
which then reduces to the following two equations \;(since $\,\la n_t\ra=\la n_{1,t}+n_{2,t}\ra=N\,$)
\beqn
\ts {d\over dt}\la n_{12,t}\ra&=&\;-\;4\vep\;\im\la q_t\ra \pS   \lbeq{5.109} \\ 
\ts {d\over dt}\la q_t\ra&\approx&\;+\;i\,\vep\,\la n_{12,t}\ra\;-\; i\,u\,(1+\cos 2ut)\,\la n_{12,t}\ra\la q_t\ra \,-\,uN\sin2ut \,\la q_t\ra  \nn
\eeqn

\medskip
Before we look at the numerical results, let's make some quick analytical considerations. 
Since from now on we are purely in the ODE framework, 
let's omit the angular brackets and summarize the system as follows:
\beqn
\dot n_{12,t}&=&\;-\;4\vep\,\im\, q_t  \nn  \\ 
\dot q_t&=&\;+\;i\,\vep\,n_{12,t}\;-\; i\,u\,(1+\cos 2\tu t)\,n_{12,t}\, q_t\,-\,uN\sin2\tu t \, q_t  \lbeq{5.110}
\eeqn
Here the $\tu$ on the right hand side of (\req{5.110}) is actually a $u$, we put this in since for $\tu=0$ this reduces to the system (\req{5.27},\req{5.28}) of section 5.1 
where we have shown that this is actually the mathematical pendulum. Thus, by switching the $\tu$ from 0 to $u$ we can interpolate between the 
mathematical pendulum without collapse and revivals and the actual case under consideration. We write
\beqn
\ddot n_{12,t}&=&\;-\;4\vep\,\im\, \dot q_t    \nn  \\ 
&=&\;-\;4\vep^2\,n_{12,t} \;+\;4\vep\,u\,(1+\cos 2\tu t)\,n_{12,t}\, \re\, q_t\;+\; 4\vep uN\sin2\tu t \, \im\, q_t \nn \\ 
&=&\;-\;4\vep^2\,n_{12,t} \;+\;4\vep\,u\,(1+\cos 2\tu t)\,n_{12,t}\, \re\, q_t\;-\; uN\sin2\tu t \, \dot n_{12,t} 
\eeqn
or 
\beqn
\ddot n_{12,t}\;+\;uN\sin2\tu t \, \dot n_{12,t} \;+\;4\vep^2\,n_{12,t}&=& \;+\;4\vep\,u\,(1+\cos 2\tu t)\,n_{12,t}\re\, q_t \pI \lbeq{5.112}
\eeqn
For $\tu=0$, the case without collapse and revivals, this reduces to 
\beqn
\ddot n_{12,t}\;+\;0 \;+\;4\vep^2\,n_{12,t}&=& \;+\;4\vep\,u\,(1+1)\,n_{12,t}\re\, q_t \pI 
\eeqn
That is, the right hand side of (\req{5.112}) is basically responsible for the difference between a harmonic pendulum and the mathematical pendulum 
and it generates the effect of self trapping or rollovers for $g>2\vep$. Since collapse and revivals do already show up for very small $g$ 
where surely the harmonic approximation should be valid, we should be able to see them already in the following equation 
\beqn
\ddot n_{12,t}\;+\;uN\sin2\tu t \, \dot n_{12,t} \;+\;4\vep^2\,n_{12,t}&=& 0 \pI 
\eeqn 
This is simply a harmonic oscillator with a time dependent friction
\beqn
\gamma_t&:=&uN\sin2\tu t  \pI
\eeqn
which can be transformed away with the Ansatz 
\beqn
n_{12,t}&=&e^{\,-\,{1\over 2}\int_0^t \gamma_s\, ds} \; y_t  \pI 
\eeqn 
which then produces the following equation for $y_t$,
\beqn
\ddot y_t \;+\; \om_t^2\, y_t &=& 0 \pI 
\eeqn
with a time dependent frequency 
\beqn
\om_t^2&=&\ts 4\vep^2\,-\,{\gamma_t^2\over 4} \,-\,{\dot\gamma_t\over 2} \;\;=\;\;\ts 4\vep^2\,-\,{(uN)^2\sin^2 2\tu t\over 4} \,-\,\tu\,uN\cos 2\tu t \pI
\eeqn
For small $g=uN$, one may put this roughly to $4\vep^2$ such that $y_t$ is identical to the exact $u=0$ solution which is 
\beqn
y_t&\approx& N\cos2\vep t  \pM 
\eeqn
Thus, collapse and revivals arise from the damping factor 
\beqn
e^{\,-\,{1\over 2}\int_0^t \gamma_s\, ds}&=&\ts \exp\bigl\{\,-\,{uN\over 2}\int_0^t \sin2\tu s\, ds \, \bigr\}  
\;\; \buildrel \tu=u\over = \;\; \ts \exp\bigl\{\,-\,{N\over 4}\,(1-\cos 2ut)\, \bigr\} \pI \lbeq{5.120}
\eeqn
and we arrive at the approximate small $g$ solution 
\beqn
n_{12,t}&\approx&\ts \exp\bigl\{\,-\,{N\over 4}\,(1-\cos 2ut)\, \bigr\} \;\times\;  N\cos2\vep t  \pI  
\eeqn
In the following pictures, the quantity $n_{12,t}/N$ obtained from exact diagonalization, in black, is plotted together with the analytical 
collapse factor (\req{5.120}), in red. We chose $N=50$, $\vep=1$ and $g$ as displayed on the plots and all 50 particles were put onto lattice site 1 at time $t=0\,$: 

\bigskip
\bigskip
\bigskip

\bigskip
\centerline{\includegraphics[width=14.5cm]{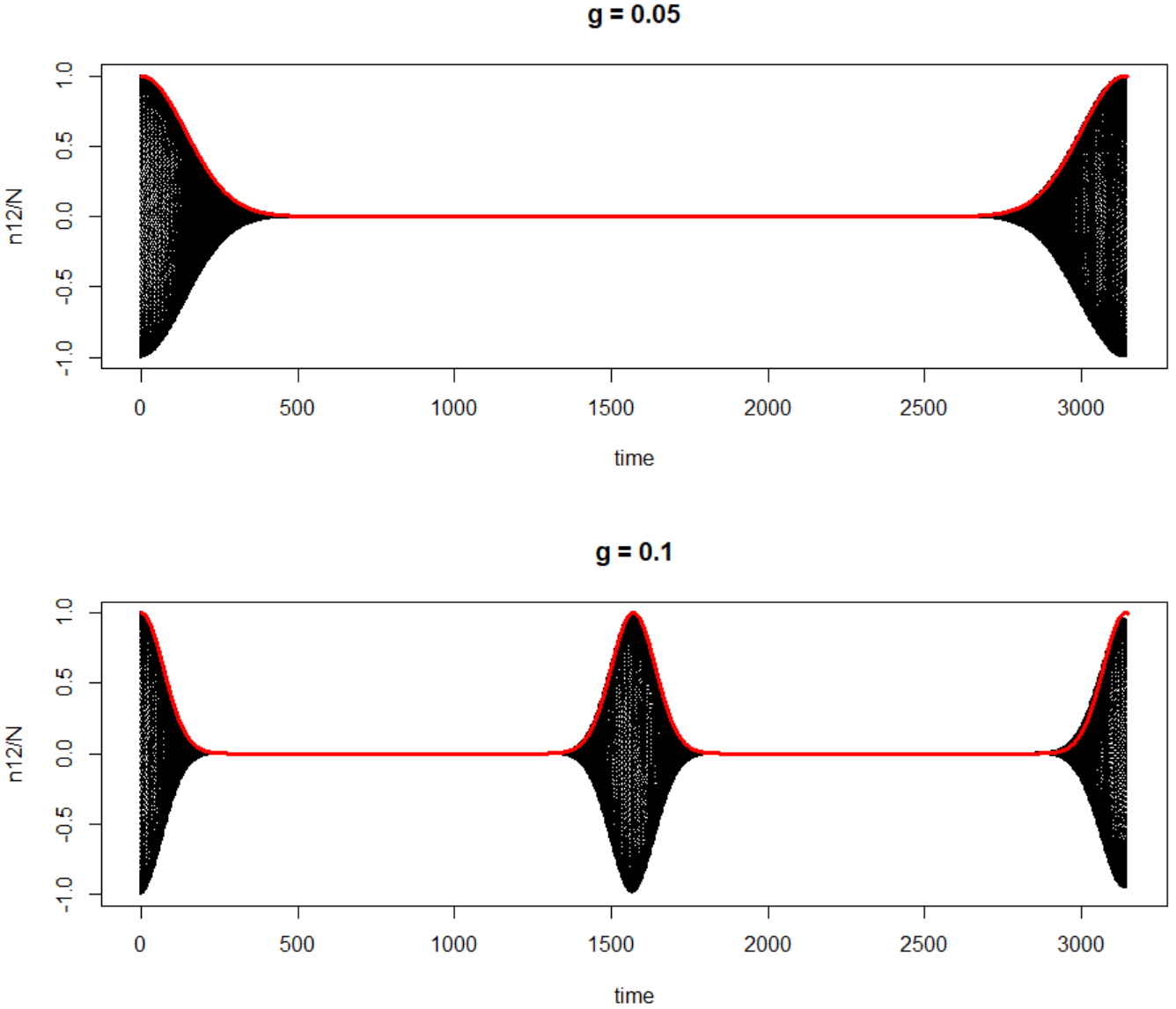}}

\medskip
\centerline{\footnotesize{Figure 5.3.1: Exact diagonalization vs.~analytical collapse factor $\ts \exp\bigl\{\,-\,{N\over 4}\,(1-\cos 2ut)\, \bigr\}$}}

\centerline{\includegraphics[width=14.5cm]{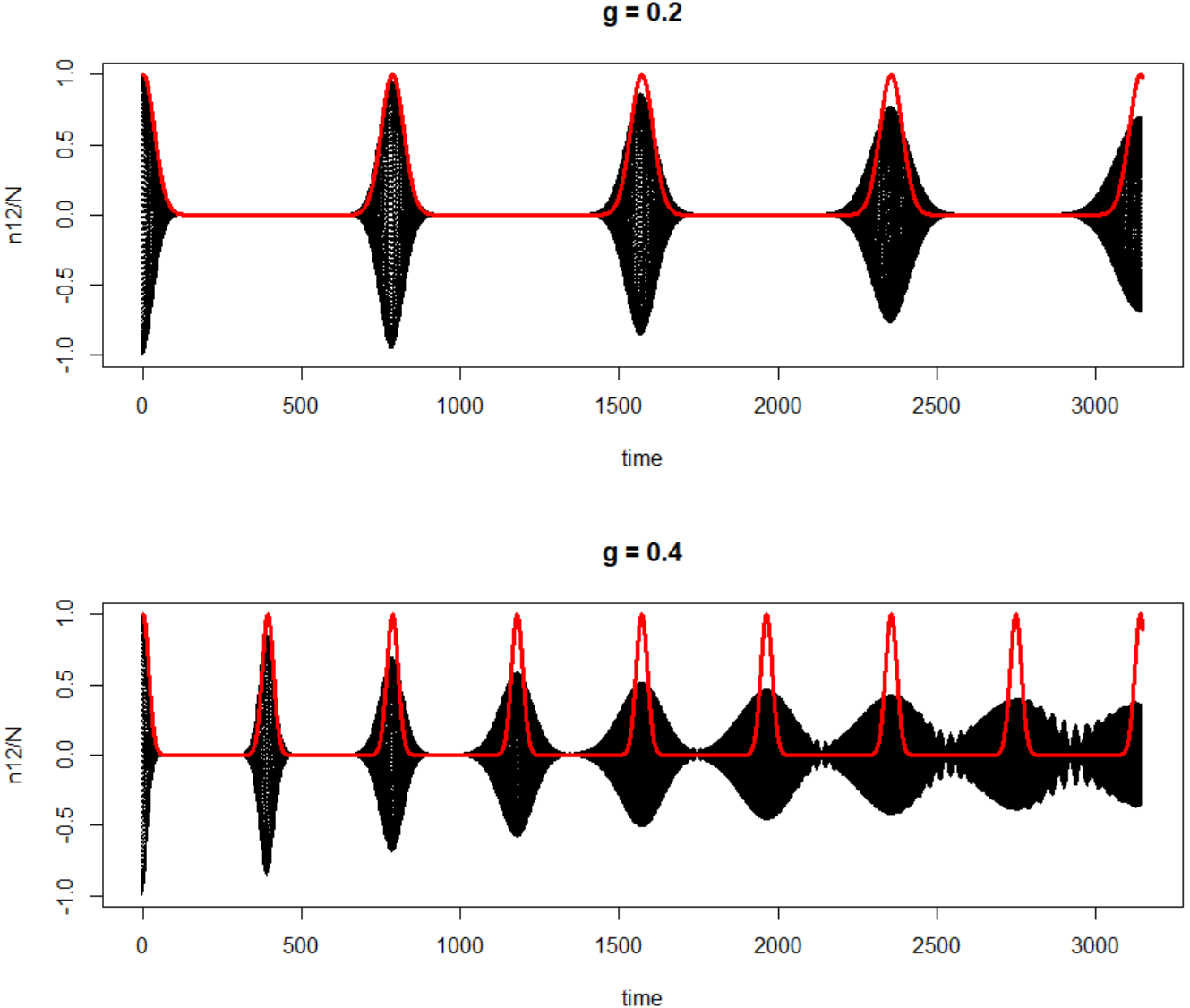}}

\medskip
\centerline{\footnotesize{Figure 5.3.2: Exact diagonalization vs.~analytical collapse factor $\ts \exp\bigl\{\,-\,{N\over 4}\,(1-\cos 2ut)\, \bigr\}$}, other $g$'s}

\bigskip
\bigskip
However, now we have to remark that the exact diagonalization numbers were produced with a number state, not with a coherent state. 
While in the large $N$ limit the dynamics of coherent states and number states are identical, this no longer 
holds in the revival region after the first collapse has occured. In the large $N$ limit, this region moves to infinity and hence is not visible there. For example, 
if we put $g=0.02$ and choose a time horizon of $T=10000\,\pi$ instead of $T=1000\,\pi$ as above, a comparison of number state and coherent state dynamics 
looks as follows,

\bigskip
\centerline{\includegraphics[width=7.5cm]{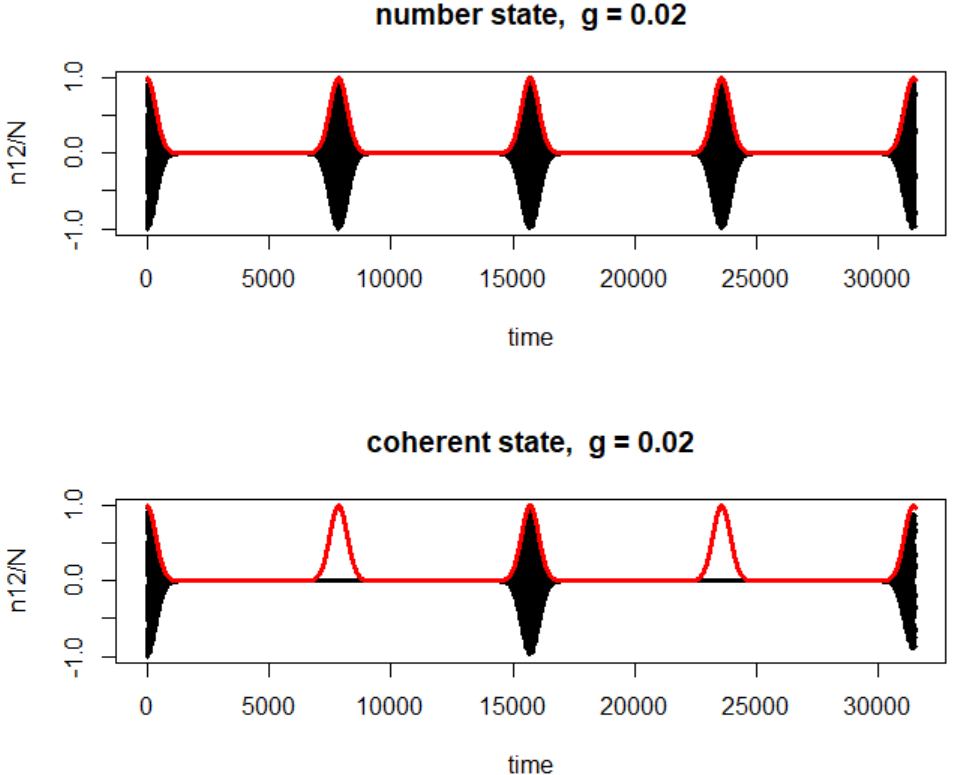}}

\bigskip
This clearly demonstrates that the problem is subtle and a more careful analysis is required. For example, in the approximation (\req{5.107}) 
we could have equally well have said, we use $e^{-i2ut}$ instead of an average $(1+e^{-i2ut})/2$, and then this would have had the effect that the analytical 
collapse factor in (\req{5.120}) would have come with an $N/2$ in the exponent instead of an $N/4$ and the revival blobs were too small. So, this really should be considered 
as some kind of a `teaser', some kind of a motivational argument, but nothing more. A proper and systematic treatment of the diffusive part is basically 
equivalent to solving the quantum mechanical many body problem and this still needs to be developed. But to do so, we believe indeed that the formalism 
presented in this paper is a very useful tool.  

\medskip
We also implemented the full approximate ODE system (\req{5.109}) and compared with exact diagonalization numbers, again using a number state, not a coherent state. 
We used a fourth order Runge-Kutta method and separated off the collapse factor to obtain a stable implementation ($N=50$ and $\vep=1$ as above, $g$ fixed to 0.1 
and zoomed in on different time windows), the exact diagonalization numbers are in black and the ODE solution is in red:

\bigskip
\bigskip
\bigskip
\centerline{\includegraphics[width=14.5cm]{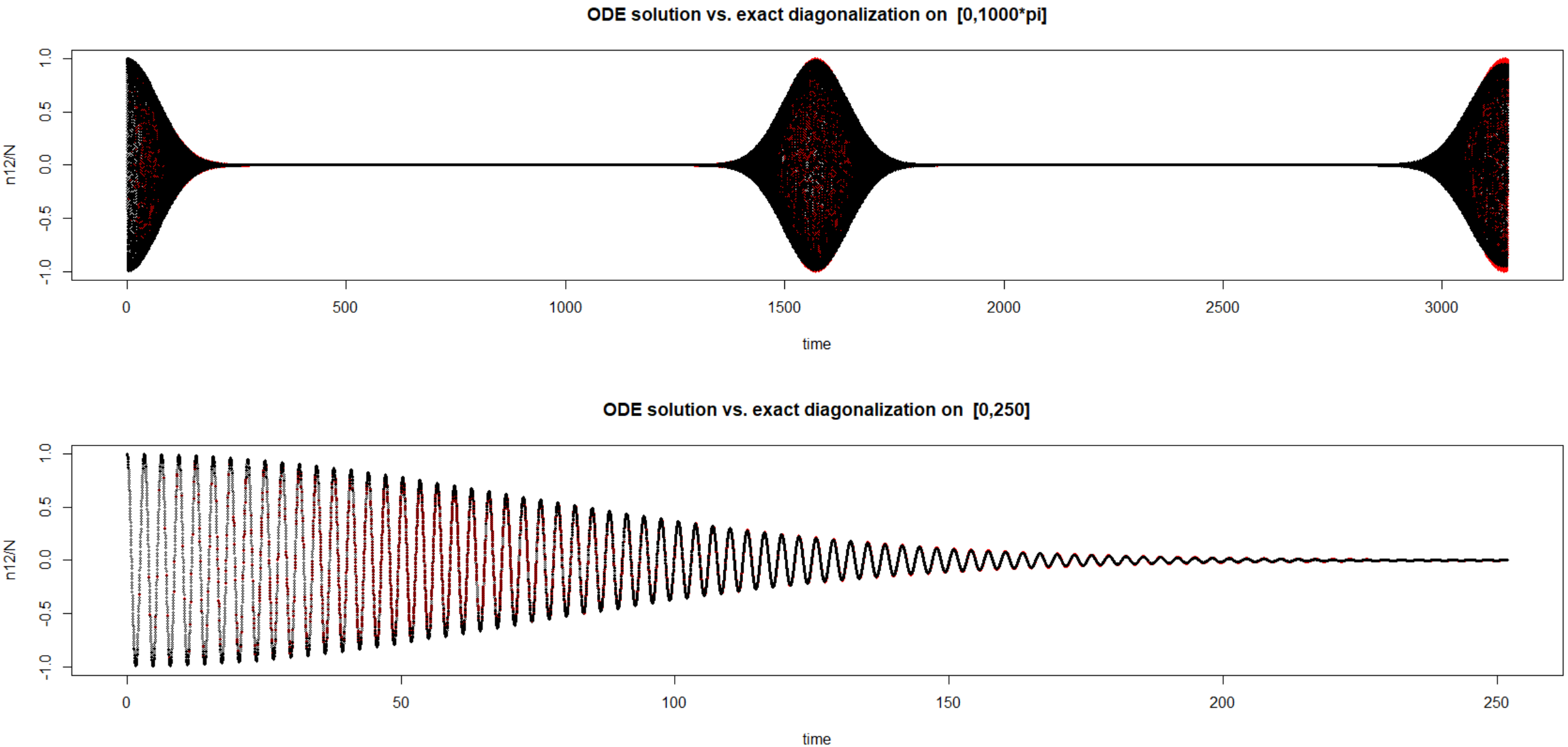}}

\bigskip
\centerline{\includegraphics[width=14.5cm]{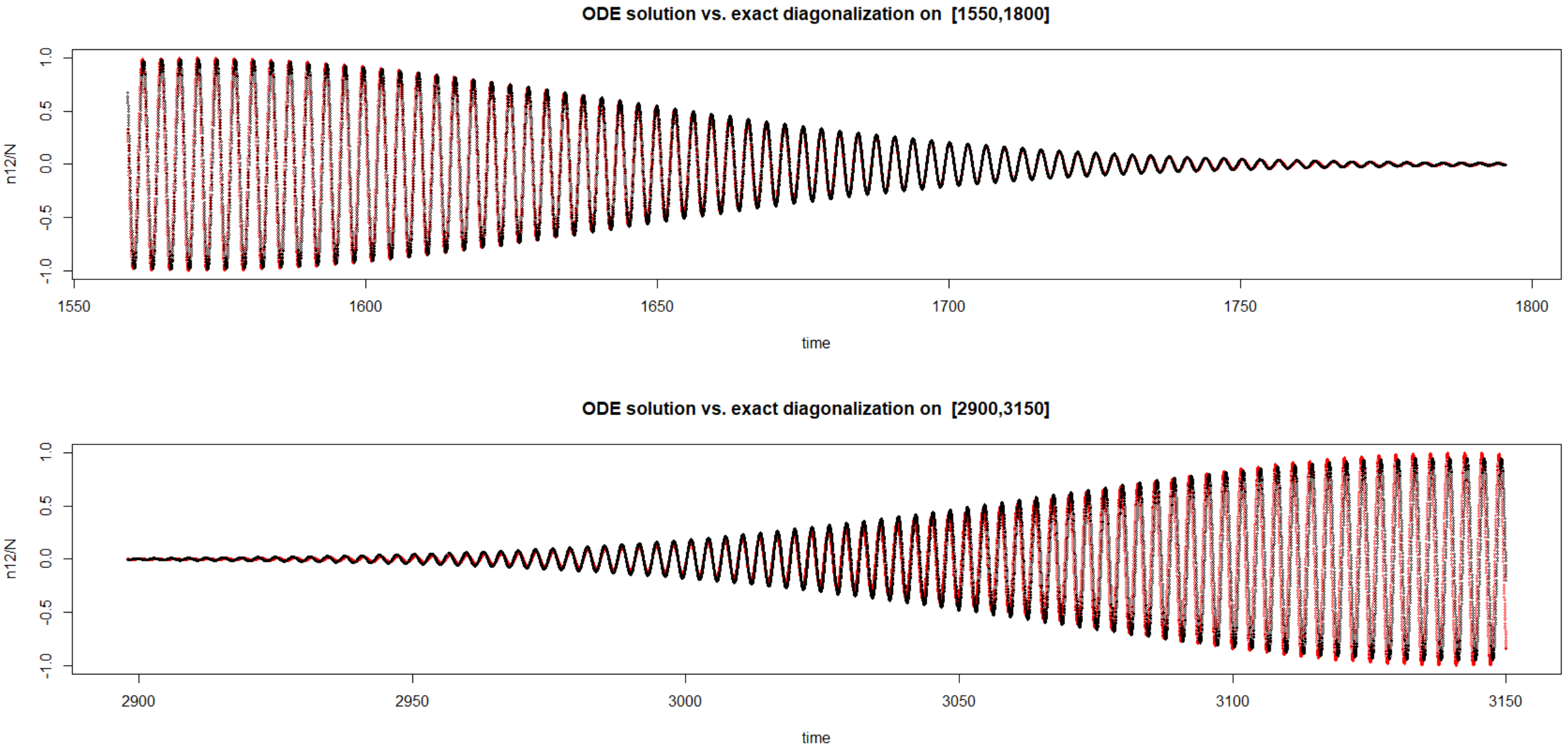}}

\medskip
\centerline{\footnotesize{Figure 5.3.3: Exact diagonalization (black) and ODE solution (red) on different time windows}}

\bigskip
\bigskip
\bigskip
\bigskip

\goodbreak

\noindent{\large\bf 6. Summary} 

\medskip
\bigskip
The paper has demonstrated that the formalism of stochastic calculus is very useful to address the dynamics of the 
Bose-Hubbard model. The fact that in the large $N$ limit the exact quantum dynamics can be obtained 
from an ODE system, the time dependent discrete GP equation, has been derived in a 
conceptually very pure and clean and transparent way. For finite $N$, the dynamics is given 
by the SDE systems of Theorems 4 and 5 of chapter 2 and the diffusive parts of those systems vanish in the large $N$ limit. 
More generally, the paper provides a technique to obtain GP-like mean field equations for an arbitrary given initial state, 
in arbitrary dimension and for an arbitrary hopping matrix.  
For the two site Bose-Hubbard model, the diffusive part has been taken into account with an approximation and 
collapse and revivals could be reproduced, numerically and also through an analytic calculation. A proper systematic 
treatment of the diffusive part is still missing and needs to be developed. It has also been shown that 
density matrix elements can be obtained from various exact parabolic second order PDEs.

\goodbreak

\bigskip
\bigskip
\bigskip
\noindent{\large\bf 7. Additional Remarks} 

\bigskip
\bigskip
By the end of 2016, the mathematics department of Hochschule RheinMain joined the Fa\-culty of Engineering and the author 
was asked by Klaus Michael Indlekofer from Electrical Engineering whether there would be some interest in joining a project 
on quantum dyna\-mics. After 9 years as a financial engineer at a bank, the author found that this would be a good opportunity 
to reenter the field and it didn't took long until it was realized that the Hubbard model is more relevant than ever due to some 
major experimental breakthroughs in the ultracold atoms area [17-21]. Working purely on the theoretical 
side, we can only humbly take notice of what is doable there [22]. 

\medskip
First attention then was drawn to phase space methods and the truncated Wigner approximation because of the very attractive idea 
to get the quantum dynamics from suitably weighted ODE trajectories. In particular, the beautiful papers of Polkovnikov [23-25], 
Polkovnikov, Sachdev and Girvin [26] and Davidson, Sels and Polkovnikov [27] served as a major motivation and inspiration for the 
current work. 

\medskip
With a theoretical and practical background in stochastic calculus from 9 years of option pricing, then it was natural to take 
a closer look to the long history of stochastic methods applied to the quantum many body problem [28,29]. In particular, the 
formalism of the Husimi Q-Function and the Positve P-Representation [30-34] was considered more closely and this, combined 
with the background of the author [35,36], then lead to the approach which is taken in this paper. 

\bigskip
Nowadays nearly taken for granted, but the almost unlimited and instantaneous access to the 
science knowledge of the planet and the people who provide it also has been critical for the 
completion of this work. There have been numerous papers, the majority of them probably not being cited here, 
where just a particular item was looked up and then the conclusion was, okay, for our purposes this does not lead in the right direction.  
Those references may not seem directly relevant to the now final version of this paper, but they have been 
critical in order to get there. In this class fall for example references for BCH like formulas and time ordered exponentials [37-40]
([39] derives very interesting formulae which are in the same spirit but more general than a formula derived by the author in chapter 10 of [36]), 
Carleman Linearization Technique and Kroenecker products of matrices (Kowalski and Steeb [41] wrote a very beautiful book on that) or 
references which relate to the author's attempt to evaluate the Fresnel integrals directly.

\medskip
With the presented formalism, the paper opens up the possibility to address a range of very interesting topics  
like fermionic models or thermodynamic quantities with an $e^{-\beta H}$ in it instead of an $e^{-itH}$, and 
the author looks very much forward to consider these issues, but with a teaching load of 18 hours 
per week at a German University of Applied Sciences, research basically has to be restricted to 
the off-term periods which are March and August and September each year.

\newpage

\noindent{\large\bf Appendix: Compact Summary Stochastic Calculus}
\numberwithin{equation}{section}
\renewcommand\thesection{A}
\setcounter{equation}{0}

\bigskip
\bigskip
\bigskip
\noindent{\bf A.1\; Standard Brownian Motion and Wiener Measure} 

\bigskip
\bigskip
A standard Brownian motion $\,x_t=x_{t_k}\,$ in discretized time $\,t=t_k=k\Delta t\,$ is the combination of inte\-gration variables
\beqn
x_{t_k}&=&\sqrt{\Delta t}\,\ts \sum_{\ell=1}^k \phi_\ell 
\eeqn
where the $\phi_\ell$ are to be integrated against Wiener measure which is simply a product of independent standard Gaussian distributions, 
\beqn
dW&:=&\pro_{\ell=1}^N e^{\,-{\phi_\ell^2\over 2} }\; {\ts {d\phi_\ell \over \sqrt{2\pi} }} \;\;=\;\; 
  \pro_{\ell=1}^N e^{\,-\,{(x_{t_\ell}-x_{t_{\ell-1}})^2\over {2\Delta t}} }\; {\ts {dx_{t_\ell} \over \sqrt{2\pi \,\Delta t} }}   \pI  \phantom{mm} \lbeq{A.2}
\eeqn
where $T=N\Delta t$ is a fixed time horizon. Basic to stochastic calculus, in particular for the Ito formula in the next section, is the 
Brownian motion calculation rule 
\beqn
(dx_t)^2&=& dt \pI \lbeq{A.3}
\eeqn
which can be motivated in several ways. Consider the discretized version of the quantity $\,\int_0^T f(t)\,(dx_t)^2\,$ which is 
\beqn
I_{\Delta t}(f)&:=&\ts \sum_{k=1}^N f(t_k)\,(x_{t_k}-x_{t_{k-1}})^2\;\;=\;\; \Delta t\sum_{k=1}^{N} f(t_k)\,\phi_k^2 \lbeq{n1}  \pI \lbeq{A.4}
\eeqn
where $f$ is an arbitrary function of one variable. Its expectation value and variance are given by 
\beqn
\E\bigl[\,I_{\Delta t}(f)\,\bigr]&=&\ts\Delta t\sum_{k=1}^{N_t} f(t_k) \;\; \buildrel \Delta t\to 0 \over \to \;\; \int_0^T f(t)\,dt  \\ 
\V\bigl[\,I_{\Delta t}(f)\,\bigr]&=&\ts 2\,(\Delta t)^2 \sum_{k=1}^{N_t} f(t_k)^2\;\;\approx\;\;2\Delta t \int_0^T f(t)^2\,dt
 \;\; \buildrel \Delta t\to 0 \over \to \;\; 0 \pI
\eeqn
Thus, with Chebyshev's inequality we get for any $\vep>0$
\beqn
\lim_{\Delta t\to 0} \ts{\sf Prob}\Bigl[ \, \bigl|\,I_{\Delta t}(f)\,-\,\int_0^T f(t)\,dt\,\bigr| \ge\vep \, \Bigr] &= & 0   
\eeqn
or more intuitively 
\beqn
{\ts \int_0^T f(t)\,(dx_t)^2 }\;\;=\;\; \lim_{\Delta t\to 0}\ts \sum_{k=1}^N f(t_k)\,(x_{t_k}-x_{t_{k-1}})^2 &=& \ts \int_0^T f(t)\,dt   \pI
\eeqn
for arbitrary $f$. The validity of this equation is usually more compactly written as 
\beqn
(dx_t)^2&=&dt \pM  \lbeq{A.9}
\eeqn
although the last equation on its own is not correct, in discretized form we have
\beqn 
(\Delta x_{t_k})^2\;=\;(x_{t_k}-x_{t_{k-1}})^2 \;=\; \bigl(\; \sqrt{\Delta t}\,\phi_k\;\bigr)^2 \;=\; \Delta t\, \phi_k^2 \;&\ne&\; \Delta t \phantom{mmmm} \lbeq{A.10}
\eeqn
and only after applying the operation 
\beqn
{\ts \int_0^Tdt\, f(t)\,\cdots }&=&\lim_{\Delta t\to 0} \;\ts \sum_{k=1}^N f({t_k})\, \cdots  
\eeqn
to the left and right hand side of (\req{A.10}) we get a valid equation. Since we cannot use Chebyshev's inequality in the complex Fresnel case, 
let us motivate the basic Brownian motion calculation rule (\req{A.3}) in a different way which directly generalizes to the Fresnel case:

\bigskip
\bigskip
\noindent{\bf  Theorem A1: }  Let $I_{\Delta t}(f)$ be defined as in (\req{A.4}) and let $\E[\,\cdot\,]$ denote the expectation with respect 
to the Wiener measure (\req{A.2}). Then 
\beqn
\lim_{\Delta t\to 0}  \E\bigl[\;e^{iq\,I_{\Delta t}(f)}\;\bigr] &=& e^{iq\,\int_0^Tf(t)dt}   \pI \lbeq{A.12}
\eeqn
such that for any $\;F(x)= \int_{\mathbb R} \hat F(q)\; e^{iqx}\;\ts {dq/(2\pi)}\;$ we obtain 
\beqn
\lim_{\Delta t\to 0} \E\Bigl[\; F\bigl(\,I_{\Delta t}(f)\,\bigr) \;\Bigr] &=& F\Bigl(\;{\ts \int_0^T f(t)\, dt }\;\Bigr) \pI
\eeqn

\bigskip
\noindent{\bf Proof:} We have
\beqn
\E\bigl[\;e^{iq\,I_{\Delta t}(f)}\;\bigr]&=&\int_{\mathbb R^N} \; \pro_{k=1}^N e^{-{1\over 2}\bigl(\,1\,-\,2iq\Delta t f({t_k})\,\bigr)\phi_k^2}
    \;\ts  {d\phi_k\over \sqrt{2\pi}}  \pI \nn\\ 
&=&\pro_{k=1}^N \ts{1\over \sqrt{\,1\,-\,2iq\Delta t f({t_k})\,} }   \pI\nn \\ 
&=&\ts \exp\Bigl\{\; -{1\over 2}\, \sum_{k=1}^N\log\bigl[\,1\,-\,2iq\Delta t f({t_k})\,\bigr] \; \Bigr\}  \pI\nn \\
&=&\ts \exp\Bigl\{\; +{1\over 2}\, \sum_{k=1}^N\,2iq\Delta t f({t_k}) \;+\;O(\Delta t) \; \Bigr\}  \pI \nn \\ 
&\buildrel \Delta t\to 0 \over \to&\ts \exp\bigl\{\, +\,i\,q\int_0^T f(t)\,dt \;\bigr\} \pI 
\eeqn
which coincides with (\req{A.12}). $\;\blacksquare$ 

\bigskip
\bigskip
Let us close this section by recalling that the calculation rule $(dx_t)^2= dt$ is accompanied by the rules 
\beqn
dx_t\,dt \;\;=\;\; dt\,dt\;\;=\;\;0 
\eeqn 
which also will be used in the next section. 

\goodbreak

\bigskip
\bigskip
\bigskip
\noindent{\bf A.2\; Ito Formula and Stochastic Integrals} 

\bigskip
\bigskip
For some arbitrary function $f=f(x)$ and $x_t$ a Brownian motion we can write
\beqn
f(x_T)-f(x_0)&=& \ts\sum_{k=1}^N \bigl\{ \; f(x_{t_k})\,-\, f(x_{t_{k-1}}) \; \bigr\}  \pI \nn \\ 
&=& \ts\sum_{k=1}^N \bigl\{ \; f(x_{t_{k-1}}+ dx_{t_k} )\,-\, f(x_{t_{k-1}}) \; \bigr\} 
\eeqn
with
\beqn
dx_{t_k}&=&x_{t_k} \,-\,x_{t_{k-1}} \lbeq{I38} 
\eeqn
Using a standard Taylor expansion,
\beqn
f(x_{t_k})&=&f\big( x_{t_{k-1}}+ dx_{t_k} \bigr) \lbeq{I39}   \\ 
&=&\ts f( x_{t_{k-1}})\,+\,f'(x_{t_{k-1}})\,dx_{t_k}  \,+\,{1\over 2}\,f''(x_{t_{k-1}})\,(dx_{t_k})^2
    \,+\,{1\over 3!}\,f'''(x_{t_{k-1}})\,(dx_{t_k})^3\,+\,\cdots  \pI  \nn 
\eeqn
and the calculation rules of a Brownian motion, 
\beqn
(dx_{t_k})^2&=&dt
\eeqn
and $(dx_{t_k})^3\;=\;(dx_{t_k})^2\,dx_{t_k}\;=\; dt\,dx_{t_k}\;=\;0\,$, we obtain in the limit $\Delta t\to 0$: 
\medskip
\beqn
f(x_{t_k})&=&\ts f( x_{t_{k-1}})\,+\,f'(x_{t_{k-1}})\,dx_{t_k}  \,+\,{1\over 2}\,f''(x_{t_{k-1}})\,dt
    \,+\,{1\over 3!}\,f'''(x_{t_{k-1}})\,\times\,0  \pM
\eeqn
or
\beqn
df(x_{t_k})\;\;:=\;\;\ts f(x_{t_k})\,-\, f( x_{t_{k-1}}) 
&=&\ts  f'(x_{t_{k-1}})\,dx_{t_k}  \;+\;{1\over 2}\,f''(x_{t_{k-1}})\,dt  \pM
\eeqn

\medskip
which is the differential version of the Ito formula. If we sum this up, 
\beqn
f(x_T)-f(x_0)&=& \ts\sum_{k=1}^N \bigl\{ \; f(x_{t_k})\,-\, f(x_{t_{k-1}}) \; \bigr\} \pI \nn \\ 
&=&\ts\sum_{k=1}^N \bigl\{\ts \;f'(x_{t_{k-1}})\,dx_{t_k}  \,+\,{1\over 2}\,f''(x_{t_{k-1}})\,dt \; \bigr\} \pI \nn \\ 
&\buildrel \Delta t \to 0 \over =&\ts {\ts \int_0^T f'(x_t)\,dx_{t} \;+\; {1\over 2}\,\int_0^T f''(x_t)\,dt }  \pI   \lbeq{I42}
\eeqn
where the first integral is refered to as an Ito integral. Thus, its definition is, we replace the $f'$ by an $f$,
\beqn
\ts \int_0^T f(x_t)\,dx_{t}&:=&\lim_{\Delta t \to 0}\ts \sum_{k=1}^N  \, f(x_{t_{k-1}})\,dx_{t_k}  \pS \lbeq{A.23}
\eeqn
What on a first sight looks a bit odd is the fact that if we replace the $x_{t_{k-1}}$ on the right hand side of (\req{A.23}) in the $f$ by, say, $(x_{t_{k-1}}+x_{t_k})/2$, 
we actually get a different limit. This is a consequence of the fact that $(dx_t)^2$ is nonzero. The following definition is refered to 
as a Stratonovich integral:
\beqn
\ts \int_0^T f(x_t)\circ dx_{t}&:=&\lim_{\Delta t \to 0} \ts\sum_{k=1}^N \ts \, f\bigl({x_{t_{k-1}}+x_{t_k}\over 2}\bigr)\,dx_{t_k}  \lbeq{I44}
\eeqn
Then there is the following relation:
\beqn
\ts \int_0^T f(x_t)\circ dx_{t}&=& {\ts \int_0^T f(x_t)\,dx_{t} \;+\; {1\over 2}\,\int_0^T f'(x_t)\,dt }  \pI   \lbeq{A.25}
\eeqn
Namely, because of
\beqn
\ts {x_{t_{k-1}}+x_{t_k}\over 2}&=&x_{t_{k-1}}\;+\; \ts {x_{t_k}-x_{t_{k-1}}\over 2} \;\;=\;\; x_{t_{k-1}}\;+\; \ts {dx_{t_k}\over 2}
\eeqn
we have, using $\,(dx_t)^2=dt\,$ and $\,dx_t\,dt=0\,$ in the third line,
\beqn
\ts f\bigl({x_{t_{k-1}}+x_{t_k}\over 2}\bigr)\,dx_{t_k}&=& f\bigl(x_{t_{k-1}}+ \ts {dx_{t_k}\over 2}\bigr)\,dx_{t_k}  \pI \\ 
&=&\ts \Bigl\{ f(x_{t_{k-1}}) \; +\; f'(x_{t_{k-1}})\ts {dx_{t_k}\over 2} \; +\; {1\over 2}\,f''(x_{t_{k-1}})\ts \bigl({dx_{t_k}\over 2}\bigr)^2 
     \Bigr\} \,dx_{t_k}  \pI \nn \\ 
& = & f(x_{t_{k-1}})\,dx_{t_k} \; +\; f'(x_{t_{k-1}})\ts {dt\over 2} 
    \; +\; {1\over 2}\,f''(x_{t_{k-1}})\,\times\,0 \pI  \nn
\eeqn
and (\req{A.25}) follows. We summarize in the following

\bigskip
\bigskip
\noindent{\bf Theorem A2:} Let $f=f(x)$ be an arbitrary function and $\{x_t\}_{t\ge 0}$ be a Brownian motion. Then: 
\beqn
f(x_T)-f(x_0)\;\;=\;\;\ts \int_0^T df &=&  {\ts \int_0^T f'(x_t)\,dx_{t} \;+\; {1\over 2}\,\int_0^T f''(x_t)\,dt }  \pM \lbeq{I46} \\ 
&=&  {\ts \int_0^T f'(x_t)\circ dx_{t} }   \pI   \lbeq{I47}
\eeqn
where the first integral in (\req{I46}) is a stochastic Ito integral and the integral in 
(\req{I47}) is a Stratonovich integral. (\req{I46}) is refered to as Ito formula or the Ito lemma.  

\medskip
\bigskip
\bigskip
\noindent Finally we want to recall a very practical property of Ito integrals, namely, that their expectation value always vanishes, 
\beqn
\E\Bigl[\,\ts \int_0^T f(x_t)\,dx_{t} \,\Bigr]&=& 0 \pI 
\eeqn
for arbitrary $f$. This follows from the fact that 
\beqn
dx_{t_k}\;=\;x_{t_k} \,-\,x_{t_{k-1}}&=&\sqrt{dt}\,\phi_k 
\eeqn
while 
\beqn
f(x_{t_{k-1}})&=&\ts f\bigl(\,\sqrt{dt}\,\sum_{\ell=1}^{k-1}\phi_\ell\,\bigr) 
\eeqn

\medskip
does not depend on $\phi_k$. Thus, $f(x_{t_{k-1}})$ and $dx_{t_k}$ are independent quantities and we obtain 
\beqn
\E\bigl[\, f(x_{t_{k-1}})\,dx_{t_k} \,\bigr]&=&\E\bigl[\, f(x_{t_{k-1}}) \,\bigr] \times \sqrt{dt}\; \E[\phi_k] \;\;=\;\; 0 \pI
\eeqn
It is exactly this property which makes Ito integrals a preferable choice over Stratonovich integrals, at least in the context of this paper, 
although just concerning the optics one may consider the Ito formula (\req{I46}) as more complicated than the Stratonovich formula (\req{I47}) 
which looks more like the standard calculus formula.  

\goodbreak

\bigskip
\bigskip
\bigskip
\noindent{\bf A.3\; Kolmogorov Backward Equation and Feynman-Kac Formula} 

\bigskip
\bigskip

\noindent A, let's say, one dimensional Ito diffusion $X_t$ is a stochastic quantity which is given by the recursion 
\beqn
X_{t_k}&=&X_{t_{k-1}}\;+\;  a(X_{t_{k-1}},t_{k-1})\,dt\;+\; b(X_{t_{k-1}},t_{k-1})\, \sqrt{dt}\,\phi_k \pI 
\eeqn
where the $\{\phi_k\}_{k=1}^N$ are to be integrated against standard Wiener measure. In continuous time, this reads 
\beqn
dX_t&=&a(X_t,t)\,dt\;+\; b(X_t,t)\, dx_t   \pI
\eeqn
with $x_t$ a Brownian motion. To each Ito diffusion we can assign a second order differential operator $A$ defined by the following equation: 
Let $f=f(x,t)$ be an arbitrary function of two variables. Then, using $\,(dx_t)^2=dt\,$ and $\,dx_t\,dt=(dt)^2=0\,$ in the fourth line, 
\beqn
df(X_t,t)&=&f(X_t,t)\;-\;f(X_{t-dt},t-dt)  \nn \pI \\
&=&\ts {\pt f\over \pt x}\,dX_t\;+\;{1\over 2}\,{\pt^2 f\over \pt x^2}\,(dX_t)^2 \;+\;{\pt f\over \pt t}\,dt\nn \\ 
&=&\ts {\pt f\over \pt x}\,[\,a\,dt\,+\,b\,dx_t\,]\;+\;{1\over 2}\,{\pt^2 f\over \pt x^2}\,[\,a\,dt\,+\,b\,dx_t\,]^2 \;+\;{\pt f\over \pt t}\,dt\nn \pI \\ 
&=&\ts {\pt f\over \pt x}\,[\,a\,dt\,+\,b\,dx_t\,]\;+\;{b^2\over 2}\,{\pt^2 f\over \pt x^2}\,dt\;+\;{\pt f\over \pt t}\,dt \nn \\ 
&=&\ts\{\, a{\pt f\over \pt x}\,+\,{b^2\over 2}\,{\pt^2 f\over \pt x^2}\,+\,{\pt f\over \pt t}\,\}\,dt \;+\; b{\pt f\over \pt x}\,dx_t  \nn \pI\\ 
&=:&\ts \{\,Af\,+\,{\pt f\over \pt t}\,\}\,dt \;\;+\;\;{\rm diffusive} \lbeq{A.36}
\eeqn
That is, 
\beqn
(Af)(x,t)&:=&\ts a(x,t)\,{\pt f\over \pt x}(x,t)\;+\;{b^2(x,t)\over 2}\,{\pt^2 f\over \pt x^2}(x,t)  \pI  \lbeq{A.37}
\eeqn
Now let $g=g(x)$ be another arbitrary function of one variable. We fix a start time $t$ and an end time $T>t$ and consider the expectation (the $f$ below 
also depends on end time $T$, but this dependency we do not make explicit in the notation)
\beqn
f(x,t)&:=&\E_t^T[ \,g(X_T^{t,x})\,]  \pI
\eeqn
which in discrete time  $t=t_k=kdt$ and $T=t_N=Ndt$ is given by 
\medskip
\beqn
f(x,t_k)\;\;=\;\;\E_{t_k}^{t_N}\bigl[ \, g(X^{t_k,x}_{t_N})\, \bigr]  
&:=&\int_{\mathbb R^{N-k}}  g\Bigl(\; X^{t_k,x}_{t_N}\bigl(\{\phi_\ell\}_{\ell=k+1}^N\bigr) \; \Bigr) \; 
  \pro_{\ell=k+1}^N e^{-{\phi_\ell^2\over 2}} \; \ts {d\phi_\ell\over \sqrt{2\pi}} \pI \phantom{mm}
\eeqn

\medskip
Here we used the notation $X_T^{t,x}$ with superscripts $(t,x)$ to indicate that the diffusion $X$ starts at time $t$ with initial value $x$, 
\beqn
X_t^{t,x}&=&x 
\eeqn
In particular, there is the identity 
\beqn
X^{t_k,X^{t_j,x_j}_{t_k}}_{t_N}&=&X^{t_j,x_j}_{t_N}  \pI
\eeqn
for arbitrary times $\,t_j\le t_k\le t_N$\,. Thus, if we define for some fixed initial values $(t_0,x_0)$ the stochastic quantity
\beqn
M_{t_k}\;\;:=\;\;f(X^{t_0,x_0}_{t_k},t_k)&=&\E_{t_k}^T\biggl[ \; g\Bigl(\, X^{t_k,X^{t_0,x_0}_{t_k}}_{t_N} \,\Bigr)\; \biggr] \;\;=\;\;
 \E_{t_k}^T\Bigl[ \; g\bigl(\, X^{t_0,x_0}_{t_N} \,\bigr)\; \Bigr]    \pI   \lbeq{9b} 
\eeqn
then this quantity is a martingale since we have for arbitrary time $t_j<t_k$ 
\medskip
\beqn
\E^T_{t_j}[\,M_{t_k}\,]&=&\E^T_{t_j}\biggl[\; \E_{t_k}^T\Bigl[ \; g\bigl(\, X^{t_0,x_0}_{t_N} \,\bigr)\; \Bigr] \; \biggr] 
  \;\; = \;\; \E_{t_j}^T\Bigl[ \; g\bigl(\, X^{t_0,x_0}_{t_N} \,\bigr)\; \Bigr] \;\;=\;\; M_{t_j} \pI 
\eeqn

\medskip
In particular, we have $\,\E^T_{t_{k-1}}[\,M_{t_k}\,]=M_{t_{k-1}}\,$ which means that we have to have
\beqn
0&\buildrel !\over =& \E^T_{t_{k-1}}[\,M_{t_k}\,]\;-\; M_{t_{k-1}}\;\;=\;\; \E^T_{t_{k-1}}\bigl[\,M_{t_k}-M_{t_{k-1}}\,\bigr] 
\;\;=\;\; \E^T_{t_{k-1}}\bigl[\,df(X^{t_0,x_0}_{t_k},t_k)\;\bigr] \pS \nn \\ 
&\buildrel (\req{A.36})\over=&\ts \E^T_{t_{k-1}}\Bigl[\;\;  \bigl\{Af\,+\,{\pt f\over \pt t}\,\bigr\}(X^{t_0,x_0}_{t_{k-1}},t_{k-1})\;dt \;\;+\;\;{\rm diffusive} \;\; \Bigr]  \nn \\ 
&=&\ts  \bigl\{Af\,+\,{\pt f\over \pt t}\,\bigr\}(X^{t_0,x_0}_{t_{k-1}},t_{k-1})\;dt \;\;+\;\; \E^T_{t_{k-1}}[\,{\rm diffusive} \,] \pS \nn \\ 
&=&\ts  \bigl\{Af\,+\,{\pt f\over \pt t}\,\bigr\}\;dt \;\;+\;\;0 \pS
\eeqn

\smallskip
That is, the expectation $\;f(x,t)=\E_t^T[ \,g(X_T^{t,x})\,]\;$ has to satisfy the second order PDE 
\beqn
\ts {\pt f\over \pt t}\,+\, Af&=&0 \pS
\eeqn
We summarize in the following 

\medskip
\bigskip
\bigskip
\noindent{\bf Theorem A3:} {\bf a)} Let $X_t$ be an Ito diffusion given by 
\beqn
dX_t&=&a(X_t,t)\,dt\;+\; b(X_t,t)\, dx_t   \pI
\eeqn
and for some initial values $(t,x)$ define the Wiener expectation 
\beqn
f(x,t)&:=&\E_t^T[ \,g(X_T^{t,x})\,] \pI 
\eeqn
Then $f$ can be obtained as the solution of the parabolic second order PDE 
\beqn
\ts {\pt f\over \pt t}\,+\, Af&=&0 \pI
\eeqn
with final condition $\,f(x,T)=g(x)\,$ and $A$ given by (\req{A.37}) above.  

\smallskip
\bigskip
\noindent{\bf b)} Let $X_t$ be a time-homogenous Ito diffusion given by 
\beqn
dX_t&=&a(X_t)\,dt\;+\; b(X_t)\, dx_t   \pI
\eeqn
with coefficients $a=a(X_t)$ and $b=b(X_t)$ which do not explicitely depend on time. Then there is the identity 
\beqn
\E_t^T[ \,g(X_T^{t,x})\,]&=&\E_0^{T-t}[ \,g(X_{T-t}^{0,x})\,]  \pS 
\eeqn
In particular, the expectation 
\beqn
f(x,t)&:=&\E_0^{t}[ \,g(X_{t}^{0,x})\,]  \pS
\eeqn
now considered as a function of the end time $t$ (start time is 0), is a solution of the PDE 
\beqn
\ts \,-\,{\pt  f\over \pt t}\,+\, A f&=&0 \pI  \lbeq{A.54}
\eeqn
with initial condition $\,f(x,0)=g(x)\,$ and $A$ given by 
\beqn
(Af)(x,t)&:=&\ts a(x)\,{\pt f\over \pt x}(x,t)\;+\;{b^2(x)\over 2}\,{\pt^2 f\over \pt x^2}(x,t) \;\;. \pI
\eeqn

\bigskip
\bigskip
\noindent Equation (\req{A.54}) is then usually refered to as Kolmogorov's backward equation, see for example Theorem 8.1.1 in the book of Oeksendal [42]. 
By a slight variation of the above argument one also obtains a PDE representation for the quantity 
\beqn
u(x,t)&:=&\E_0^{t}[ \,e^{\,-\,\int_0^t r(X_s^{0,x})ds}\,g(X_{t}^{0,x})\,]  \pI 
\eeqn
which is then the Wiener measure version of the Feynman-Kac formula, this is Theorem 8.2.1 in Oeksendal [42], and it reads 
\beqn
\ts \,-\,{\pt u\over \pt t}\,+\, Au\,-\,ru&=&0 \pS \pM\\ 
 u(x,0)&=&g(x) \;\; \nn
\eeqn
In this paper we do not use it, neither the Wiener nor the Fresnel version, we only use the Fresnel version of part (b) of the theorem above 
and this version we write down in the next section.

\goodbreak

\bigskip
\bigskip
\bigskip
\bigskip
\noindent{\bf A.4\; Stochastic Calculus with Respect to Fresnel Measure}

\bigskip
\bigskip
\noindent{\bf  Fresnel Brownian Motion and Fresnel Measure} 

\medskip
\bigskip
A Fresnel Brownian motion or Fresnel BM in discretized time $t=t_k=k\Delta t$ we define as the combination of integration variables
\beqn
x_{t_k}&=&\sqrt{\Delta t}\,\ts \sum_{\ell=1}^k \phi_\ell 
\eeqn
where the $\phi_\ell$ are to be integrated against Fresnel measure which is given by 
\beqn
dF&:=&\pro_{\ell=1}^N e^{\,i\,{\phi_\ell^2\over 2} }\; {\ts {d\phi_\ell \over \sqrt{2\pi i} }} \;\;=\;\; 
  \pro_{\ell=1}^N e^{\,i\,{(x_{t_\ell}-x_{t_{\ell-1}})^2\over {2\Delta t}} }\; {\ts {dx_{t_\ell} \over \sqrt{2\pi i \,\Delta t} }}   \pI  \phantom{mm}
\eeqn
Let's consider again the discretized version of the quantity $\,\int_0^T f(t)\,(dx_t)^2\,$ which is  
\beqn
I_{\Delta t}(f)&:=&\ts \sum_{k=1}^N f(t_k)\,(x_{t_k}-x_{t_{k-1}})^2\;\;=\;\; \Delta t\sum_{k=1}^{N_t} f(t_k)\,\phi_k^2 \lbeq{n1}  \pI
\eeqn
Then the Fresnel analog of equation (\req{A.12}) is 
\beqn
\lim_{\Delta t\to 0}  \E\bigl[\;e^{iq\,I_{\Delta t}(f)}\;\bigr] &=& e^{\,-\,q\,\int_0^Tf(t)dt}   \pI  \lbeq{A.62}
\eeqn
which then leads to the following basic calculation rule for Fresnel Brownian motions: 
\beqn
(dx_t)^2&=&i\,dt \pI  \lbeq{A.63}
\eeqn
and $\,dx_t\,dt = dt\,dt=0\,$. The proof of (\req{A.62}) is as follows:  
\beqn
\E\bigl[\;e^{iq\,I_{\Delta t}(f)}\;\bigr]&=&\int_{\mathbb R^N} \; \pro_{k=1}^N e^{\,{i\over 2}\bigl(\,1\,+\,2q\Delta t f({t_k})\,\bigr)\phi_k^2}
   \;\ts  {d\phi_k\over \sqrt{2\pi i}}  \pI \nn \\ 
&=&\pro_{k=1}^N \ts{1\over \sqrt{\,1\,+\,2q\Delta t f({t_k})\,} }   \pI \nn \\ 
&=&\ts \exp\Bigl\{\; -{1\over 2}\, \sum_{k=1}^N\log\bigl[\,1\,+\,2q\Delta t f({t_k})\,\bigr] \; \Bigr\} \nn \\
&=&\ts \exp\Bigl\{\; -{1\over 2}\, \sum_{k=1}^N\,2q\Delta t f({t_k}) \;+\;O(\Delta t) \; \Bigr\}  \pI \nn \\ 
&\buildrel \Delta t\to 0 \over \to&\ts \exp\bigl\{\, -\,q\int_0^T f(t)\,dt \;\bigr\} \pI  
\eeqn
and this coincides with (\req{A.62}). 

\goodbreak

\medskip
\bigskip
\bigskip
\noindent{\bf The Fresnel Version of the Ito Formula} 

\medskip
\bigskip
As in section A.2, we can write for some arbitrary function $f=f(x)$ and $x_t$ now being a Fresnel BM 
\beqn
f(x_T)-f(x_0)&=& \ts\sum_{k=1}^N \bigl\{ \; f(x_{t_k})\,-\, f(x_{t_{k-1}}) \; \bigr\}  \pS 
\eeqn
with
\beqn
df(x_{t_k})\;\;:=\;\;\ts f(x_{t_k})\,-\, f( x_{t_{k-1}}) &=&\ts  f'(x_{t_{k-1}})\,dx_{t_k}  \;+\;{1\over 2}\,f''(x_{t_{k-1}})\,(dx_{t_k})^2  \pS \nn \\ 
&\buildrel (\req{A.63}) \over=&\ts  f'(x_{t_{k-1}})\,dx_{t_k}  \;+\;{i\over 2}\,f''(x_{t_{k-1}})\,dt 
\eeqn
Summing this up, 
\beqn
f(x_T)-f(x_0)&=& \ts\sum_{k=1}^N \bigl\{\ts \;f'(x_{t_{k-1}})\,dx_{t_k}  \,+\,{i\over 2}\,f''(x_{t_{k-1}})\,dt \; \bigr\}  \nn \pS \\ 
&\buildrel dt \to 0 \over =:&\ts {\ts \int_0^T f'(x_t)\,dx_{t} \;+\; {i\over 2}\,\int_0^T f''(x_t)\,dt }  \pS  
\eeqn

\medskip
where the first integral again has the property that its expectation value always vanishes, 
\beqn
\E\Bigl[\,\ts \int_0^T f(x_t)\,dx_{t} \,\Bigr]&=& 0 \pI 
\eeqn
since as in the Wiener case the quantities $f(x_{t_{k-1}})$ and $dx_{t_k}$ are independent and we obtain 
\beqn
\E\bigl[\, f(x_{t_{k-1}})\,dx_{t_k} \,\bigr]&=&\E\bigl[\, f(x_{t_{k-1}}) \,\bigr] \times \sqrt{dt}\; \E[\phi_k] \;\;=\;\; 0 \pI
\eeqn
Concerning the last equality, one probably should make the definition 
\beqn
\E[\,\phi\,]&=&{\ts \int_{-\infty}^{+\infty}\, \phi\;e^{\,i\,{\phi^2\over 2}}\;{d\phi\over \sqrt{2\pi i}} } 
\;\;:=\;\;\lim_{R\to \infty} \ts \int_{-R}^{+R} \,\phi\;e^{\,i\,{\phi^2\over 2}}\;{d\phi\over \sqrt{2\pi i}} \;\;=\;\;0  \pI
\eeqn

\goodbreak

\bigskip
\bigskip
\noindent{\bf  Fresnel Version of Kolmogorov's Backward Equation} 

\bigskip

\noindent We proceed as in section A.3 and define a Fresnel diffusion $X_t$ as a stochastic quantity which is given by the recursion 
\beqn
X_{t_k}&=&X_{t_{k-1}}\;+\;  a(X_{t_{k-1}},t_{k-1})\,dt\;+\; b(X_{t_{k-1}},t_{k-1})\, \sqrt{dt}\,\phi_k \pI 
\eeqn
where the $\{\phi_k\}_{k=1}^N$ are to be integrated against Fresnel measure. In continuous time, we write  
\beqn
dX_t&=&a(X_t,t)\,dt\;+\; b(X_t,t)\, dx_t   \pI
\eeqn
To each Fresnel diffusion, we assign the second order operator $A$ through  ($f$ again denotes an arbitrary function of two variables) 
\medskip
\beqn
df(X_t,t)&=&\ts {\pt f\over \pt x}\,dX_t\;+\;{1\over 2}\,{\pt^2 f\over \pt x^2}\,(dX_t)^2 \;+\;{\pt f\over \pt t}\,dt\nn \\ 
&=&\ts {\pt f\over \pt x}\,[\,a\,dt\,+\,b\,dx_t\,]\;+\;{1\over 2}\,{\pt^2 f\over \pt x^2}\,[\,a\,dt\,+\,b\,dx_t\,]^2 \;+\;{\pt f\over \pt t}\,dt\nn \pI \\ 
&=&\ts {\pt f\over \pt x}\,[\,a\,dt\,+\,b\,dx_t\,]\;+\;i\,{b^2\over 2}\,{\pt^2 f\over \pt x^2}\,dt\;+\;{\pt f\over \pt t}\,dt \nn \\ 
&=&\ts\{\, a{\pt f\over \pt x}\,+\,i\,{b^2\over 2}\,{\pt^2 f\over \pt x^2}\,+\,{\pt f\over \pt t}\,\}\,dt \;+\; b{\pt f\over \pt x}\,dx_t  \nn \pI\\ 
&=:&\ts \{\,Af\,+\,{\pt f\over \pt t}\,\}\,dt \;\;+\;\;{\rm diffusive} \lbeq{A.73}
\eeqn
That is, 
\beqn
(Af)(x,t)&:=&\ts a(x,t)\,{\pt f\over \pt x}(x,t)\;+\;i\,{b^2(x,t)\over 2}\,{\pt^2 f\over \pt x^2}(x,t)  \pS \lbeq{A.71n}
\eeqn
and we still have 
\beqn
\E[\,{\rm diffusive}\,]&=& 0 \pM
\eeqn

\medskip
We consider the Fresnel expectation 
\beqn
f(x,t)&:=&\E_t^T[ \,g(X_T^{t,x})\,]  \pI
\eeqn
which in discrete time  $t=t_k=kdt$ and $T=t_N=Ndt$ is given by 
\medskip
\beqn
f(x,t_k)\;\;=\;\;\E_{t_k}^{t_N}\bigl[ \, g(X^{t_k,x}_{t_N})\, \bigr]  
&:=&\int_{\mathbb R^{N-k}}  g\Bigl(\; X^{t_k,x}_{t_N}\bigl(\{\phi_\ell\}_{\ell=k+1}^N\bigr) \; \Bigr) \; 
  \pro_{\ell=k+1}^N e^{\,i\,{\phi_\ell^2\over 2}} \; \ts {d\phi_\ell\over \sqrt{2\pi i}} \pI \phantom{mm}  \lbeq{A.75}
\eeqn

\medskip
For some fixed initial values $(t_0,x_0)$, we define the stochastic quantity
\beqn
M_{t_k}\;\;:=\;\;f(X^{t_0,x_0}_{t_k},t_k)&=&\E_{t_k}^T\biggl[ \; g\Bigl(\, X^{t_k,X^{t_0,x_0}_{t_k}}_{t_N} \,\Bigr)\; \biggr] \;\;=\;\;
 \E_{t_k}^T\Bigl[ \; g\bigl(\, X^{t_0,x_0}_{t_N} \,\bigr)\; \Bigr]    \pI   \lbeq{9b} 
\eeqn
which again is a martingale since for $t_j<t_k$
\beqn
\E^T_{t_j}[\,M_{t_k}\,]&=&\E^T_{t_j}\biggl[\; \E_{t_k}^T\Bigl[ \; g\bigl(\, X^{t_0,x_0}_{t_N} \,\bigr)\; \Bigr] \; \biggr] 
  \;\; = \;\; \E_{t_j}^T\Bigl[ \; g\bigl(\, X^{t_0,x_0}_{t_N} \,\bigr)\; \Bigr] \;\;=\;\; M_{t_j} \pI 
\eeqn

\medskip
In particular, we have $\,\E^T_{t_{k-1}}[\,M_{t_k}\,]=M_{t_{k-1}}\,$ which means that we have to have
\beqn
0&\buildrel !\over =& \E^T_{t_{k-1}}[\,M_{t_k}\,]\;-\; M_{t_{k-1}}\;\;=\;\; \E^T_{t_{k-1}}\bigl[\,M_{t_k}-M_{t_{k-1}}\,\bigr] 
\;\;=\;\; \E^T_{t_{k-1}}\bigl[\,df(X^{t_0,x_0}_{t_k},t_k)\;\bigr] \pS \nn \\ 
&\buildrel (\req{A.73})\over=&\ts \E^T_{t_{k-1}}\Bigl[\;\;  \bigl\{Af\,+\,{\pt f\over \pt t}\,\bigr\}(X^{t_0,x_0}_{t_{k-1}},t_{k-1})\;dt \;\;+\;\;{\rm diffusive} \;\; \Bigr]  \nn \\ 
&=&\ts  \bigl\{Af\,+\,{\pt f\over \pt t}\,\bigr\}(X^{t_0,x_0}_{t_{k-1}},t_{k-1})\;dt \;\;+\;\; \E^T_{t_{k-1}}[\,{\rm diffusive} \,] \pS \nn \\ 
&=&\ts  \bigl\{Af\,+\,{\pt f\over \pt t}\,\bigr\}\;dt \;\;+\;\;0 \pS
\eeqn

\smallskip
That is, the Fresnel expectation $\;f(x,t)=\E_t^T[ \,g(X_T^{t,x})\,]\;$ has to satisfy the PDE 
\beqn
\ts {\pt f\over \pt t}\,+\, Af&=&0 \pI
\eeqn
with final condition $\,f(x,T)=g(x)\,$ and second order operator $A$ given by $(\req{A.71n})$.    

\bigskip
\noindent The time-homogenous case, the analog of part (b) of Theorem A3, then reads as follows: Let $X_t$ be a time-homogenous Fresnel diffusion given by 
\beqn
dX_t&=&a(X_t)\,dt\;+\; b(X_t)\, dx_t   \pI
\eeqn
with coefficients $a=a(X_t)$ and $b=b(X_t)$ which do not explicitely depend on time and let $g=g(x)$ be an arbitrary function 
of one variable. Then the Fresnel expectation 
\beqn
f(x,t)&:=&\E_0^{t}[ \,g(X_{t}^{0,x})\,]  \pI 
\eeqn
is a solution of the PDE 
\beqn
\ts {\pt  f\over \pt t}\;\;=\;\;Af&=&  \ts a(x)\,{\pt f\over \pt x}\;+\;i\,{b^2(x)\over 2}\,{\pt^2 f\over \pt x^2}  \pI  \lbeq{A.83}
\eeqn
with initial condition
\beqn
f(x,0)&=&g(x) \;\;.
\eeqn

\bigskip
\bigskip
\noindent Finally, let us recall that in general Wiener or Fresnel expectations can be calculated through the following 

\bigskip
\bigskip
\noindent{\bf Theorem A4:} Consider $m$ times $0<t_1<t_2<\cdots<t_m\le T$ and let $x_{t_j}$ be a standard or Fresnel BM observed at time $t_j$. Let 
\beqn
F&=&F(x_{t_1},\cdots,x_{t_m}) 
\eeqn
be an arbitrary function of $m$ variables and let $\,\E[F]=\E_0^T[F]\,$ denote its Wiener or Fresnel expectation value. Then, with $t_0:=0$ and $x_0:=0$, 
\beqn
\E[\,F\,]&=& \int_{\R^m}  F(x_{t_1},\cdots,x_{t_m}) \;\pro_{j=1}^m p_{t_j-t_{j-1}}(x_{t_{j-1}},x_{t_j})\,dx_{t_j}  \pI \lbeq{A.86}
\eeqn

\medskip
with Gaussian or Fresnel kernels given by 
\medskip
\beqn
p_{\tau}(x,y)&:=&\begin{cases} \ts {1\over \sqrt{2\pi\tau}} \; e^{\,-\,{(x-y)^2\over 2\tau}}  &{\rm for\;Wiener\;expectations} \\ 
  \ts {1\over \sqrt{2\pi i\tau}} \; e^{\,i\,{(x-y)^2\over 2\tau}}  &{\rm for\,\;Fresnel\;expectations\,.}   \end{cases} 
\eeqn

\newpage

\noindent{\bf\large References} 

{\small


\bigskip
\bigskip
\begin{itemize}
\item[{\rm[1]} ] Brian C.~Hall, {\sl Holomorphic Methods in Analysis and Mathematical Physics}, in {\sl First Summer School in Analysis and Mathematical 
  Physics: Quantization, the Segal-Bargmann Transform and Semiclassical Analysis}, edited by Salvador Perez-Esteva and Carlos Villegas-Blas, AMS Series in 
  Contemporary Mathematics, Volume 260, 2000.
\item[{\rm[2]} ] Niels Benedikter, Marcello Porta and Benjamin Schlein, {\sl Effective Evolution Equations from Quantum Dynamics}, Springer Briefs in 
  Mathematical Physics 7, 2016.
\item[{\rm[3]} ] Niels Benedikter, Gustavo de Oliveira and Benjamin Schlein, {\sl Quantitative Derivation of the Gross-Pitaevskii Equation}, 
  Communications on Pure and Applied Mathematics, Volume 68, Issue 8, p.1399-1482, August 2015. 
\item[{\rm[4]} ] Peter Pickl, {\sl Derivation of the Time Dependent Gross-Pitaevskii Equation Without Positivity Condition on the Interaction}, Journal 
  of Statistical Physics, Volume 140, p.76-89, May 2010. 
\item[{\rm[5]} ] Peter Pickl, {\sl Derivation of the Time Dependent Gross-Pitaevskii Equation with External Fields}, Reviews in Mathematical Physics, 
  Volume 27, No.1, p.1550003, March 2015. 
\item[{\rm[6]} ] Maximilian Jeblick, Nikolai Leopold and Peter Pickl, {\sl Derivation of the Time Dependent Gross-Pitaevskii Equation in Two Dimensions}, 
  Communications in Mathematical Physics, Volume 372, p.1-69, 2019. 
\item[{\rm[7]} ] J.~Schachenmayer, A.J.~Daley and P.~Zoller, {\sl Atomic Matter-Wave Revivals with Definite Atom Number in an Optical Lattice}, 
  Physical Review A, Volume 83, p.043614, April 2011. 
\item[{\rm[8]} ] Shouryya Ray, Paula Ostmann, Lena Simon, Frank Grossmann and Walter T.~Strunz, {\sl Dyna\-mics of Interacting Bosons Using the Herman-Kluk 
  Semiclassical Initial Value Representation}, Journal of Physics A: Mathematical and Theoretical, Volume 49, Issue 16, p.165303, 2016. 
\item[{\rm[9]} ] S.~Raghavan, A.~Smerzi, S.~Fantoni and S.R.~Shenoy, {\sl Coherent Oscillations Between Two Weakly Coupled Bose-Einstein Condensates: 
  Josephson Effects, Pi-Oscillations and Macroscopic Quantum Self-Trapping}, Physical Review A, Volume 59, Nr.1, January 1999. 
\item[{\rm[10]} ] Jon Links, Angela Foerster, Arlei Prestes Tonel and Gilberto Santos, {\sl The Two-Site Bose-Hubbard Model}, Annales Henri Poincare, Volume 7, 
  p.1591-1600, 2006. 
\item[{\rm[11]} ] Christian Gross und Markus Oberthaler, {\sl Ultrakalte Quantenpendel}, Physik Journal 9, Seiten 29-34, Februar 2010. 
\item[{\rm[12]} ] Eva-Maria Graefe, Hans J\"urgen Korsch and Martin P.~Strzys, {\sl Bose-Hubbard Dimers, Viviani's Windows and Pendulum Dynamics}, Journal of 
  Physics A: Mathematical and Theoretical, Volume 47, p.085304, February 2014. 
\item[{\rm[13]} ] Lena Simon, {\sl Semiklassische Dynamik Ultrakalter Bose-Gase}, Dissertation unter der Anleitung von Walter T.~Strunz an der Technischen Universit\"at 
  Dresden, November 2012.
\item[{\rm[14]} ] Shmuel Fishman, Hagar Veksler, {\sl Semiclassical Analysis of Bose-Hubbard Dynamics}, New Journal of Physics, Volume 17, p.053030, May 2015.
\item[{\rm[15]} ] Alexandra Bakman, Shmuel Fishman and Hagar Veksler, {\sl Collapse and Revival for a Slightly Anharmonic Hamiltonian}, Physics Letters A, Volume 381, 
Issue 29, p.2298, 2017.
\item[{\rm[16]} ] Lena Simon and Walter T.~Strunz, {\sl Analytical Results for Josephson Dynamic of Ultracold Bosons}, Physical Review A 86, p.053625, November 2012.
\item[{\rm[17]} ] D.~Jaksch, C.~Bruder, J.I.~Cirac, C.W.~Gardiner and P.~Zoller, {\sl Cold Bosonic Atoms in Optical Lattices}, Physical Review Letters 81, 
  Issue 15, p.3108, October 1998.
\item[{\rm[18]} ] Markus Greiner, Olaf Mandel, Tilman Esslinger, Theodor W.~H\"ansch and Immanuel Bloch, {\sl Quantum Phase Transition from a Superfluid to a 
Mott Insulator in a Gas of Ultracold Atoms}, Nature, Volume 415, p.39, January 2002.
\item[{\rm[19]} ] Immanuel Bloch, Jean Dalibard and Wilhelm Zwerger, {\sl Many-Body Physics with Ultracold Gases}, Reviews of Modern Physics, Volume 80, p.885, July 2008.
\item[{\rm[20]} ] Maciej Lewenstein, Anna Sanpera and Veronica Ahufinger, {\sl Ultracold Atoms in Optical Lattices, Simulating Quantum Many-Body Systems}, 
Oxford University Press, 2012. 
\item[{\rm[21]} ] I.M.~Georgescu, S.~Ashhab, F. Nori, {\sl Quantum Simulation}, Reviews of Modern Physics, Volume 86, p.153, January 2014.
\item[{\rm[22]} ] William D.~Phillips, {\sl The Coldest Stuff in the Universe: From Quantum Clocks to Quantum Simulators}, Presentation given at the Annual Meeting 
of the Simons Collaboration on Localization of Waves, Slides and a passionate Video available at the Simons Foundation under \\
{\scriptsize \url{https://www.simonsfoundation.org/event/simons-collaboration-on-localization-of-waves-annual-meeting-2022}}
\item[{\rm[23]} ] Anatoli Polkovnikov, {\sl Quantum Corrections to the Dynamics of Interacting Bosons: Beyond the Truncated Wigner Approximation}, Physical Review A, Volume 68, 
  p.053604, November 2003. 
\item[{\rm[24]} ] Anatoli Polkovnikov, {\sl Evolution of the Macroscopically Entangled States in Optical Lattices}, Physical Review A, Volume 68, 
  p.033609, September 2003. 
\item[{\rm[25]} ] Anatoli Polkovnikov, {\sl Phase Space Representation of Quantum Dynamics}, Annals of Physics, Volume 325, p.1790-1852, 2010. 
\item[{\rm[26]} ] Anatoli Polkovnikov, Subir Sachdev and S.M.~Girvin, {\sl Nonequilibrium Gross-Pitaevskii Dynamics of Boson Lattice Models}, Physical Review A, Volume 66, 
  p.053607, November 2002.
\item[{\rm[27]} ] Shainen M.~Davidson, Dries Sels and Anatoli Polkovnikov, {\sl Semiclassical Approach to Dyna\-mics of Interacting Fermions}, Annals of Physics, Volume 384, 
  p.128-141, September 2017. 
\item[{\rm[28]} ] C.W.~Gardiner, {\sl Handbook of Stochastic Methods, for Physics, Chemistry and the Natural Sciences}, Springer Series in Synergetics, 2nd Edition, 1985. 
\item[{\rm[29]} ] C.W.~Gardiner and P.~Zoller, {\sl Quantum Noise, A Handbook of Markovian and Non-Markovian Quantum Stochastic Methods with Applications 
  to Quantum Optics}, Springer Series in Synergetics, Second Enlarged Edition, May 1999. 
\item[{\rm[30]} ] Wolfgang P.~Schleich, {\sl Quantum Optics in Phase Space}, Chapter 12: Phase Space Functions, Wiley-VCH Verlag Berlin GmbH, 2001.
\item[{\rm[31]} ] F.~Trimborn, D.~Witthaut and H.J.~Korsch, {\sl Exact Number-Conserving Phase-Space Dynamics of the M-Site Bose-Hubbard Model}, Physical Review A, 
  Volume 77, p.043631, April 2008. 
\item[{\rm[32]} ] C.W.~Gardiner and P.D.~Drummond, {\sl Ten Years of the Positive P-Representation}, in {\sl Recent Developments in Quantum Optics}, edited by R.~Inguva, 
  Plenum Press, New York, 1993. 
\item[{\rm[33]} ] A.~Gilchrist, C.W.~Gardiner and P.D.~Drummond, {\sl Positive P-Representation: Application and Validity}, Physical Review A, Volume 55, p.3014, April 1997. 
\item[{\rm[34]} ] S.~W\"uster, J.F.~Corney, J.M.~Rost and P.~Deuar, {\sl Quantum Dynamics of Long-Range Interac\-ting Systems Using the Positive-P and Gauge-P 
  Representations}, Physical Review E, Volume 96, p.013309, July 2017. 
\item[{\rm[35]} ] Detlef Lehmann, {\sl Pricing and Hedging in the Presence of Stochastic Volatility and Stochastic Interest Rates}, Chapter 18 of the Financial Mathematics 
  Lecture Notes at Hochschule RheinMain, available under \;{\scriptsize \url{http://hsrm-mathematik.de/WS201516/master/option-pricing/Chapter18.pdf}}
\item[{\rm[36]} ] Detlef Lehmann, {\sl Mathematical Methods of Many-Body Quantum Field Theory}, Chapman and Hall/CRC Research Notes in Mathematics Series 436, August 2004. 
\item[{\rm[37]} ] C.S.~Lam, {\sl Decomposition of Time-Ordered Products and Path-Ordered Exponentials}, Journal of Mathematical Physics, Volume 39, Issue 10, p.5543, 1998.
\item[{\rm[38]} ] Masuo Suzuki, {\sl Decomposition Formulas of Exponential Operators and Lie Exponentials with some Applications to Quantum Mechanics and 
  Statistical Physics}, Journal of Mathematical Physics, Volume 26, p.601, April 1985. 
\item[{\rm[39]} ] P.L.~Giscard, S.J.~Thwaite and D.~Jaksch, {\sl Evaluating Matrix Functions by Resummations on Graphs: The Method of Path-Sums}, SIAM Journal on 
  Matrix Analysis and Applications, Volume 34, Issue 2, p.445-469, 2013. 
\item[{\rm[40]} ] Alexander Van-Brunt, Matt Visser, {\sl Explicit Baker-Campbell-Hausdorff Expansions}, Mathematics, Volume 6, Issue 8, p.135, August 2018. 
\item[{\rm[41]} ] Krzysztof Kowalski and Willi-Hans Steeb, {\sl Nonlinear Dynamical Systems and Carleman Li\-nearization}, World Scientific Publishing Company, 1991. 
\item[{\rm[42]} ] Bernt Oeksendal, {\sl Stochastic Differential Equations, An Introduction with Applications}, Fifth Edition, Corrected Printing, Springer-Verlag, May 2000.

\end{itemize}


}

\vfill

\end{document}